\documentclass[a4paper,10pt]{elsarticle}
\usepackage[english]{babel}
\usepackage[nottoc]{tocbibind}

\usepackage{graphicx}  
\usepackage{hyperref}
\usepackage{sidecap}
\usepackage{amsmath}
\usepackage{amssymb}
\usepackage{xcolor}

\begin{document}
\title{The ANTARES detector: two decades of neutrino searches in the Mediterranean Sea}
\author[IPHC,UHA]{A.~Albert}
\author[IFIC]{S.~Alves}
\author[UPC]{M.~Andr\'e}
\author[UPV]{M.~Ardid}
\author[UPV]{S.~Ardid}
\author[CPPM]{J.-J.~Aubert}
\author[APC]{J.~Aublin}
\author[APC]{B.~Baret}
\author[LAM]{S.~Basa}
\author[APC]{Y.~Becherini}
% ...

\address[IPHC]{\scriptsize{Universit\'e de Strasbourg, CNRS,  IPHC UMR 7178, F-67000 Strasbourg, France}}
\address[UHA]{\scriptsize Universit\'e de Haute Alsace, F-68100 Mulhouse, France}
\address[IFIC]{\scriptsize{IFIC - Instituto de F\'isica Corpuscular (CSIC - Universitat de Val\`encia) c/ Catedr\'atico Jos\'e Beltr\'an, 2 E-46980 Paterna, Valencia, Spain}}
\address[UPC]{\scriptsize{Technical University of Catalonia, Laboratory of Applied Bioacoustics, Rambla Exposici\'o, 08800 Vilanova i la Geltr\'u, Barcelona, Spain}}
\address[UPV]{\scriptsize{Institut d'Investigaci\'o per a la Gesti\'o Integrada de les Zones Costaneres (IGIC) - Universitat Polit\`ecnica de Val\`encia. C/  Paranimf 1, 46730 Gandia, Spain}}
\address[CPPM]{\scriptsize{Aix Marseille Univ, CNRS/IN2P3, CPPM, Marseille, France}}
\address[APC]{\scriptsize{Universit\'e Paris Cit\'e, CNRS, Astroparticule et Cosmologie, F-75013 Paris, France}}
\address[LAM]{\scriptsize{Aix Marseille Univ, CNRS, CNES, LAM, Marseille, France }}
% ...
\cortext[cor1]{maurizio.spurio@unibo.it}
\fntext[fn1]{Deceased}


\begin{abstract}
Interest for studying cosmic neutrinos using deep-sea detectors has increased after the discovery of a diffuse flux of cosmic neutrinos by the IceCube collaboration and the possibility of wider multi-messenger studies with the observations of gravitational waves.
The ANTARES detector was the first neutrino telescope in seawater, operating successfully in the Mediterranean Sea for more than a decade and a half. All challenges related to the operation in the deep sea were accurately addressed by the collaboration. 
Deployment and connection operations became smoother over time; data taking and constant re-calibration of the detector due to the variable environmental conditions were fully automated.
A wealth of results on the subject of astroparticle physics, particle physics and multi-messenger astronomy have been obtained, despite the relative modest size of the detector, paving the way to a new generation of larger undersea detectors.
This review summarizes the efforts by the ANTARES collaboration that made the possibility to operate neutrino telescopes in seawater a reality and the results obtained in this endeavor.
\end{abstract}
\maketitle
\bigskip
\tableofcontents


\section{Introduction\label{sec:intro}}
The neutrino interaction cross section is very small: this allows them to escape dense astrophysical regions and reach the Solar system also from cosmological distances. 
Neutrinos are  electrically neutral, and are not deflected by galactic and extragalactic magnetic fields.
On the other hand, the small interaction probability is a drawback, as their detection requires a large target mass. 

The idea of a neutrino telescope based on the detection of the secondary particles produced in neutrino interactions was first formulated in  1960 by M. Markov \cite{Markov:1960}. 
He proposed \textit{to install detectors deep in a lake or  in the sea and to determine the direction of the charged particles with the help of  Cherenkov radiation}. 

Since the first idea of Markov, it took about 15 years until the scientific community started the first R\&D efforts (see \S  \ref{sec:early}).
The long evolution toward detectors with enough sensitivity to be able to make observations started in the seventies, with the pioneering works in the Pacific Ocean close to Hawaii and in Lake Baikal in Siberia, followed by the beginning of investigation in the South Pole. Refer to \cite{Spiering:12} for a full review of the main milestones.

After a long qualification campaign, the ANTARES (\textit{Astronomy with a Neutrino Telescope and Abyss environmental RESearch}) detector was the first operational neutrino telescope deployed in seawater, starting data taking in 2006.
The main reason why it took such a long time to develop these experiments is that abyssal seawater constitutes a very hostile environment. 
First, salt water is a conductor and is highly corrosive for most materials, so that a suitable choice must be made for all offshore components of the detector, since a long operation time is needed to collect large data samples.
Second, the availability of vessels and underwater vehicles to perform deep-sea operations is necessary. 
Furthermore, since the detector elements must be attached to flexible lines anchored to the seabed, these will be constantly moving by the sea currents and the position of the individual detector elements must be continuously monitored. 
Finally, to construct and operate an offshore detector, a complex architecture must be implemented as well as an efficient Quality Assurance/Quality Control system. 
The infrastructure required to power and control the apparatus includes the onshore buildings to house the electronics for monitoring and data acquisition, a main electro-optical cable providing the electrical power and the data link between the detector and the shore, at least one main junction box (and possibly secondaries junction boxes), the interlink cables to distribute the power and the optical fibers to the detector units, and additional instruments to perform environmental measurements.

On the other hand, deep seawater is highly transparent and homogeneous, which is ideal for detecting high-energy neutrinos. 
At sufficient depth the overlying layer of water not only represents a very effective shield against daylight but it also absorbs significantly atmospheric muons: 
indeed at a depth of $\sim~1$ km under the water, the flux of atmospheric muons is reduced by more than 5 orders of magnitude (see  \S \ref{sec:atmunu}).

\subsection{Early projects in seawater\label{sec:early}}
The history of neutrino telescopes started in 1976 with the project for the construction of the \textit{Deep Underwater Muon And Neutrino Detection} (DUMAND), to be placed at a 4800 m depth in the Pacific Ocean off Keahole Point, on the Big Island of Hawaii \cite{Roberts.92}. A prototype vertical string of instruments suspended from a special ship was employed to demonstrate the technology, and to measure the cosmic ray muon flux at  various depths (2000--4000 m, in steps of 500 m) in deep ocean \cite{DUMAND:90}. 
A major operation took place in December 1993,  with a partial success. 
However, in 1995 the U.S. DOE canceled further efforts on DUMAND.

The DUMAND initiative provided valuable experience for all following undersea neutrino telescopes.
Some reasons for the long DUMAND development time were the: $i)$ huge depth of the chosen site; $ii)$ lack of advanced optical-fiber technology for data transmission; $iii)$ lack of reliable pressure-resistant underwater connectors; $iv)$ lack of remotely operated vehicles for underwater connections; $v)$ limited funding. These aspects were very instructive for future projects.
In addition,  theoretical predictions on astrophysical neutrino fluxes were not reliable. Pessimistic and optimistic estimates differed by 2--3 orders of magnitude, predicting very different energy spectra, thus making  the choice of the detector configuration difficult \cite{Roberts.92}. 

The neutrino telescope must be close to scientific and logistic infrastructures onshore, far enough from shelf-breaks, in a position that  minimizes the cost of operations for,  e.g., the average conditions of sea waves and winds. Thus, in general, Pacific Ocean environment is much more hostile than the Mediterranean Sea. 
The latter was started to be considered an optimal location.
In 1991, a Greek/Russian collaboration performed a first test at a depth of 4100 m at a site close to Pylos, on the West coast of the Peloponnese, effectively starting  the NESTOR (\textit{Neutrino Extended Submarine Telescope with Oceanographic Research}) project \cite{NESTOR:05}. 
In a short time, the collaboration grew with different  European and U.S. institutions. 
The NESTOR design consisted of seven {towers}, six on the edges of a hexagon and one in the center. 
In  2003, the NESTOR  collaboration successfully deployed  a test floor of one detector tower, fully equipped with 12 optical modules, final electronics and associated environmental sensors. The  detector  worked for more than a month, but its operation terminated due to a failure of the cable to shore, which caused the end of the NESTOR history.

The \textit{NEutrino Mediterranean Observatory} (NEMO) was a project of the Italian research institute for nuclear and sub-nuclear physics (\textit{INFN}) \cite{NEMO:08}. 
The activities focused on: the search and characterization of an optimal site for the detector installation with more than 20 sea campaigns in the Mediterranean Sea;  on the development of key technologies for a km$^3$-scale underwater telescope; and on a feasibility study for such a detector, which included the analysis of all construction and installation issues and optimization of the detector geometry by means of numerical simulations. 
A deep site  with proper features in terms of  depth and water optical properties  was  identified at  a  depth of  3500 m  about 85  km offshore from Capo Passero at the East coast of Sicily: this is today one of the sites for the KM3NeT detector \cite{KM3NeT:LoI}.

\subsection{ANTARES: the first operational telescope in the sea\label{sec:ANTAintro}}

The ANTARES collaboration was created in 1996 by French institutes devoted to high energy physics, astrophysics, and sea science. The initial activity was mainly focused on site exploration and characterization with autonomous lines, with more than 30 deployments between 1996 and 1999. The first proposal for a demonstration line towards a large detector \cite{ANTprop:97} was signed by groups from France, Spain, and U.K. and the activities reported in \cite{ANTARES:01}.  
The demonstrator line was deployed and connected to an existing electro-optical cable offshore Marseille and operated from November 1999 to June 2000. 
The final proposal for a detector consisting of $\sim$1000 photomultipliers on 13 strings \cite{ANTprop:99} also included institutes from the Netherlands and Russia. Italian and German institutes joined in 2000 and 2003, respectively.
The construction for the 0.02 km$^3$-scale detector started in 2006 and was completed in 2008.

The expected duration of the experiment was 10 years, with decommissioning of the apparatus foreseen in 2016. However, the observation of the first gravitational wave event in 2015 (GW150914) modified the scenario. 
At that time, the ANTARES detector was still the largest neutrino telescope in the Northern hemisphere. The possibility of multi-messenger observations of transient astrophysical events had suddenly increased the interest of the scientific community. 
Detectors like neutrino telescopes with a high duty cycle and very wide sky coverage have a key role in this field.
This motivated the extension of the activities until 2022, when the KM3NeT detector reached an equivalent photocathode area to the ANTARES one.

In this review, after introducing the neutrino detection principle, with its advantages and drawbacks with respect to other astrophysical probes \S~\ref{sec:detection}, the first operating and successful seawater neutrino telescope, the ANTARES detector, is described.
The main features of the detector, which allowed to overcome all the challenges linked to an installation in such a remote location, are detailed in \S~\ref{sec:antares}. The document includes also a brief description of the measurements performed on the properties of the Mediterranean Sea water \S~\ref{sec:water}, a description of the ANTARES software \S~\ref{sec:soft}, 
the acoustic systems  and other oceanographic instruments used both for neutrino physics and sea science \S~\ref{sec:ESS}, a summary of the main results on neutrino physics studies and searches for exotic particles \S~\ref{sec:ppresults}, an outline of the contributions to astroparticle physics \S~\ref{sec:apresults}, and on multi-messenger astrophysics \S~\ref{sec:MM}, and finally the legacy of the ANTARES experiment for future seawater experiments in the world \S~\ref{sec:legacy}.
These ANTARES activities have been detailed in more than 100 papers published in peer-reviewed journals.

\section{Cosmic neutrino observation \label{sec:detection}} 
\begin{SCfigure}
  \centering
\includegraphics[width=0.65\textwidth]%
    {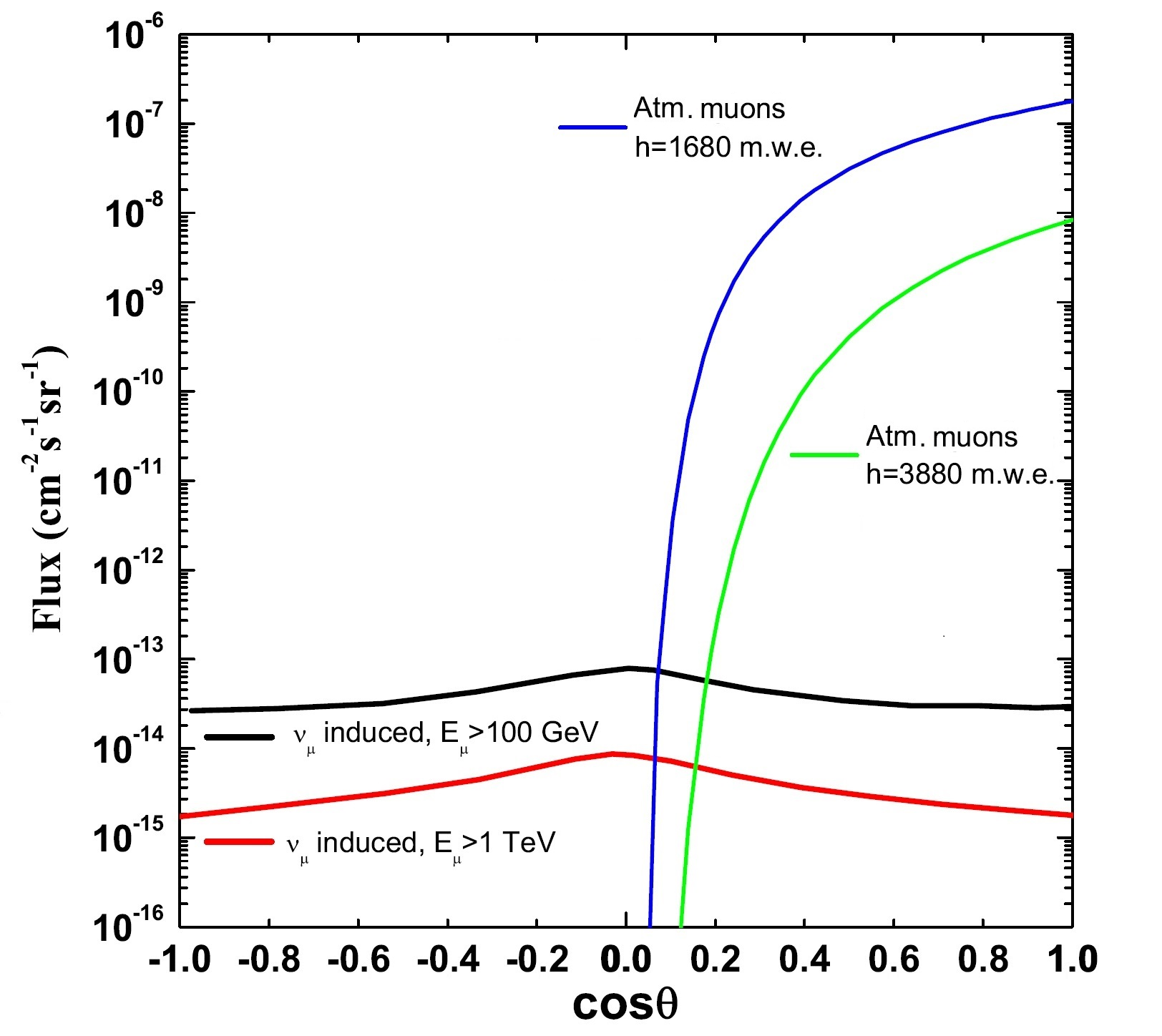}
  \caption{\small Different contributions (as a function of the cosine of the zenith angle $\theta$) of: $i)$ atmospheric muons (computed according to \cite{PARA:06}) for two different depths (m.w.e. is meters of water equivalent); ii) muons induced by atmospheric neutrinos (from \cite{Bartol}), for two different muon energy thresholds.}
    \label{fig:atmunu}
\end{SCfigure}

The basic idea for a neutrino telescope is to build a matrix of light detectors inside a transparent medium. This medium, such as deep ice or water:
\begin{itemize}
\item offers a large volume of free targets for neutrino interactions;
\item provides shielding against secondary particles produced by cosmic rays;
\item allows transmission of Cherenkov photons induced in the medium by relativistic charged particles produced by the neutrino interaction.
\end{itemize}
Cosmic neutrino detectors are not background-free. Showers induced by cosmic ray interactions  with the Earth's atmosphere produce {\it atmospheric muons} and {\it atmospheric neutrinos}. Muons can penetrate the atmosphere and up to several kilometers of ice/water. As a consequence, neutrino detectors must be located deeply under a large amount of shielding in order to reduce this background. 
The flux of downward-going atmospheric muons exceeds that induced by atmospheric neutrino interactions by many orders of magnitude, decreasing with increasing detector depth, as shown in Fig. \ref{fig:atmunu}.

Charged particles induced by a neutrino interaction travel through the medium until they either decay or interact. The mean distance traveled by each particle is the \textit{path length}, and depends on its energy loss in the medium.
In a high-energy neutrino detector, one can distinguish between two main event topologies:  events with a \textit{track}, and events with a  \textit{cascade} (or \textit{shower}) of  particles when the  trajectories of individual particles are too short to be resolved.
\textcolor[rgb]{0,0,0}{However, the longitudinal development of the cascade could sometimes be long enough to provide information about neutrino direction.}
Relativistic charged particles induce Cherenkov radiation when passing through a transparent medium. 
A three-dimensional array of photo-detectors (usually  photomultiplier tubes, PMTs) embedded in this medium can sample this Cherenkov radiation. From the number of detected photons and their arrival time, some of the properties of the neutrino (flavor, direction, energy) can be inferred \cite{Spiering:12}. 

The role of neutrino telescopes and of cosmic neutrino observations in the context of  multi-messenger astrophysics and of particle physics is widely discussed in \cite{MS:18,Gaisser:16,Vissani:22}.

\subsection{Neutrino interactions}
\begin{figure}[th]
  \centering
\includegraphics[width=0.95\textwidth]%
    {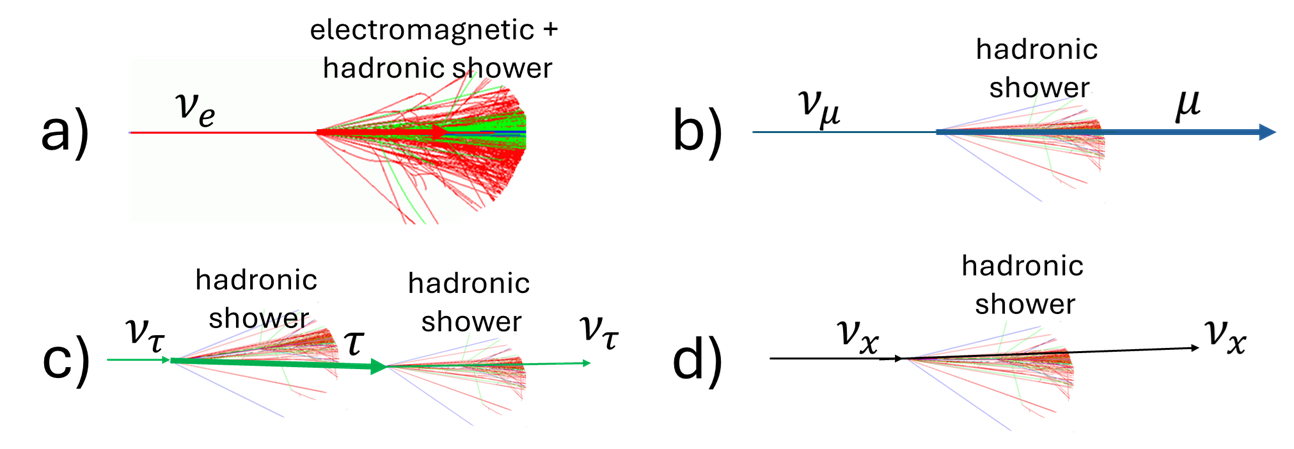}
  \caption{\small Main event signature topologies for different neutrino flavors and interactions: 
  a) CC interaction of $\nu_e$ that produces both an electromagnetic and a hadronic shower;
 b) CC interaction of a $\nu_\mu$ that produces a muon and a hadronic shower; c) CC interaction of a $\nu_\tau$ that produces a tau that subsequently decays; d) a NC interaction produces a hadronic shower. Particles and antiparticles cannot be distinguished in neutrino telescopes.}
    \label{fig:topo}
\end{figure}

Neutrino and antineutrino interaction in a detector without magnetic field, such as ANTARES, are not distinguishable. Thus, in the following, neutrino will be used to refer both to neutrinos and antineutrinos.
A high-energy neutrino  of flavor $l=e,\mu,\tau$ interacts with a nucleon $N$ of the nucleus, via either charged current (CC) weak interaction
\begin{equation}
\nu_l + N \rightarrow l+X \ ,
\label{cc}\end{equation}
\noindent where $l$ is the charged lepton that conserves the leptonic number, and $X$ the final state hadronic system, or neutral current  (NC) weak interaction
\begin{equation}
\nu_l + N \rightarrow \nu_l +X \ .
\label{nc}\end{equation}
Schematic views of a $\nu_e, \nu_\mu$ and $\nu_\tau$ CC interactions and of a NC interaction are shown in Fig. \ref{fig:topo}.

A high-energy electron resulting from a $\nu_e$  CC interaction has a high probability to radiate a photon via brems\-strah\-lung after a few tens of cm of water/ice (the water radiation length is $\sim~36$ cm). The development of an electromagnetic (EM) shower continues for a few meters until the energy of the constituents falls below the critical energy $E_c$.
The shower length in water for a 10 TeV electron is only $\sim$7.5 m; its lateral extension is of the order of tens of centimeters and therefore negligible compared to the longitudinal one.   
This EM size is small compared to the spacing of the PMTs in neutrino telescopes. 

\begin{SCfigure}
  \centering
\includegraphics[width=0.65\textwidth]%
    {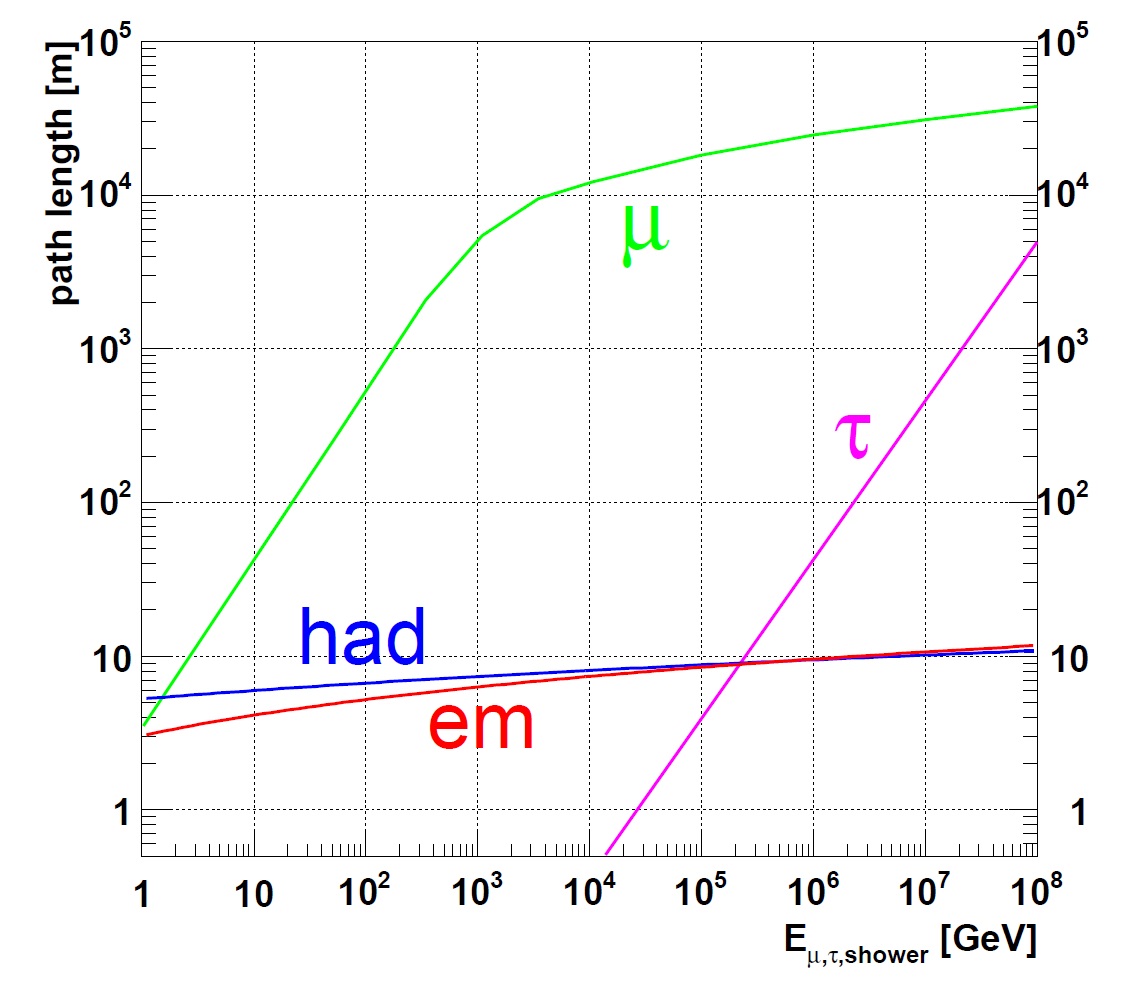}
  \caption{\small Path length of particles produced by neutrino interactions in water: muons ($\mu$), taus ($\tau$), electromagnetic (\textit{em}) and hadronic (\textit{had}) showers, as a function of their respective energy. Details on the computation are in \cite{hartmann:06}.}
    \label{fig:nupath}
\end{SCfigure}

In  $\nu_\mu$ CC  interactions,  only the muon track is detected in most cases, as the path length of a muon in water exceeds that of a shower by more than three orders of magnitude for energies above 2 TeV, see Fig. \ref{fig:nupath}. Therefore, such an event might very well be detected even if the interaction has taken place several kilometers outside the instrumented volume, provided that the muon traverses the detector. 
This gives a clean experimental signal which allows an accurate reconstruction of the muon direction, which is closely correlated with the neutrino direction.
The angular connection between the parent neutrino and the muon is essential for the concept of a neutrino telescope. Since neutrinos are not deflected by (extra)galactic magnetic fields, it is possible to trace the muon back to the neutrino source. This is equivalent to traditional astronomy where photons point back to their source.

Tau neutrinos are particularly important because they can essentially be only of cosmic origin. In $\nu_\tau$  CC interactions, the  $\tau$-lepton travels some distance (depending on its energy) before it decays and produces a second shower. 
In a large detector, below 1 PeV the  pattern produced by $\nu_\tau$ CC interaction yields a cascade (except when the tau decays into a muon).
Above 1 PeV, see Fig. \ref{fig:nupath}, the $\tau$ path could be long enough to distinguish between the primary interaction of the $\nu_\tau$ and the $\tau$ decay. 
The ANTARES detector was unfortunately too small to observe this topology.

The NC channel (eq. \ref{nc}) gives the same signature for all neutrino flavors. Here, a fraction of the interaction energy is always carried away unobserved by the outgoing neutrino, and therefore the uncertainty on the reconstructed energy of the primary neutrino increases accordingly. 
Even though EM and hadronic showers are in principle different from each other,  the $\nu_e$ CC and the $\nu_x$ NC channels are practically not distinguishable.

In order to behave as a neutrino \textit{telescope}, a neutrino \textit{detector} must be able to reconstruct as accurately as possible the direction of the incoming neutrino and extract a possible signal excess over the background.
Thus, neutrino telescopes must have the same capability as GeV-TeV $\gamma$-ray experiments to detect new sources or to associate signals to objects already known in other electromagnetic bands.
An angular resolution of a fraction of degree can only be achieved by  $\nu_\mu$ CC interactions. Directional information obtained using other flavors, or using NC interactions, are usually so poor that there is no possibility to identify sources or to perform associations. 
\textcolor[rgb]{0,0,0}{A case of the use of cascades for studying extended regions by the IceCube collaboration is presented in \S \ref{sec:agaldiffuse}.
}

However, a high-energy neutrino detector is also motivated by discovery and must be designed to reveal neutrinos of all flavors over a wide energy range and with the best energy resolution. 
Propagation effects must be considered: neutrino oscillations change the source admixture from $\nu_e : \nu_\mu : \nu_\tau = 1: 2 : 0$ ratio in the standard neutrino production scenario to $\nu_e : \nu_\mu : \nu_\tau= 1 : 1 : 1$   when reaching the Earth \cite{Beacom:03,Nuratio:06}.

\subsection{Neutrino absorption from the Earth}
In the golden channel of CC $\nu_\mu$ interactions, and contrary to what happens for electromagnetic radiation observatories,  neutrino telescopes are "looking downward". Indeed, upward-going muons can only be produced by interactions of upward-going neutrinos. From the bottom hemisphere, the neutrino signal is thus almost background-free.
Only atmospheric neutrinos may have traversed the Earth, and represent an irreducible background for the study of cosmic neutrinos.
The rejection of this background depends upon the accuracy of the telescope in the track reconstruction and upon the possibility to estimate the event energy.

The neutrino nucleon interaction cross section increases with increasing neutrino energy $E_\nu$ and at high energy the Earth is not transparent anymore. 
For instance, 50\% of the incoming neutrinos interact in the Earth diameter when $E_\nu = 40$ TeV. 
At PeV energies, only neutrinos going downwards or in the near-horizontal direction can arrive at the detector site.
Ultra high-energy downward-going neutrino candidates can be extracted from the atmospheric muon background with the requirement that the interaction vertex is contained within the instrumented  volume of the medium, i.e. without any signal on the PMTs located on the top or sides of the detector. 
The peripheral layers of large volume telescopes, as the IceCube or KM3NeT detectors, can be used for vetoing downward-going atmospheric events \cite{VetoGaisser:14}. 
This veto technique relies on the fact that a high-energy atmospheric neutrino has a large probability of being accompanied by a downward-going atmospheric muon produced in the same cascade. 
The ANTARES telescope was too small in size to use this technique to select cosmic neutrinos from the upper hemisphere. 

\section{The ANTARES detector\label{sec:antares}}
\begin{figure}
  \centering
\includegraphics[width=0.95\textwidth]%
    {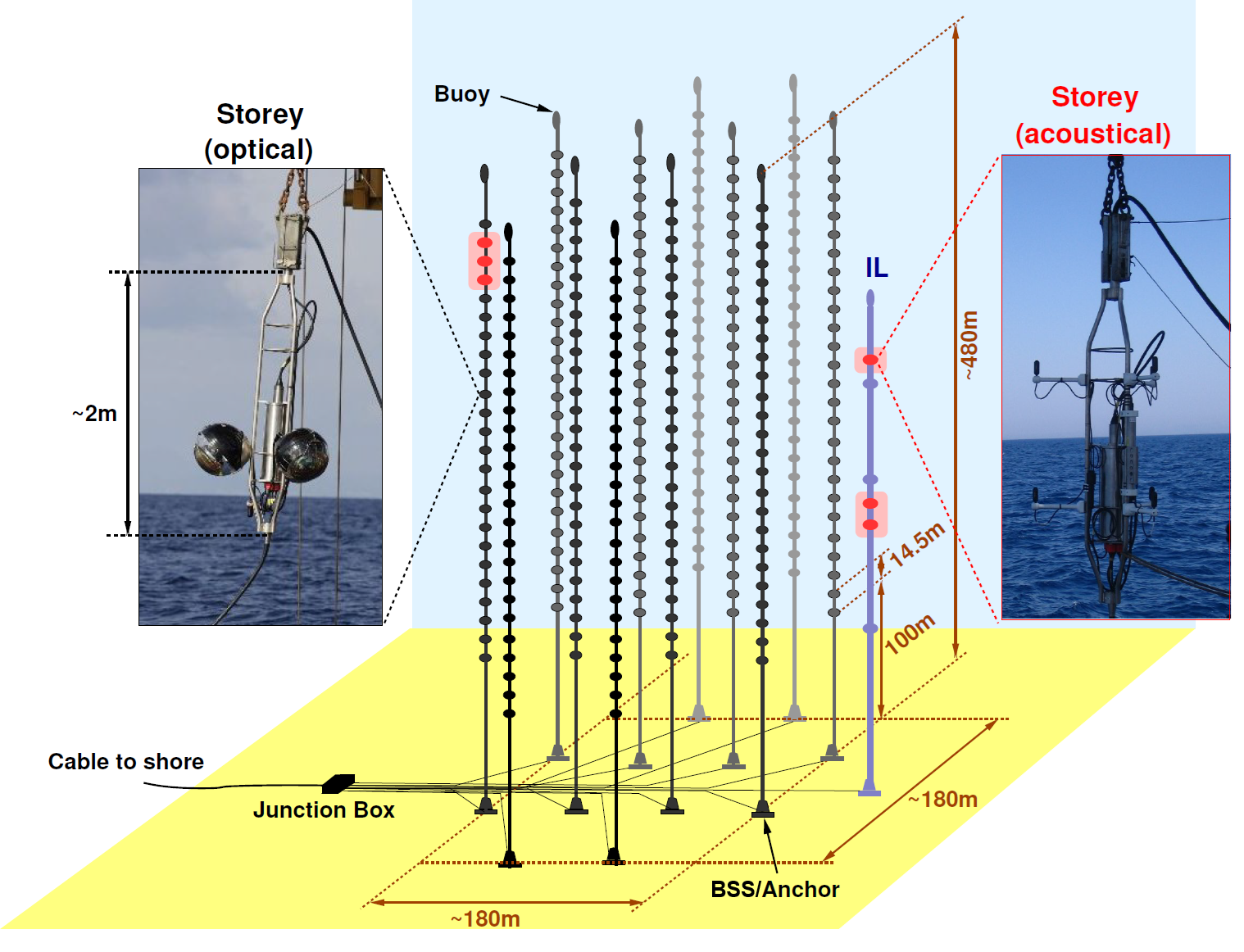}
  \caption{\small Schematic view of the ANTARES detector with 12 lines at a distance of about 60 m between them. The distance between consecutive optical modules is about 14.5 m. The main components described in the text are shown.  The inset on the left shows a standard \textit{storey} with three optical modules, the inset on the right an acoustic storey of the AMADEUS system. The locations of the six acoustic storeys are highlighted. }
      \label{fig:ANTARES}
\end{figure}

After an R\&D phase from about 1998 to 2005 \cite{ANTARES:09}, the first ANTARES line \cite{ANTARES:10} was deployed on February 14th, 2006 in a site located 40~km offshore from Toulon (France) and anchored at 2475~m depth. 
The detector was completed on May 29th, 2008 and took data until  February 12th, 2022, making it for a long time the largest neutrino telescope in the Northern hemisphere and the first to operate in the deep  sea \cite{ANTARES:18}. 
The telescope infrastructure consisted of 12 flexible detector lines, holding  photomultiplier tubes in glass spheres.
The lines were arranged on the seabed in an octagonal configuration, as illustrated in Fig. \ref{fig:ANTARES}. 
A detection line was the assembly of a Base Support Structure sitting on the seabed, and 25 \textit{storeys} each spaced by 14.5 m  linked by electro-optical mechanical cable segments. A top buoy completed the arrangement.
 The {storey} (inset of Fig. \ref{fig:ANTARES})  is the assembly of a mechanical structure with a frame supporting three optical modules looking downwards at 45$^\circ$, and a titanium container, the Local Control Module, housing the offshore electronics and embedded processors \cite{ANTARES:12}. 
To limit the number of single point failures, each line was divided in 5 independent sectors for power distribution and the data transmission.
ANTARES comprised a 13th line, dubbed Instrumentation Line (IL in Fig. \ref{fig:ANTARES}) \cite{ANTARES:06}, which was equipped with instruments for monitoring the environment with different configurations along the operational period, \S \ref{sec:ESS}.
In its configuration from 2007 until its decommissioning in 2014, the IL held six storeys where for two pairs of consecutive storeys, the vertical distance was increased to 80 m.

\begin{figure}
  \centering
\includegraphics[width=0.95\textwidth]%
    {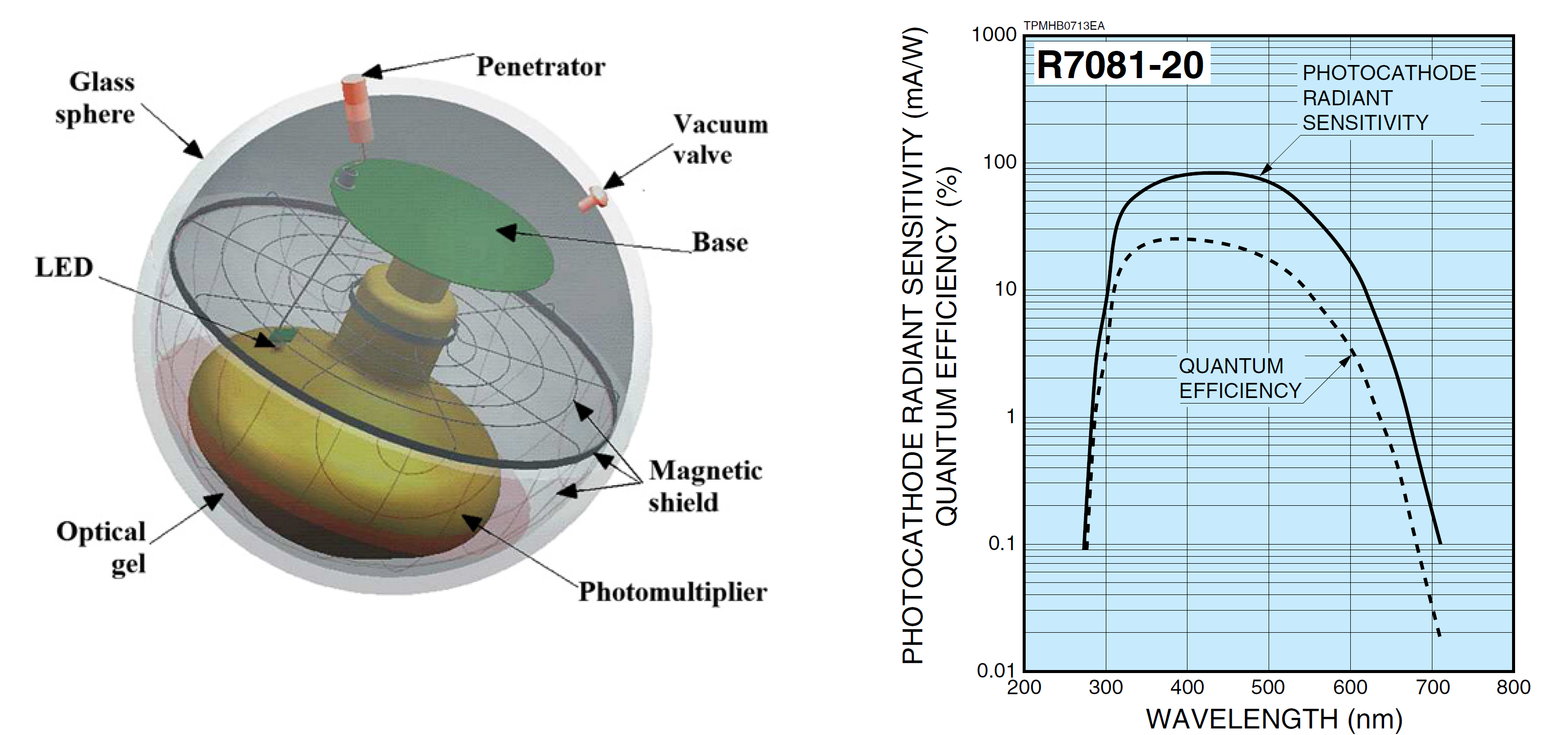}    
  \caption{\small Left:  Sketch of an ANTARES optical module \cite{ANTARES:18}. The large hemispherical (10$^{\prime\prime}$ in diameter) PMT is protected by a pressure-resistant glass sphere. The outer diameter of the sphere is 43.2 cm. A mu-metal cage protects the PMT from the Earth’s magnetic field. An internal LED is used for the calibration. {Right: the quantum efficiency (dashed line) versus wavelength for the PMTs used in ANTARES (from Hamamatsu).}}
  \label{fig:OM}
\end{figure}

The key element of the detector was the Optical Module (OM) \cite{ANTARES:02}, Fig. \ref{fig:OM}.
The protective envelope was a glass sphere of 17$^{\prime\prime}$ diameter that contained a 10$^{\prime\prime}$ hemispherical PMT from Hamamatsu of the type R7081-20 \cite{ANTARES:05}, the interfacing optical gel, a magnetic shield cage, an internal LED, the HV power supply, and a cabled link with the electronics container. In total, 885 OMs were installed. 
A further part of the ANTARES detector was the AMADEUS system, designed for the investigation of techniques for acoustic detection of neutrinos in the deep sea, \S \ref{sec:amadeus}. 

The onshore building that housed the electronics for monitoring and data acquisition, and the PC farm to process data, was located at the Institute Michel Pacha, at La Seyne-sur-Mer.  Although there was the possibility for shifters to be hosted in such a fascinating building facing the beautiful Bay of Toulon, full control of the detector was for a significant fraction of the time performed remotely from all institutes participating in the experiment by means of a remote controlling software. 
Other infrastructures required to power and control the offshore detector were  the 40 km long main electro-optical cable and a network of submarine connections, including a junction box and interlink cables, to distribute power and  optical fibers to the 12 detector lines.
Devices for positioning and timing calibrations were essential components as well. 

\subsection{Construction and sea operations\label{sec:construction}}
The construction of the ANTARES detector started in 2001 with the installation of the long distance electro-optical cable followed by the deployment of the underwater junction box in 2002. 
Several prototype lines were then deployed and operated in situ, allowing the validation and optimization of the detector design. 
The first detection line was installed in early 2006 and the last two lines of the apparatus were put into operation in May 2008. 
The production, integration, and tests of optical modules and of the different mechanical parts and electronics boards were performed by a number of laboratories in different countries in a European-wide effort. 
This comprises the optical module frames, the local control modules which host the readout electronics, the vertical electro-optical cables and the structures to anchor the lines to the seabed with the capability of a recovery. 
 Details on the construction and integration of detector components are available in \cite{ANTARES:18}. 

\begin{figure}[tbh]
  \centering
\includegraphics[width=0.98\textwidth]%
    {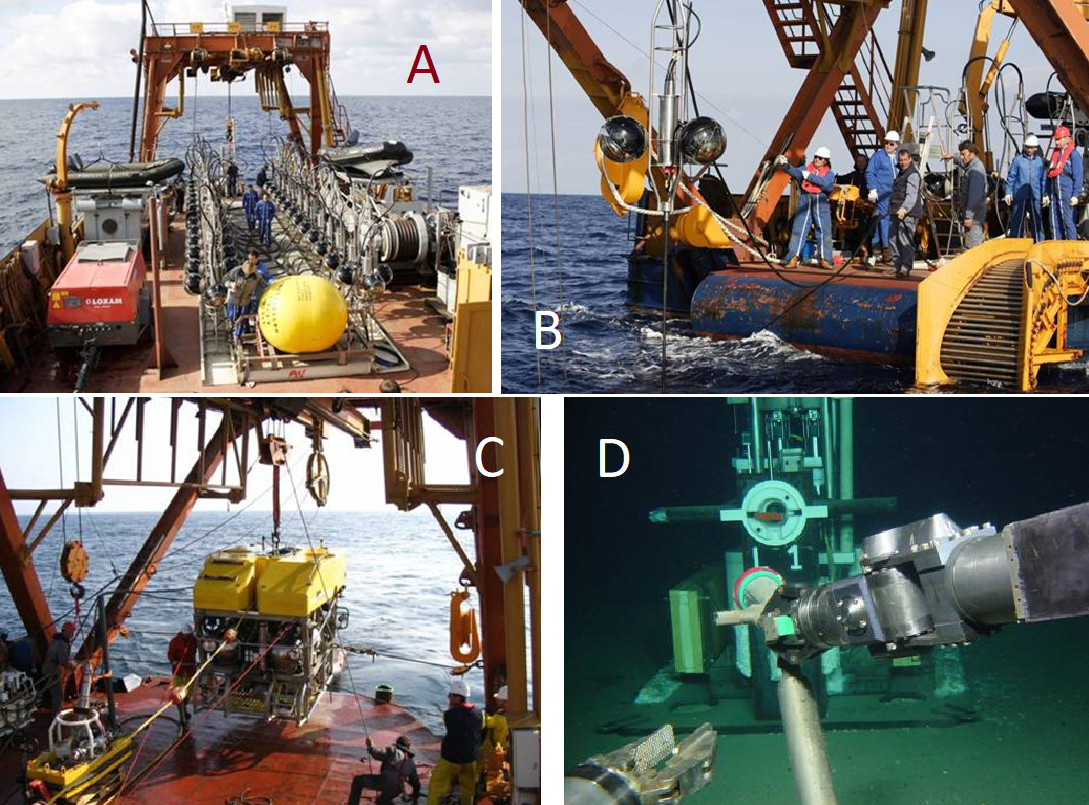}
  \caption{\small Four different stages of the deployment and connection of one detector line. See text for the description.}
    \label{fig:ope}
\end{figure}

The final and most delicate step of the installation of a detector line was its full integration onshore, and the subsequent deployment at the deep sea site followed by the connection to the seabed infrastructure.
In the case of ANTARES, these final steps of detector integration started in a dedicated hangar at the Bregaillon port area of La Seyne-sur-Mer.
Here, before deployment, the storeys were arranged on wheeled carts, equipped with the optical modules and the instrumentation and moved onto a deployment pallet. 
An integrated line was arranged on a single pallet which was then installed on the deck of the ship for the deployment, see Fig. \ref{fig:ope}A.

For the deployment of all ANTARES lines, the  \textit{Castor 02} vessel of the Foselev Marine Company was used. 
After reaching the deployment site, the ship adjusted its position using its GPS-controlled dynamic positioning capabilities during the entire deployment operation.
First, the Base Support Structure (BSS) was lowered into the sea containing a dead weight to keep the line anchored to the sea floor, a power module to supply the complete line, a connection base to accommodate the connector of the interlink cable, the acoustic emitter/receiver used for relative positioning, and two lithium battery powered transponders with an acoustic release mechanism. 
Some of the BSS were equipped with additional calibration devices such as laser beacons or devices to  measure seawater properties.
Then, the 25 storeys were put into the water (Fig. \ref{fig:ope}B)  until the top buoy of the line. 
In preparation for the detector installation, an array of acoustic beacons were placed and calibrated around the detector site \cite{ANTARES:26}. 
The exchange of acoustic signals between these beacons, the ship and the transponders mounted on the BSS were used to ensure a precise placement of the BSS on the target location. 
This procedure allowed the positioning of the lines within a few meters from their planned positions.

 The subsequent actions required the use of an underwater vehicle.
Almost all connection operations were performed by means of the remote operated vehicle (ROV) \textit{Victor} of IFREMER 
(Fig.~\ref{fig:ope}C).
The connection between the junction box (JB) and the lines was made with electro-optical cables of suitable length (from 120 to 350 m), equipped with a wet-mateable connector at each end,  Fig. \ref{fig:ope}D.  
These interlink cables were prepared on turrets deposited on the seabed. The ROV moved the turret close to the JB  and connected one end of the cable to a free output of the JB. 
Once a good connection was established at the JB, the ROV moved the turret towards the base of the line to be connected, while routing the cable on the seabed. Finally, the connection to the BSS of the line was performed. 
Each operation was monitored from the shore station and the electrical and optical connections verified.

Specific weather conditions were necessary to allow safe operations of the vessel and of the ROV. A swell of less than 1~m was required and the seabed conditions had also to be acceptable, since operation of the ROV becomes difficult when the sea current exceeds 10 cm/s. The ROV was also used for other tasks, such as inspections and tests of the outputs of the junction box, the survey of optical modules, and the installation of various sea science equipment. 

While no routine maintenance of the offshore apparatus was scheduled, the possibility of recovering lines was foreseen in case of severe functionality problems. 
During the ANTARES livetime, three detection lines were recovered, repaired, upgraded with new optical calibration devices, and successfully reinstalled. Further the instrumentation line, hosting supplementary sea science devices, was recovered in 2010, refurbished and redeployed in 2013.
Each recovery operation started with issuing a release command to the acoustic transponders in the BSS. 
Once the release had opened, the buoyancy of the buoy pulled out the active parts of the BSS from the dead weight thereby disconnecting the interlink cable. The whole detector line to be recovered rose to the surface in about one hour.

On February 12th, 2022 the detector was definitively switched off.
Due to the extended livetime with respect to the 10 years of the nominal running period of the detector, the lithium batteries on the BSS were no longer working, making the above-described line-recovery procedure impractical. An alternative procedure was used. In a first operation, the interlink cables were disconnected with submarine support. In two followup sea campaigns in May and June 2022, all the lines were recovered by hooking  each BSS to the deep sea cable of the \textit{Castor~02} vessel by means of the ROV. This method allowed to successfully recover the lines together with their dead weights. 

\subsection{DAQ, triggering and detector control\label{sec:DAQ}}
An essential attribute of the ANTARES infrastructure was the permanent connection to shore with high-bandwidth acquisition capacity. 
This allowed the Data Acquisition (DAQ) system to rely on the \textit{all-data to-shore} concept \cite{ANTARES:07}: all signals from the PMTs  passing a pre-set threshold (typically 0.3 single photoelectron, p.e.) were digitized in a custom built ASIC chip. 
The basic digitized information was a {\it hit}, also called {\it L0 hit}, consisting of a 24 bit time stamp of a 20~MHz global GPS-controlled clock system measuring the threshold crossing time, an 8 bit TVC value to provide a nanosecond timing information and an 8 bit AVC value which encoded the time integrated signal amplitude.  

Data transmission was done through a bidirectional optical fiber to a specific Local Control Module located every fifth storeys, equipped with an Ethernet switch which gathers the data of the {sector} of 5 storeys. 
In turn, the switch of each sector was connected via a pair of unidirectional fibers to a Dense Wavelength Division Multiplexing system situated at the bottom of each line. 
Then, this system was connected to the detector junction box on the seabed via the interlink cables,  and finally the data stream was gathered from the junction box onto the main Electrical-Optical Cable and sent to the shore station. 
Offshore, the data were packed into arrays of hits of predefined time frame duration of $\sim$100 ms.
The data were then sent to shore and there de-multiplexed. The data collected for the full detector for the same time frame were sent to a single data filter process in the onshore data processing farm. About 100 of such processes running in parallel were able to digest the complete data flow of the detector.
The data flow was between few Gb s$^{-1}$ to several tens of Gb s$^{-1}$, depending on the level of  bioluminescence.

The {data filter} \cite{ANTARES:07} extracted physics events from the data stream using fast algorithms. Each  algorithm  was based on a different trigger criterion, including a muon trigger, a directional trigger, 
a minimum bias trigger for monitoring the data quality, and dedicated triggers for multi-messenger investigations.
The seeds for a trigger were formed around {\it L1 hits}.
One L1 hit was either a coincidence of two L0 hits on neighboring PMTs in the same {storey} in a time window of 20 ns  or the occurrence of a L0 hit with a large amplitude ($> 3$ p.e.) in a single PMT. 
The muon trigger criteria consisted either in requesting a small number of L1 hits in adjacent or next-to-adjacent storeys within $\sim$70 ns or into a combination of a number of causally related L1 and L0 hits with respect to a muon track hypothesis.
The causality relation required that the  difference between the time of L1 hits must be smaller than the distance of the two involved {storeys}, divided by  the speed of light.
Directional triggers were used to maximize the detection efficiency of tracks coming from predefined directions, as for instance the direction of the Galactic center. 
The ANTARES onshore data processing system was linked to the Gamma-ray bursts Coordinates Network (GCN) \footnote{Now called General Coordinates Network \url{https://gcn.nasa.gov/}} \cite{ANTARES:19}. For each alert, all raw data were saved to disk for a predefined period of 2 min and events  with a relaxed causality condition were searched for.

One CPU processed a frame of $\sim$100 ms of raw data  with concurrent software triggers in about 500 ms. 
A typical physics event contained all hits in a time window of 2--4 $\mu$s, longer than the time needed for a muon to cross the whole apparatus.
The event selection reduced the data flow by a factor of  $\sim$$10^4$. The observed trigger rate was dominated by downward-going atmospheric muons and amounts to 5--10 Hz (depending on the trigger conditions). The standard trigger algorithms were able to operate with hit rates up to about 250 kHz per PMT. 
The filtered data were written to disk in ROOT \footnote{\url{https://root.cern/}} format  and backed up every night to the computer center in Lyon. 

The DAQ system involved about 300 data handling and 300 slow control offshore processes, plus 120 processes running on the onshore computers for data processing and filtering, monitoring and user interface. 
Monitoring programs allowed the operators to have a detailed view of the working conditions of the apparatus at a glance.
This was very important for an undersea apparatus since, depending on the optical background conditions, the operator had to choose the best data-taking configuration.

\subsection{Calibrations (positioning, time, charge)\label{sec:cali}}
Time and position calibrations are fundamental tasks for a seawater telescope whose detector lines  are continuously moving by  the sea currents and need to be flexible.

In the ANTARES experiment, an essential element to achieve the required time precision was a 20~MHz master clock system located onshore, which distributed a common reference time to all the offshore electronics. 
This system delivered a timestamp derived from GPS time via a optical fiber network from the shore station to the junction box, then to each line base, and finally to each Local Control Module (LCM) that managed a storey.

The time calibration of the detector was first performed during the construction of the lines, previous to their deployment, by means of a laser set up on the OMs. Once underwater, it was continually verified and adjusted during operation on a weekly basis. 
The master clock system measured the time delays between the shore station and the LCMs. 
The short delays between the electronics in the LCM and the photon arrival at the PMTs photocathode required further calibration. 
To perform this task, the in situ calibration used systems of external light sources: a series of LED beacons (four per line) and a laser beacon at the bottom of one of the central detector lines \cite{ANTARES:08}.  
In addition, an LED inside each optical module enabled to monitor changes in the transit time of the PMT. 
These devices  allowed to determine time offsets between PMTs belonging to the same line (intra-line calibration) or to different lines (inter-line calibration) \cite{ANTARES:16}. 
Furthermore,  the relative time offsets between PMTs at a precision level of 0.5 ns were obtained using the reconstructed trajectories of the downward-going atmospheric muons \cite{ANTARES:45}.

The  prerequisite for the ANTARES construction was to monitor the relative positions of OMs  with an accuracy better than 20 cm, corresponding to a 1 ns uncertainty on timing.
To attain such an accuracy, a constant monitoring with two independent systems was performed~\cite{ANTARES:26}:
$i)$ a high-frequency long baseline acoustic system, yielding the 3D position of 5 hydrophones placed along each line, using a triangulation method from emitters anchored in the base of the line plus autonomous transponders on the sea floor; 
$ii)$ a set of tiltmeter-compass sensors giving the local tilt angles (pitch and roll) of each storey with respect to the vertical line  as well as its orientation with respect to the Earth magnetic North (heading).
\begin{figure}
  \centering
\includegraphics[width=0.99\textwidth]%
    {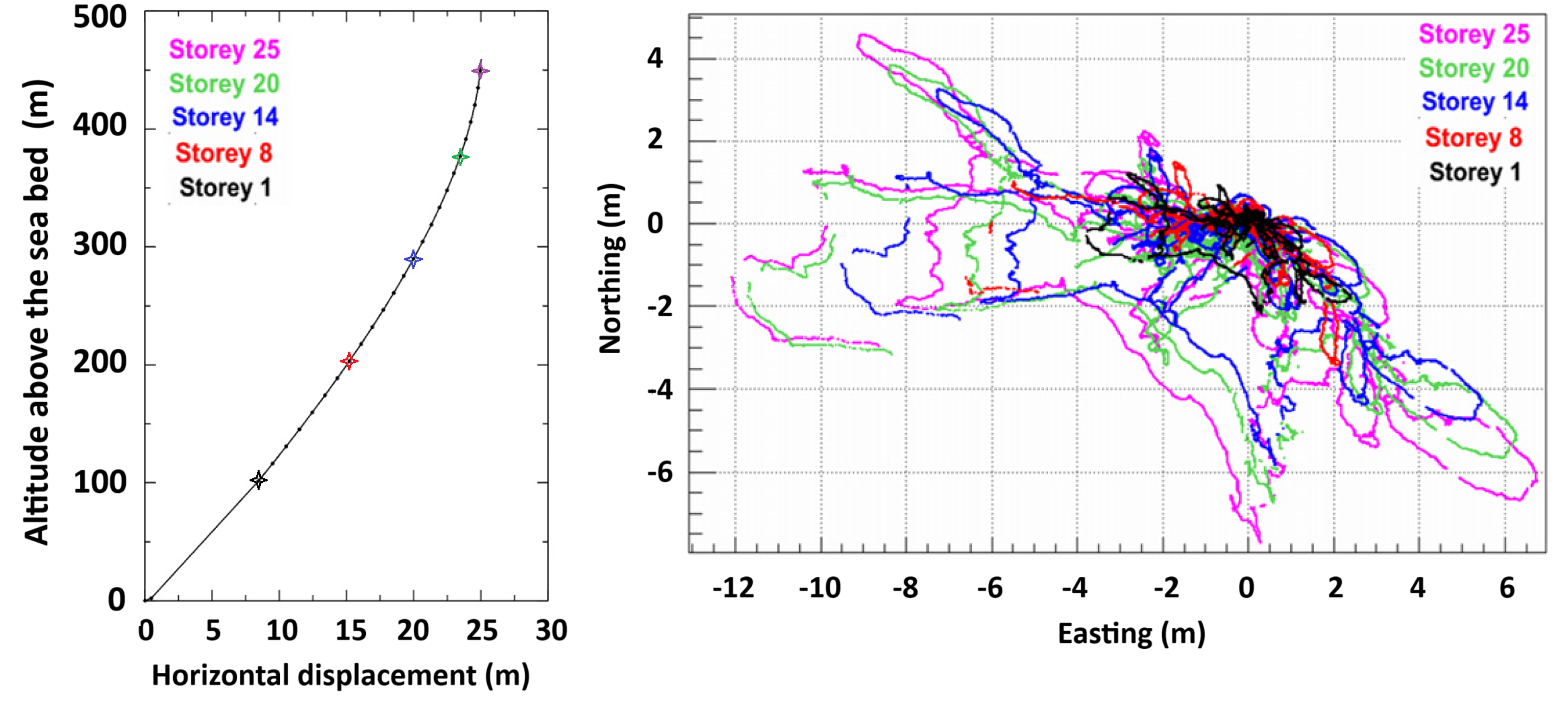}
  \caption{\small Left: Line shape for a sea current speed of 25 cm/s. The horizontal scale is enhanced for the sake of visibility of the line curvature. Right: Displacements of the hydrophones of a given detector line over a period of six months (from July to December 2007) in the horizontal plane as determined by the positioning system. The Base Support Structure position  position is at (0,0,0).}
    \label{fig:cali}
\end{figure}

Combining these information, the shape of each line was reconstructed by performing a fit based on  the coordinates coming from the acoustic positioning system, the headings provided by the compasses and the tilt angles provided by the tiltmeters. 
These measurements were performed every 2 minutes. The relative positions of the OMs were then deduced from the reconstructed line shape, as in the left panel of Fig. \ref{fig:cali}. 
The right panel of the figure shows the \textit{x–y} displacement in the horizontal plane of the five hydrophones located at different heights along a line as a function of time for a period of 6 months. 
The reconstruction of the line shape was based on a model which predicts the mechanical behavior of the line under the influence of the seawater flow taking into account the weight and drag coefficients of all elements of the line \cite{ANTARES:18}. The obtained spatial resolution was better than the 20 cm specification.

In addition to the weekly time calibrations, also amplitude calibrations of each PMT  were routinely performed during special runs. The single photoelectron  peak was studied with minimum bias events. 
The knowledge of the position of the single photoelectron peak and of the pedestal was used to determine the charge conversion over the full dynamical range of the ADC. 

Finally, a precise absolute orientation of the whole detector was needed in order to find potential neutrino point sources in the sky.
The method used by cosmic-ray experiments to evaluate the detector angular resolution and pointing accuracy is to make use of the Moon or Sun \textit{shadow} effect. 
The shadow is the deficit of the atmospheric muon flux in the direction of the Moon/Sun caused by the absorption of the primary cosmic rays. 
Using this method, ANTARES observed a reduction of the muon flux from the Sun \cite{ANTARES:87} (Moon \cite{ANTARES:74}) direction  with a statistical significance of 3.7$\sigma$ (3.5$\sigma$), and with an estimated angular resolution of  $0.6^\circ\pm 0.1^\circ$ for downward-going muons. 
The pointing accuracy was found to be consistent with the expectations and no evidence of systematic pointing shifts was observed.

{An independent check for the pointing accuracy was performed with a surface array detector composed of a set of 15 liquid scintillator detection units \cite{ANTARES:74}. The device was temporarily onboard of the Castor 02 vessel circulating around the position of the telescope, synchronized to a GPS reference.
The pointing accuracy of the ANTARES detector was inferred by combining
the data from the surface array and the atmospheric muons reaching the undersea telescope.
The results of the surface array study were in good agreement with the pointing performance obtained with the shadows of the Moon and of the Sun.}

\subsection{Detector efficiency as a function of time\label{sec:evst}}
\begin{SCfigure}
  \centering
\includegraphics[width=0.65\textwidth]%
    {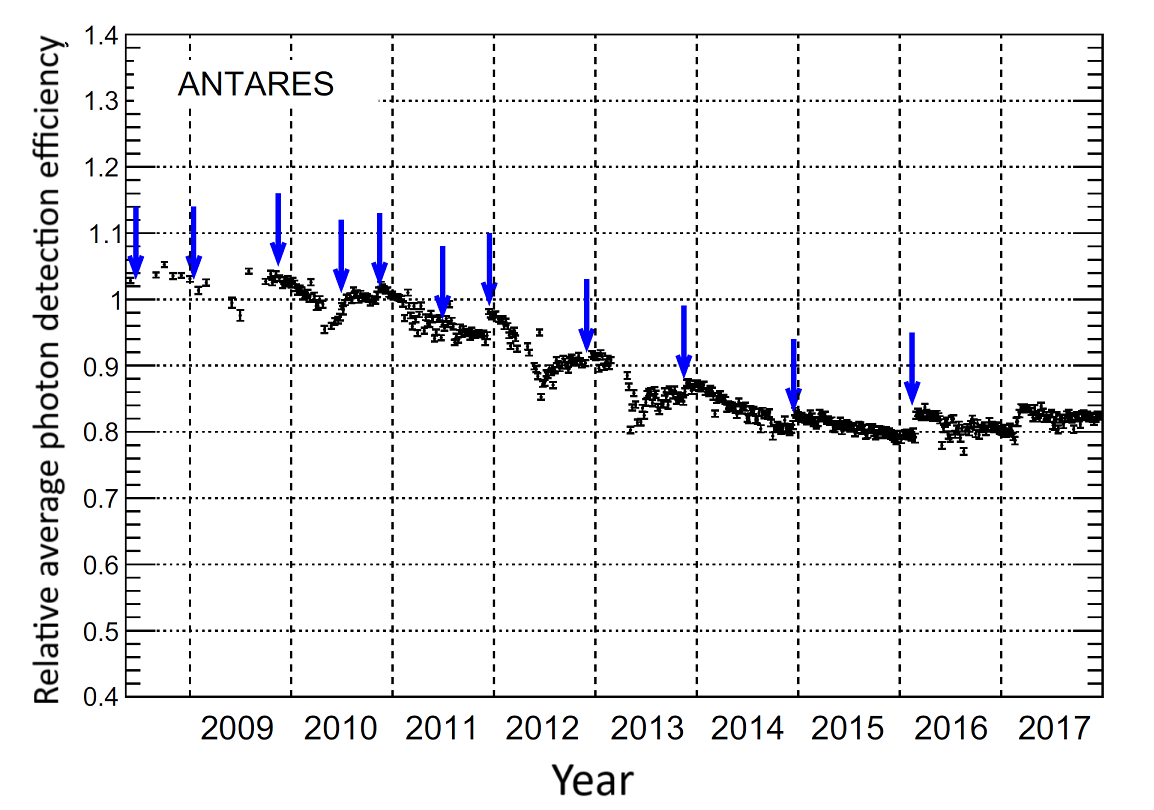}
  \caption{\small Relative OM efficiency averaged over the whole detector versus time. The blue arrows indicate when a high voltage tuning (HVT) of the PMTs was performed. The HVT procedure adjusts the effective gain of individual PMTs to the nominal one. Error bars indicate the statistical error on the mean efficiency.}
    \label{fig:k40}
\end{SCfigure}

For a neutrino telescope operating for such a long time,  monitoring the stability of the optical sensors as a function of time is mandatory.
This monitoring is necessary to preserve the precision on the reconstruction of neutrino-induced events and to guarantee a reliable estimate of the energy released by charged particles.
A seawater detector has the possibility to use a natural calibration tool offered by the presence of dissolved $^{40}$K, see \S~\ref{sec:k40}.
If a $^{40}$K nucleus decays near a storey, the resulting Cherenkov photons can be recorded by two adjacent OMs almost simultaneously. 
The  coincidence signal due to $^{40}$K is practically unaffected by water properties such as the variation in the absorption length, because the decay must occur very close to a storey to give rise to this particular signature.
ANTARES used such coincidences to derive  the time evolution of the photon detection efficiency, $\epsilon$, of optical modules~\cite{ANTARES:73}. Every month, $\epsilon$ was computed for each individual OM. Fig. \ref{fig:k40} illustrates the average $\epsilon$ as function of time. After a smooth decrease over several years, the OM photon detection efficiency finally stabilized over the last years. An overall modest detection efficiency loss of 20\% was observed over the whole analyzed time period. 
In 2010, 2012 and 2013 a particular pattern was observed: the average $\epsilon$  dropped by 5–10\% in spring and partly recovered in the second half of the year. This was attributed to the formation of dense deep seawater through a process known as \textit{open-sea convection}, see \S \ref{sec:ESS}. 
As a consequence of such an exchange of deep seawater, sedimentation as well as biofouling processes have impacted the photon detection efficiencies of OMs in these periods.

These results  served as input for detailed Monte Carlo simulations, \S \ref{sec:runbyrun}, which included a realistic simulation of the OM efficiencies in each data-taking run. 
The conclusion from the final analyses of ANTARES data is that the effect of PMT aging is surely present, but it can be mitigated by periodically tuning the high voltage of the PMT.  These results demonstrate that future underwater experiments can remain in operation for a timescale of at least a decade without major efficiency degradation.

\section{Mediterranean Sea water properties\label{sec:water}}
The detection of particles in ANTARES and other neutrino telescopes exploits the Cherenkov effect to reconstruct particle properties such as direction and energy. 
Charged particles traveling through seawater induce the emission of Cherenkov light in the medium whenever the velocity of the particle exceeds that of light in water.
Thus, a detailed understanding of seawater in the chosen site is fundamental to model correctly the details of Cherenkov light production and propagation.  
Seawater properties (in particular, refractive index) affect the quantity of produced photons as well as the expansion speed and geometry of the Cherenkov cone. 
Photon propagation over macroscopic distances is also affected by seawater inherent optical properties, i.e. \textit{absorption} and \textit{scattering} which needs to be measured in situ.
Absorption reduces  the intensity  of  the  Cherenkov  wavefront,  while scattering changes the direction of propagation of the Cherenkov photons and the distribution of their arrival time on the PMTs; this latter effect degrades the measurement of the direction of the incoming neutrino.

\subsection{Cherenkov photon emission}
The seawater refractive index $n(\lambda)$ as a function of the wavelength depends on the water's physical and chemical properties, { i.e.} its salinity, temperature, and pressure. These values are typically monitored regularly within the neutrino telescope itself. 

The Cherenkov photons are emitted at a characteristic angle $\theta_C$ with respect to the particle direction such that
\begin{equation}
 \cos\theta_C = \frac{1}{\beta n_p} \ ,
\end{equation}
and sum coherently on a conical surface.
Here, $\beta$ is the velocity of the particle relative to the speed of light in vacuum and the index of refraction,  $n_p$, corresponds to the ratio between the speed of light in vacuum and the phase velocity of light in water. 
For the ultra-relativistic limit ($\beta\approx 1$) in water, $\theta_C\simeq 42.2^\circ$.
The number of Cherenkov photons $N$ per flight path $x$ {[m]} and wavelength $\lambda$ {[m]} produced by a charged particle propagating in a medium with refractive index $n(\lambda)$ is given by the Frank-Tamm formula (below, in the SI units):
\begin{equation}
    \frac{d^2N}{dxd\lambda} = \frac{2\pi\alpha}{\lambda^2}
    \Biggl[1-\frac{1}{[\beta n_p(\lambda)]^2}\Biggr] \quad
    \textrm{with} \quad \alpha=\frac{c e^2\mu_0}{2\pi h}\simeq \frac{1}{137} \quad 
    \label{eq:FT}
\end{equation}
where $e$ is the electric charge of the relativistic particle. In the ultra-relativistic approximation  a minimal-ionizing particle produces $\sim$350 Cherenkov photons per cm in the relevant wavelength range of 300--600~nm, see Fig. \ref{fig:OM}.

After being emitted, individual photons travel through the water at the group velocity, $v_g$.  
The group velocity depends on the wavelength $\lambda$ of the photons (chromatic dispersion).
The refraction index related to the photon propagation, $n_g$, corresponds to the ratio between the speed of light in vacuum and the group velocity of light in water. Considering the relation between phase and group velocities, $n_g$ is given by
\begin{equation}\label{eq:ng}
    n_g = \frac{n_p}{1+\frac{dn_p}{d\lambda}\frac{\lambda}{n_p}} \ .
\end{equation}
Since the PMTs cannot measure the photon wavelength, the variation of the photon emission angle and the group velocity due to chromatic dispersion cannot be accounted for on the individual photon level. Nevertheless, the average effect of the wavelength dependencies are accounted for in the algorithm used to reconstruct the particle trajectory \cite{ANTARES:17,ANTARES:29}.

\begin{SCfigure}
  \centering
\includegraphics[width=0.65\textwidth]%
    {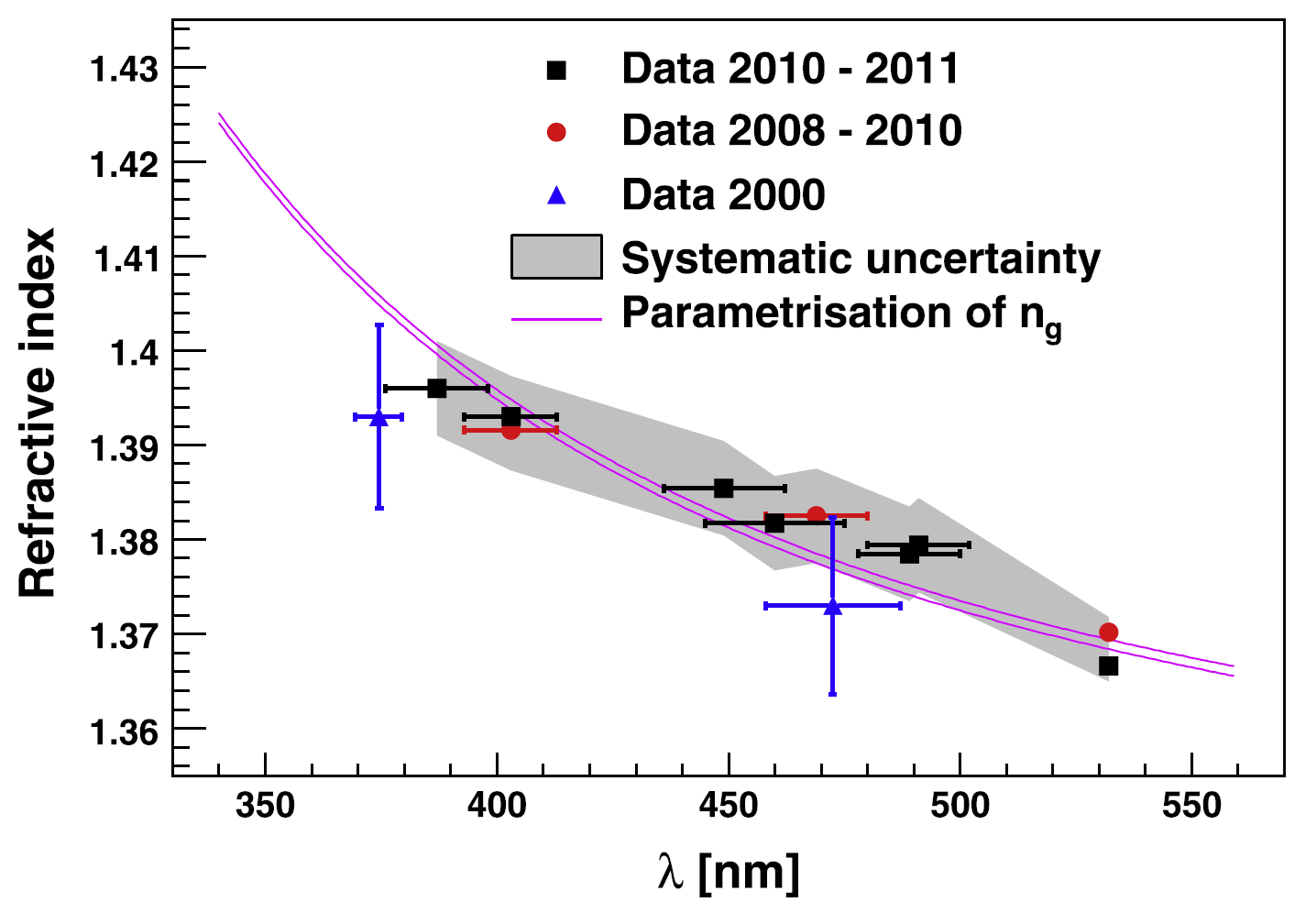}
  \caption{\small Index of refraction $n_g(\lambda)$ vs. wavelength as measured in \cite{ANTARES:23} (black and red dots) and \cite{ANTARES:04} (blue). 
  The two solid lines correspond to a parametrization of $n_g(\lambda)$ evaluated at  pressure values of 200 atm and 240 atm. See details in \cite{ANTARES:23}.}
\label{fig:ng}
\end{SCfigure}

In ANTARES, the measurement of the group velocity \cite{ANTARES:23} was made using the optical beacon system, \S \ref{sec:cali}. 
The refractive index  was deduced from the recorded time of flight distributions of photons at different distances from the sources for eight different wavelengths between 385 nm and 532 nm, allowing the determination of $n_g(\lambda)$, as shown in Fig. \ref{fig:ng}.

\subsection{Absorption, scattering and attenuation}
The propagation of light in seawater  depends, for a given wavelength $\lambda$, on the medium  inherent  optical  properties:  absorption, scattering, and attenuation.
The light attenuation length $ L_{\rm att}$ is empirically defined as \cite{ANTARES:04}:
\begin{equation}
I(r,\lambda)=I_0(\lambda) \frac{A}{4\pi r^2} e^{-r/L_{\rm att}} \ ,
\end{equation}
where $I$ is the light intensity of  wavelength $\lambda$  detected at a distance $r$ from the source by a receiver of area $A$  (photomultiplier) from an isotropic source of photons with intensity $I_0$.
The attenuation length can be written as a function of the absorption $L_{abs}$ and scattering $L_{\rm sca}$ lengths: an accurate measurement of $L_{\rm sca}$ is, indeed, difficult.

Scattering processes involve the deviation of a photon from a straight line. 
Its complete description requires, in addition to the geometric scattering length $L_{\rm sca}$, the knowledge of the scattering angular  distribution. 
Gustav Mie developed (1908) an analytical solution of the Maxwell equations for scattering of electromagnetic radiation by spherical particles, which is appropriate for modeling light scattering in transparent media.
Scattering of particles which are much larger than the photon wavelength is more difficult to model as it depends on the sizes and density of the suspended particles in water. 
Both components needs to be measured and modeled correctly for an optimal understanding of the signals from Cherenkov light.

\begin{figure}[htbp]
\includegraphics[width=0.95\textwidth]%
    {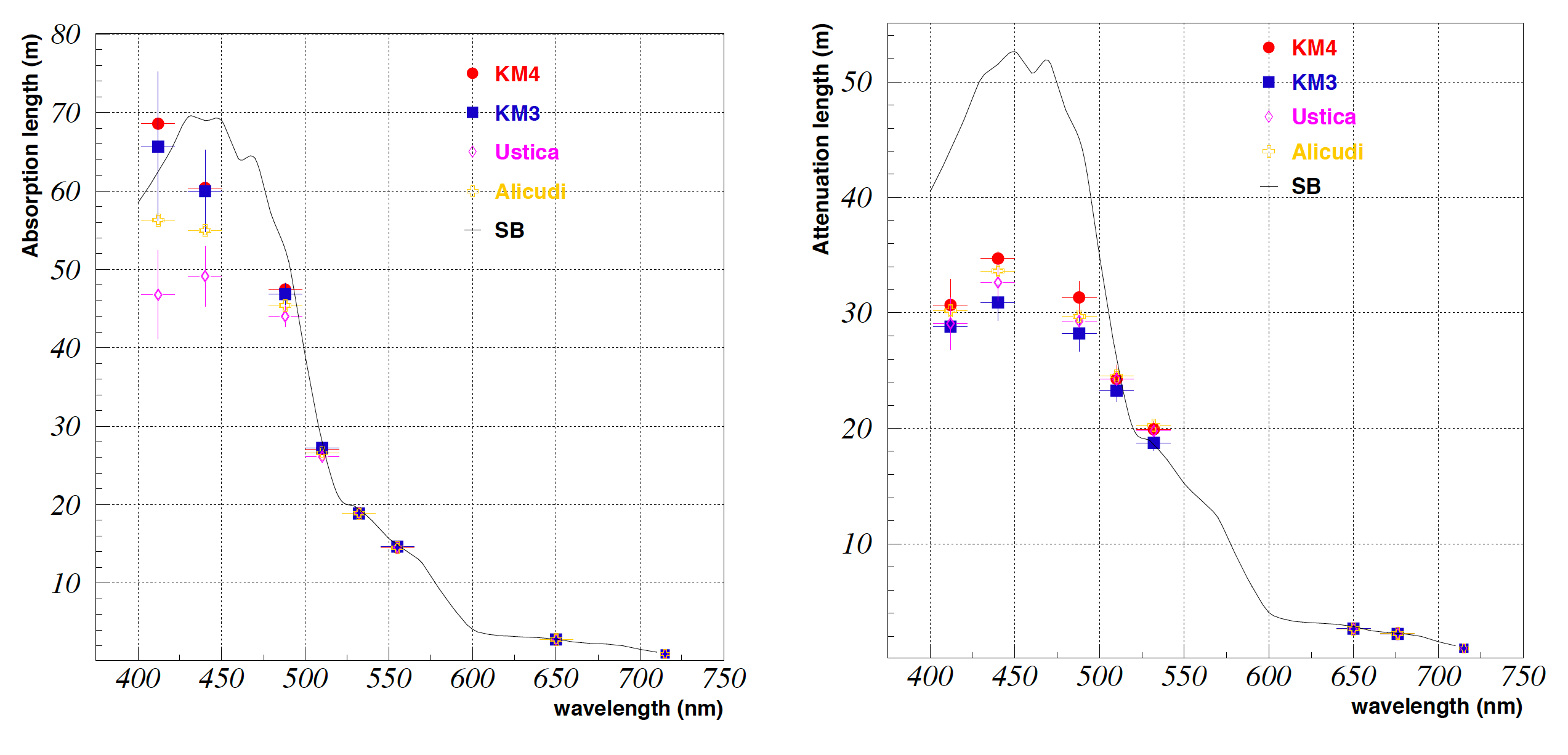}
  \caption{\small Absorption (left) and attenuation (right) lengths measured at different locations in the central Mediterranean Sea
  at a depth of about 3000 m for different wavelengths~\cite{nemowater}. The solid black line indicates optically pure water according to~\cite{smith_baker}.}
  \label{fig:Water_abs}
\end{figure}

A photon can be scattered multiple times before it reaches an optical sensor.
For this reason, experimental measurements are generally expressed in terms of the \textit{effective} light scattering length 
\begin{equation}
L_{\rm sca}^{\rm eff} = \frac{L_{\rm sca}}{1-\bigl \langle \cos\theta\bigr \rangle}, 
\end{equation}
which takes into account the averaged value of the scattering angle.
A compilation of experimental results of different campaigns and in different locations is reported in \cite{ANTARES:04,nemowater}: Fig. \ref{fig:Water_abs} shows some of the measurements of absorption (left) and attenuation (right) lengths in central Mediterranean Sea.

In the comparison between media used for neutrino telescopes \cite{Spurio:10}, seawater has a smaller $L_{abs}(\lambda)$ with respect to the more transparent ice. The same instrumented volume of ice corresponds to a larger effective volume with respect to seawater. 
On the other hand, the effective scattering length $L_{\rm sca}^{\rm eff}$ for ice (that contains air bubbles and dust particles) is smaller than water. This is a cause of a larger degradation of the angular resolution of the detected neutrino-induced events in ice with respect to water.

\subsection{Optical background in seawater\label{sec:k40}}
Optical modules in seawater suffer  background from the natural radioactivity of elements and from the luminescence produced by organisms living in the deep sea (\textit{bioluminescence}). 
The $^{40}$K is by far the dominant radioactive isotope present in natural seawater. 
Its decay channels are:
\begin{eqnarray}    
^{40}_{19}\textrm{K} &\rightarrow & ^{40}_{20}\textrm{Ca}\;+\;e^-\;+\;\bar{\nu}_e \; (\mbox{BR = 89.3}\%)\nonumber\\
^{40}_{19}\textrm{K}\;+\;e^-&\rightarrow & ^{40}_{18}\textrm{Ar}\;+\;\nu_e\;+\;\gamma \; \ (\mbox{BR = 10.7}\%)\nonumber
\end{eqnarray}
and both channels contribute to the production of  optical  noise. The maximum energy for $\beta^-$ is 1.31~MeV and a  large fraction  of produced electrons is above the Cherenkov threshold. 
The  photon  originating  in the  second reaction  has an energy  of 1.46~MeV and  can lead (through Compton  scattering) to  electrons  above the Cherenkov threshold.
The  intensity of   Cherenkov  light from  $^{40}$K radioactive  decays depends mostly on its concentration in sea  water. Since salinity in  the Mediterranean  Sea has small  geographical variation, this  optical noise is largely site independent.

In addition to the background rate due to $^{40}$K decays, a continuum bioluminescence rate, and random bursts of a few seconds duration that are only correlated in time over distances of the order of a meter, were observed.
Bioluminescence (see also \S \ref{sec:ESS}) is ubiquitous in seawater and is produced by two mechanisms: long-duration and diffuse glows produced by clouds of bacteria and intense flashes of photons produced by macroscopic organisms. 
These can  give rise to an optical  background that can be occasionally much more intense than the one due to $^{40}$K. 
Bursts observed in the counting rates are probably due to the passage of light-emitting organisms close to a given PMT.
The typical spectrum  of bioluminescence  light is centered  around 470--480 nm,  the   wavelength  of  maximal transparency  of water.  
The distribution of luminescent organisms in deep sea varies with location, depth,  and time but there is a general pattern of decrease in abundance with depth. 

\begin{figure}[htbp]
\includegraphics[width=0.95\textwidth]%
    {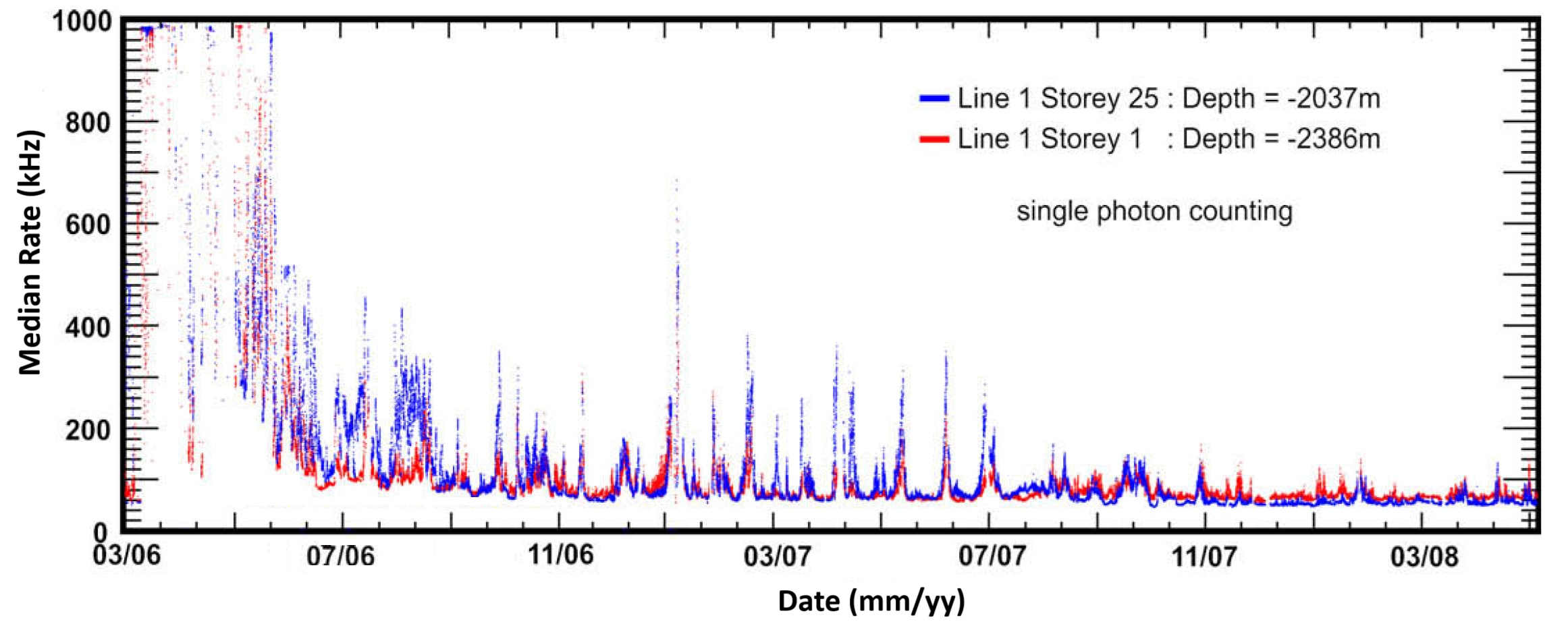}
  \caption{\small Median rates (in kHz) measured during the construction period (from March 2006 to May 2008) with the 10$^{\prime\prime}$ PMTs of the ANTARES experiment on optical modules at two different depths (2037 m and 2386 m) \cite{ANTARES:10}. The contribution of the $^{40}$K decay is evaluated to be almost constant of $\sim~35$ kHz.}
      \label{fig:rates}
\end{figure}

The counting rate on each 10$^{\prime\prime}$ diameter PMT of the detector was continuously monitored all along the data taking \cite{ANTARES:73}.
The measurements yield an average counting rate due to $^{40}$K of $(35 \pm 8)$ kHz on each PMT, as shown in  Fig. \ref{fig:rates} for the initial period from March 2006 to May 2008 which was particularly noisy. 
The two optical background components (with periodic bursts due to bioluminescence) are clearly visible. 
The long-term observation showed that the \textit{baseline component} (the plateau of the counting rate) is neither correlated with sea current, nor with burst frequency. Long-term variations of the baseline were observed, however these variations are not correlated with periods of high burst activity, suggesting that each contributions is caused by a different population of  pelagic organisms able to emit light, as discussed in \S \ref{sec:ESS}.

\subsection{Sedimentation and biofouling\label{sec:biofouling}}
The presence of organic or inorganic particulate in seawater affects the telescope response both worsening the light transmission, as discussed before, and as a factor of detector aging.
Due to biofouling  and  sediments  sticking   on  the  optical  sensors,  efficiency of the photon detection can be degraded. 
Environmental  parameters  may  vary significantly,  for each  marine  site,  as a  function  of depth  and time. Moreover, seasonal effects like the increase of surface biological activity (typically during spring)  or the precipitation of sediments transported  by rivers,  enlarge the  amount of  dissolved and suspended particulate, worsening the water transparency.

Initial measurements of sedimentation and biofouling on the optical modules \cite{ANTARES:03} were made using blue light transmission through glass spheres over several months.
It was found that the loss of transmissivity decreased steadily with increasing zenith angle with a tendency to saturate with time. For the vertical glass surface it was estimated to be $\sim~2\%$ after one year. 
Despite a fairly large accumulation rate at the ANTARES site, the slow growth of the transparent biofilm substrate implied a very loose adhesion of the sediments to the glass surfaces.
The light-absorbing particulates accumulation was found only significant for surfaces facing upwards. 
For the downward oriented glass surface facing the PMT photocathodes the loss of transmissivity due to the fouling was small even after almost 16 years \cite{ANTARES:73}.

\section{Software\label{sec:soft}}

The raw data from the ANTARES telescope consisted of \textit{events}, which represent a list of hits collected in a time window of about 2 microseconds before and after a given trigger condition is met. 
This was complemented by calibration data such as the positions and orientations of all detector elements. 
Several dedicated reconstruction algorithms were employed to extract information about the type, direction,  and energy of the particle(s) which caused the recorded hit pattern (track or cascade) in each event. Data and simulations were processed in the same way.

\subsection{Physics event generation\label{sec:generator}}
\begin{SCfigure}
  \centering
\includegraphics[width=0.70\textwidth]%
    {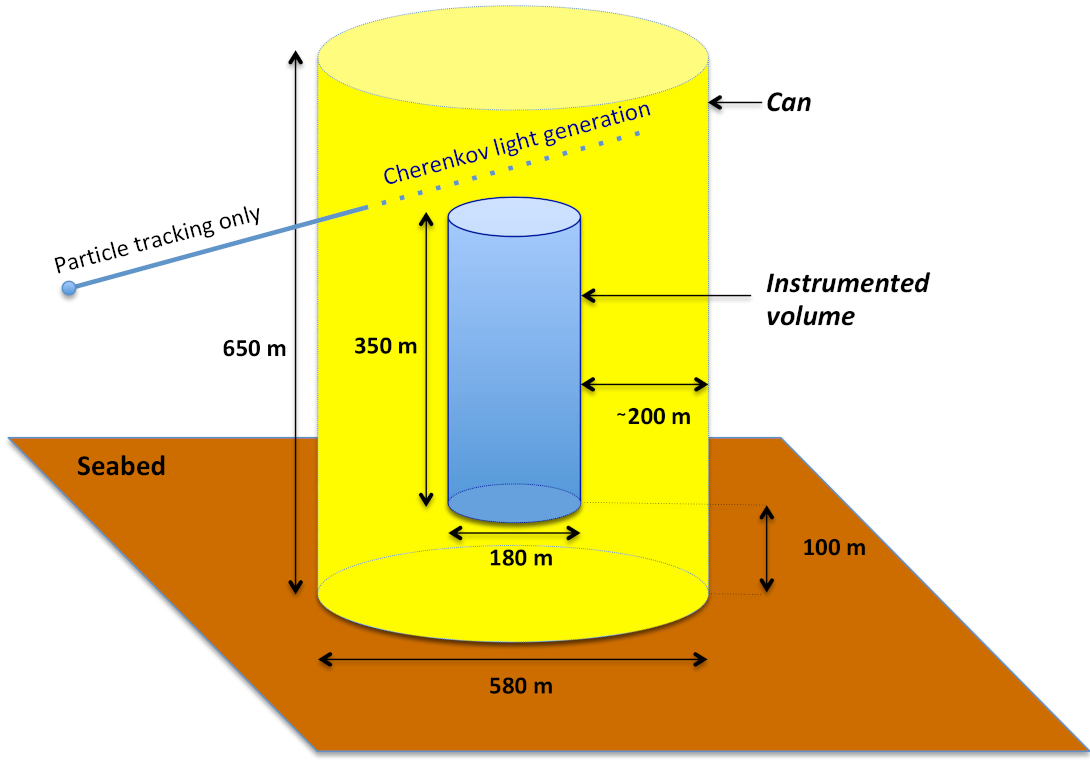}
  \caption{\small Schematic view of the ANTARES \textit{can} (in yellow), anchored to the seabed (in brown) and containing the detector instrumented volume (in blue). From \cite{ANTARES:89}.}
\label{fig:can}
\end{SCfigure}

Monte Carlo simulations started with the simulation of physics events, i.e. the generation of the kinematic information of each detectable particle, either given by neutrino interactions or from the passage of background muons.
All charged particles that can induce Cherenkov photons arriving at the optical sensors of the detector were considered. 
The sensitive volume of the detector (a cylinder bounding the water region that hosts the PMTs extended by two attenuation lengths of light in water) was called the \textit{can}, see Fig. \ref{fig:can}.
The \textit{can} defined the volume where the Cherenkov light was generated and propagated in the simulation. 
Outside this volume, only energy losses of long-tracking particles (i.e. muons and taus) were considered.
Interactions can occur either in water close to the \textit{can} volume or in the rock below the detector. 
A detailed description of the software used in ANTARES is given in~\cite{ANTARES:89}.

The generation of the kinematics of atmospheric and cosmic neutrino interactions, from the sub-GeV energy range up to $10^9$ GeV, took into account the different neutrino flavors  ($\nu_e,\nu_\mu, \nu_\tau$), chiralities ($\nu, \bar\nu$), and interaction channels (CC and NC interactions). 
Deep inelastic scattering, dominant at high energy, was simulated using the LEPTO package \cite{Lepto:97}. Above 10 TeV, an extrapolation applying the CTEQ6D \cite{Pumplin:02} parton distribution functions was used to calculate cross sections and interaction kinematics. 

Individual neutrinos were injected into the code according to a power law energy spectrum $dN/dE \propto E_\nu^{-\gamma}$, where $\gamma$ can be set by the user to have an event sample with an adequate statistical significance across the considered energy range. Afterwards, events were weighted according to a specific flux model: atmospheric neutrinos, point-like sources, diffuse fluxes, etc.

A generation volume ($V_{\rm gen}$), whose size depends on the neutrino interaction type and on the neutrino flavor and energy, was considered.
Every neutrino was generated as interacting within this volume.
When the $\nu$ interaction occurs inside the \textit{can} volume, the kinematic information of all final-state products (charged leading lepton, if present, and all charged particles) was stored and became the input to the program simulating the Cherenkov light. If the vertex is outside the \textit{can}, only long-tracking leptons were considered for the following steps. 
For $\nu_\mu$ and  $\bar \nu_\mu$ CC events, a $V_{\rm gen}$ much larger than the \textit{can} size was defined. The leading $\mu^-$ ($\mu^+$) was propagated if its direction was intercepting the \textit{can}, evaluating the energy loss occurring during the path. The distance between the vertex and the \textit{can} and, consequently, the $V_{\rm gen}$ size were calculated according to the neutrino energy.
Due to the limited longitudinal extension of either hadronic or EM showers, for these cascades the $V_{\rm gen}$ was coincident with the \textit{can}.

The $\nu_e$ and $\nu_\mu$ propagation through the Earth was not considered, and the neutrino energy at the interaction was the energy of the neutrino when it entered the Earth. The probability of Earth absorption was accounted for in the final weight. For $\nu_\tau$ interactions the propagation through the Earth was fully considered with the $\nu_\tau$ regeneration effect \cite{Halzen_tau}. 
This required considering  the $\nu_\tau \rightarrow \tau \rightarrow \nu_\tau$ decay chain, producing a lower-energy $\nu_\tau$ in the final state.

The situation is different for the simulation of atmospheric muons produced in the interactions of cosmic rays (CRs) in the upper atmosphere.  
Despite the shielding effect of the water overburden, atmospheric muons with TeV energy at sea level represent the majority of the physical  events in any deep-sea neutrino telescope \cite{PARA:06}.
Due to the large detection volumes, triggered events are generally associated with bundles of atmospheric muons instead of individual particles.

The ANTARES telescope was too small for vetoing downward-going atmospheric muons with the external layers of the detector.  In the search for neutrino-induced candidates, the atmospheric muon background was removed by focusing the analysis on upward-going events. However, since the rate of atmospheric muons exceeds that of atmospheric neutrinos by 4 to 5 orders of magnitude, a small contamination from track-like events incorrectly reconstructed as upward-going remains. To address this, mis-reconstructed muons were filtered out using reconstruction quality criteria, \S \ref{sec:reco}. Therefore, an accurate simulation of atmospheric muons was essential for properly estimate the background and optimize the event selection in each data analysis. Although atmospheric muons constitute the primary background for a neutrino telescope, they provide a nearly constant and stable flux of particles that are essential for time calibration and absolute positioning of the detector. Additionally, they enable monitoring of the detector's efficiency over time (see also \S \ref{sec:atmunu}).

Atmospheric muon bundles at the detector can be reproduced either using a complete simulation of the atmospheric showers induced by the arrival of a primary CR or by evaluating the underwater muon flux with a set of parametric formulae. 
One characteristics of detectors in seawater is that the flux of TeV  muons at sea level is the same at any location when neglecting the small dependency on the temperature profile of the atmosphere. 

An example of the first method is provided by the CORSIKA program \cite{corsika}. 
Computationally expensive, the code allows a broad flexibility in the choice of its input parameters such as the atmospheric model, the parametrization of the hadronic interactions, the chemical composition and energy spectrum of the primary CR flux.
Thus, the same simulation at sea level can be used by any undersea detector, which can eventually share this initial (computationally heavy) data sample. This input can be used to propagate muons until the detector and to fold the kinematics of surviving events with the  features of the apparatus.

A faster alternative was developed for the needs of the ANTARES experiment but it is available for any detector located at a depth between 1500 and 5000 meters of water. 
The  \texttt{MUPAGE}  package \cite{MUPAGE:08} is based on a set of parametric formulae extracted from a full simulation of events at sea level, tuned according to the underground measurements performed with the MACRO experiment at the Gran Sasso Laboratory \cite{MACRO:NIM02} and extrapolated to a deep-sea location.
The software provides the angular and energy distribution of muons at different depths as a function of the muon bundle multiplicity. The usage of parametric formulae allows the fast production of a large number of Monte Carlo events at a cylindrical  surface surrounding the underwater instrumented volume.
This approach lacks flexibility in the definition of the input parameters related to the CR composition and interaction models. 
However, despite this limitation and considering the large uncertainties on the description of the hadronic interactions and on the CR composition, the multi-year ANTARES experience has shown that this fast parametric simulation produces a reliable estimate of the atmospheric muon background. 
Comparisons between atmospheric muon data and the \texttt{MUPAGE} parametrization are available in almost all papers describing ANTARES results. The code is also widely used in the KM3NeT collaboration.

\subsection{Propagation of Cherenkov photons}
About 350 Cherenkov photons per cm of path length are induced in the medium (using eq. \ref{eq:FT}) by a relativistic charged particle in the wavelength window in which PMTs are sensitive.
Their individual propagation at distances of the order of hundreds of meters is out of question, given the excessive computational time to obtain such detailed simulation.
Thus, Cherenkov photons induced by high-energy muons and other charged particles need to be simulated using dedicated software packages.  

For the ANTARES case, customized {photon tables} were calculated, which contain the Cherenkov photon yield for a number of standardized particle types at a given energy as a function of the distance (1 parameter), the relative orientation between the particle and the PMT (3 parameters),  the photon travel time (1 parameter), and its wavelength (1 parameter). To obtain these 6-dimensional tables, individual Cherenkov photons have been traced including the effects of water absorption and scattering~\cite{ANTARES:89}.

A custom program was developed for the propagation of the particles and of the light through the \textit{can} volume. Muons were transported using the MUSIC package \cite{Music:97}. 
At each step along the muon path (1 meter long) all energy-loss processes were considered and the muon energy loss was treated as continuous,  or discrete/stochastic if the energy loss exceeded 300 MeV/m. In the case of  discrete energy losses, an independent electromagnetic shower was generated along the segment length and the number of photons was extracted from the corresponding tables. An energy dependent scaling of the amount of light was thus considered. 

\begin{SCfigure}
  \centering
\includegraphics[width=0.68\textwidth]%
    {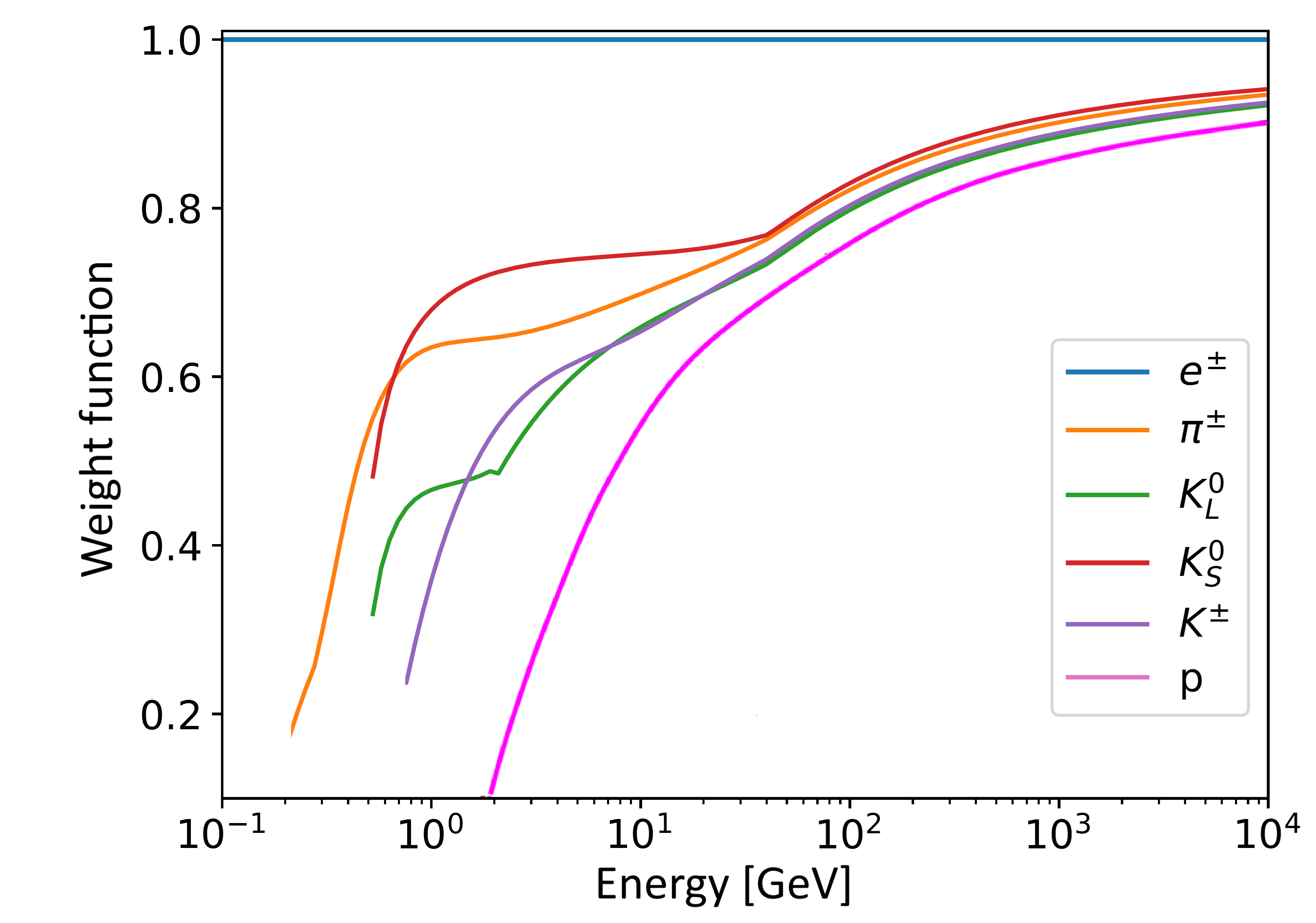}
  \caption{\small Weight function used in the simulation of particles produced in hadronic showers. The light yield from the electron at a given energy is assumed to be equal 1.}
\label{fig:weigth}
\end{SCfigure}

The treatment of hadronic cascades induced by any neutrino interaction is relevant as well. 
While  electromagnetic cascades present small variations, the hadronic ones contain many charged hadrons with large variations among individual cascades.
The production of photon tables for each single particle would require an event-by-event simulation and a huge amount of computational time. 
The solution adopted was to assign an energy-dependent weight to the light output of each hadron relative to that yield by electrons or positrons, while the detailed spatial modeling of the shower was kept unchanged. The weights reported in Fig. \ref{fig:weigth} were pre-determined with an accurate Monte Carlo simulation.

\subsection{Time-dependent detector response\label{sec:runbyrun}}
Environmental conditions in seawater undergo significant variations on different timescales that directly affect data acquisition.  
In addition, not all detector elements take data continuously, because of temporary or permanent malfunctioning of optical modules (OMs) or lack of connection to some parts of the apparatus, occasionally producing no signal from some components. 
Finally, environmental conditions affect the choice of the trigger algorithms that are applied during the onshore processing of the raw data stream. 
In the long ANTARES lifetime, conditions changed significantly. In order to reproduce the detector response under the specific conditions of each individual data taking run (with typical length a few hours), physics events were simulated following a strategy denoted as \textit{run-by-run}.
The temporarily or permanently non-operational OMs were masked in the simulation. 
The optical background, which might vary due to bioluminescence, was extracted using Poisson statistics directly from short segments of the data stream.
A connection to the database interface allowed retrieval of information on the DAQ status of each detector element, on the active trigger setup and on the detector configuration in the run.
Finally, other inefficiencies on longer timescales (in particular,  OM transparency and PMT gain) were taken into account using the information described in \S \ref{sec:evst}. 
The simulated pattern of photons on the photocathode of a PMT was then transformed into a digital {\it hit} by first, applying the probability that the photon yields a photoelectron and then, converting the analogue PMT signal into the digital time and amplitude output of the dedicated readout chip. The digitized information was formatted in exactly the same way as real data.

At the end of the full chain of simulation with this \textit{run-by-run strategy}, an archive of simulated Monte Carlo events was created containing a set of files for each run of the real data acquisition and stored on disk. They were then processed with the same reconstruction algorithms and analysis procedures used for the corresponding data files.

\subsection{Track and cascade reconstruction\label{sec:reco}}
Reconstruction methods developed in ANTARES were used both on Monte Carlo events and data. For real-time applications (to answer to an external trigger or sending an alert within a few seconds, \S \ref{sec:MM}) a fast and well-qualified reconstruction method is necessary. 
ANTARES developed an algorithm which can cope with a trigger rate of 100 Hz running on a single PC \cite{ANTARES:17}. 
The online reconstruction used an ideal static detector,  not including the knowledge of the dynamical positioning and the precise charge and time calibration sets.  The detailed geometry of the storey was ignored as the signals of the three PMTs within one storey were combined.
All events were later reconstructed offline to determine the muon trajectory or cascade properties with higher accuracy, using multistage fitting procedures and the detailed calibrated information, as the real-time  shape of each line. 

For track-like events induced by muons, the offline reconstruction consisted of a maximum likelihood fit of the measured photon arrival times on individual PMTs. 
A quality parameter for the fit, denoted as $\Lambda$, was determined based on the final value of the likelihood function \cite{ANTARES:29}. 
The estimated angular uncertainty on the direction of the reconstructed muon track, denoted as $\beta$, was provided as well. 
The event selection for each specific analysis was optimized using these two parameters.
With this reconstruction method, the direction of incoming neutrinos with an $E_\nu^{-2}$ spectrum was reconstructed with a precision better than 0.4$^\circ$ in 50\% of the cases \cite{ANTARES:65}.

In parallel, an algorithm  for the reconstruction of the energy and direction of cascades was also applied \cite{ANTARES:66}. For an EM shower resulting from a $\nu_e$ CC interaction, the algorithm reconstructed its position with a precision of about 1 m, and the neutrino direction with a median angular resolution of $\sim3^\circ$ in the 1--1000 TeV energy range, with an uncertainty on the deposited energy of the shower of about 5\%--10\%.

\section{Acoustic detectors and sea science\label{sec:ESS}}

\subsection{Underwater infrastructures for deep-sea technology and science\label{sec:dsinfra}}

The ANTARES detector also comprised a designed system for the investigation of techniques for acoustic detection of neutrinos, and a number of instruments to monitor the environmental conditions, mostly located on the dedicated Instrumentation Line  \cite{ANTARES:06} (see Fig. \ref{fig:ANTARES}).

The study of deep-sea environment needs specialized oceanographic instruments collecting data during periods possibly longer than one year in order to characterize the sites in different seasons. 
Usually deep-sea scientific instruments are equipped with batteries and the data collected on local memories are downloaded by scientists only after the recovery of the instruments.
On the contrary, a mandatory request for a neutrino observatory is the permanent connection between a shore station and the deep-sea detector to power-up the sensors and electronics and allow high data-rate transmission to shore. 
These same infrastructures constitute also a great opportunity for installation of Earth and sea science nodes, allowing long-term and real-time access for oceanographic, geophysical and biological instrumentation.

The ANTARES infrastructure offered one of the first examples of synergy between scientists pursuing different objectives, providing to the Earth and sea science community a direct connection to their instrumentation on the Instrumentation Line. 
The synergy was complete, because the same instrumentation provided important information for the telescope calibration: measurement of water optical and oceanographic properties, behavior of bioluminescent organisms, measurement of sea currents, and identification of acoustic noise sources.

\subsection{The ANTARES modules for the acoustic detection \label{sec:amadeus}}

For the detection of cosmic neutrinos with $E_\nu \gtrsim $100 PeV, approaches beyond the detection of Cherenkov light need to be pursued to survey increasing volumes for a neutrino flux decreasing with energy. 
One approach is to measure acoustic pressure pulses produced by the particle cascades that evolve when neutrinos interact with nuclei in water (or ice). 
The  energy deposition resulting from the interaction, concentrated in a cylindrical volume of a few cm in radius and several meters in length, leads to a local heating \cite{ASKARIYAN:79,Learned:79} that induces an expansion or contraction of the medium. 
The accelerated motion of the heated volume leads to a pressure pulse of bipolar shape which propagates in the surrounding medium within a flat disk-like volume in the direction perpendicular to the axis of the particle cascade. 
The pulse has a characteristic frequency spectrum that is expected to peak around 10 kHz with an attenuation length of about 5\,km in sea water. 

The AMADEUS (\textit{ANTARES Modules for the Acoustic Detection Under the Sea}) system \cite{ANTARES:14}  was conceived to perform a feasibility study for a potential future large-scale acoustic neutrino detector \cite{Lahmann:ARENA2018}. 
For this purpose, a dedicated array of acoustic sensors was integrated into the ANTARES neutrino telescope in the form of \textit{acoustic storeys}. These were modified versions of standard ANTARES storeys with the OMs replaced by custom-designed acoustic sensors.
The system comprised six acoustic storeys: three on the Instrumentation Line and three on a detection line (cf. Fig. \ref{fig:ANTARES}).
The acoustic sensors employed piezo-electric elements for the broad-band recording of signals with frequencies ranging up to about 100 kHz.
Dedicated electronics was used for the amplification, digitization and pre-processing of the analogue signals.
The distances between the acoustic storeys ranged from 14.5 m to 340 m to measure both transient signals and ambient noise in the deep sea, and to localize acoustic point sources. 
The concept of piezo ceramics glued to the inside of a glass sphere was adopted for the position calibration of the optical modules of KM3NeT.
Details on the propagation of sound waves inside the glass sphere of the ANTARES acoustic modules are reported in \cite{ANTARES:108}.
For results of the studies with the AMADEUS system concerning the feasibility of acoustic neutrino detection,  refer to \cite{Lahmann:icrc2013,Kiessling:ARENA2016}.

The AMADEUS system provided an audio data stream that allowed to identify  the presence of the Ligurian Sea sperm whales through the hourly tracking of their long-range echolocation behavior \cite{ANTARES:55}.
Even though dedicated research has been carried out to adequately map the distribution of the sperm whale in the Mediterranean Sea, the species population status is still presently uncertain. The year-round presence of sperm whales reported with the ANTARES data is probably associated with the availability of cephalopods in the region as food for the whales.  
Interestingly, the same analysis indicated that the occurrence of surface shipping noise  apparently does not affect the foraging behavior of the sperm whales in the area, since shipping noise was almost always present when sperm whales were acoustically detected.
The continuous presence of the sperm whale in the region (see also \cite{ONDE:15}) confirms the ecological value of the Mediterranean Sea  and the importance of underwater detector as ANTARES and KM3NeT to help monitoring its ecosystems.

\subsection{The Acoustic Doppler Current Profiler \label{sec:IL}}
In addition to the AMADEUS system, the Instrumentation Line included an Acoustic Doppler Current Profiler (ADCP)  \cite{ANTARES:20} to monitor the intensity and direction of the underwater flow.
This instrument provided a unique opportunity to compare high-resolution acoustic and optical observations at a depth of about  2400 m. 
\begin{figure}
  \centering
\includegraphics[width=0.85\textwidth]%
    {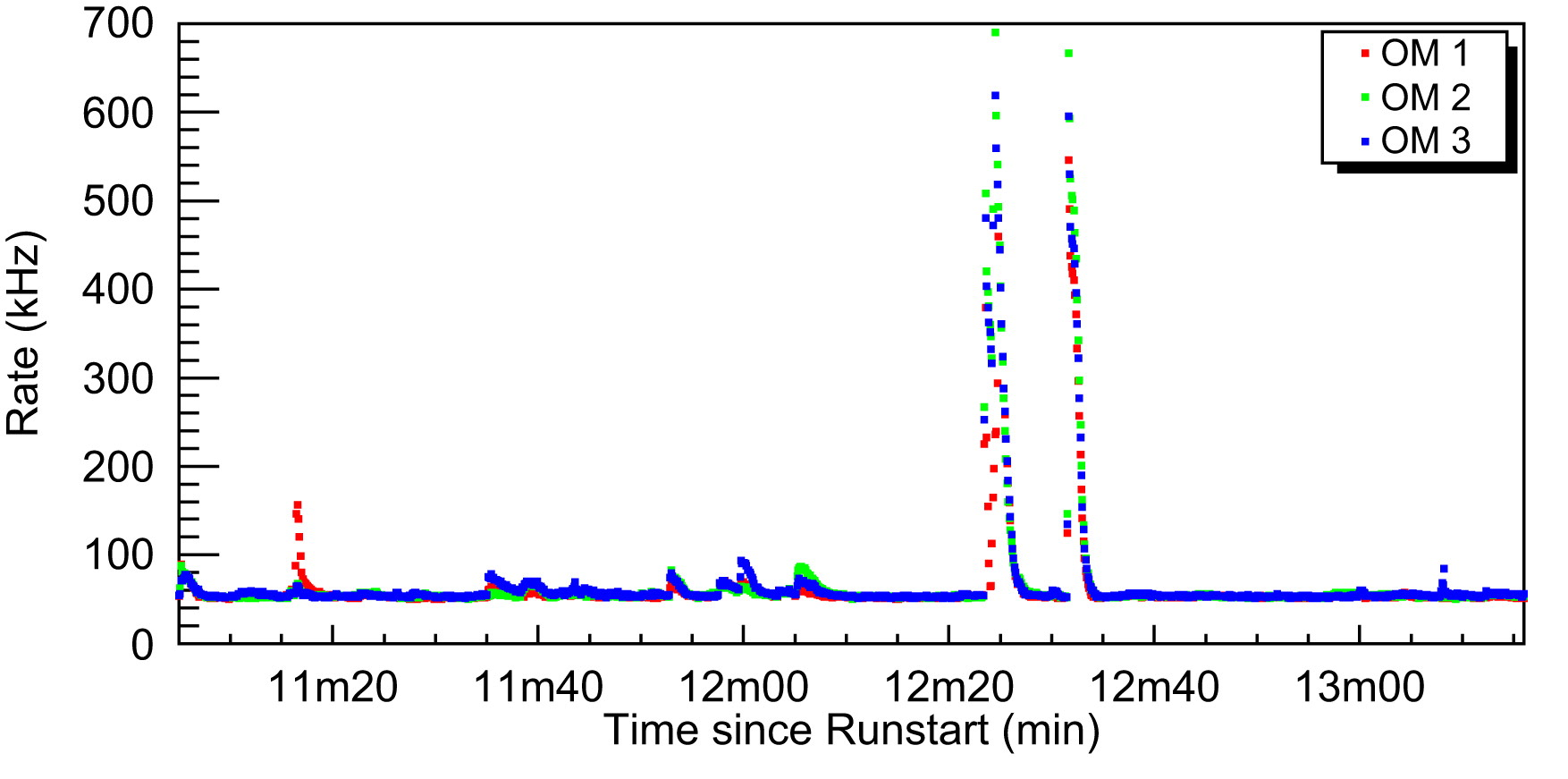}
  \caption{\small Instant counting rates of the three optical modules of one particular floor during a randomly chosen period of 2 min. The \textit{baseline} rate and episodic \textit{bursts} can be extracted from these data streams.}
\label{fig:Base}
\end{figure}

To define the biological activity the counting rate of all optical modules was permanently monitored. 
Figure \ref{fig:Base} shows the observed background rate for the three optical modules of a randomly chosen storey over a period of 2 min. The \textit{baseline} rate is the most probable counting rate of a given OM. A common  baseline to all OMs of $\sim$60 kHz can be identified, which includes the dark noise of the PMTs (4 kHz), the Cherenkov light from radioactive $^{40}$K decays (35 kHz), and spontaneous bioluminescence (likely from bacteria).   
Additionally \textit{bursts} of activity were seen at random times, sometimes uncorrelated even between neighbored modules. 
The \textit{burst fraction} in the detector represented the fraction of time during which the instantaneous counting rate was 20\% higher than the baseline rate. The burst fraction varied strongly with time. 

\begin{SCfigure}
  \centering
\includegraphics[width=0.65\textwidth]%
    {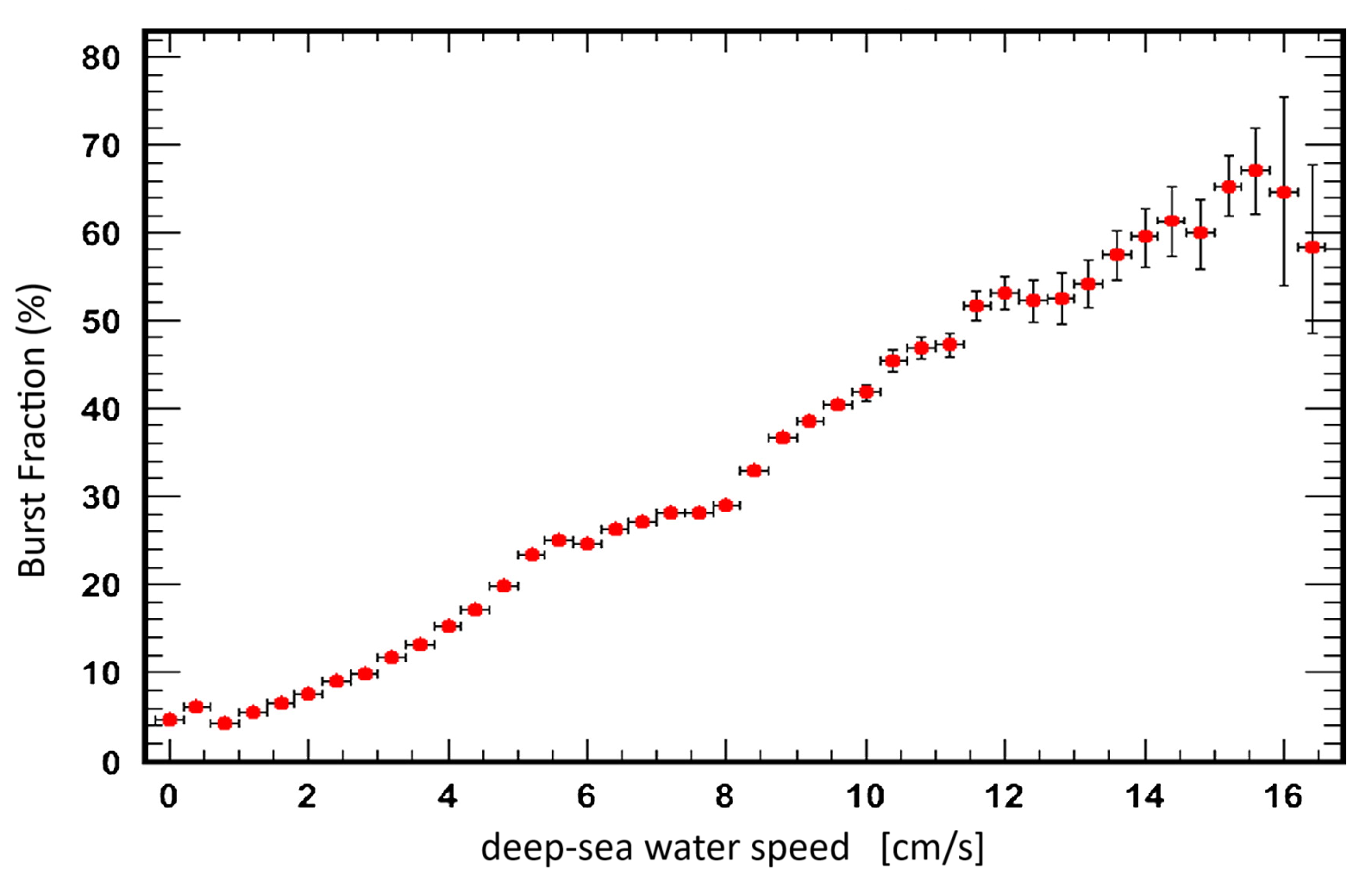}
  \caption{\small Observed correlation between horizontal sea current and burst fraction in the detector.}
\label{fig:BF}
\end{SCfigure}

The bioluminescence bursts are thought to originate from macroscopic animals which are stimulated to emit bioluminescent light, for instance when suffering from turbulent water or at contact with the detector structure  \cite{ANTARES:104}. This hypothesis was supported by the observation of a high correlation between burst fraction and sea current, as illustrated in Fig. \ref{fig:BF}.

The ADCP system was used also for a detailed study \cite{ANTARES:37} of high-frequency internal wave motions of periods down to 20 min, mainly in vertical currents. Such high-frequency internal waves are commonly observed much closer to the sea surface where the vertical density stratification is more stable than in the deep sea.
The combined observation of ADCP vertical and horizontal currents (enhanced levels of acoustic reflection, and high levels in the counting rate of the OMs of the telescope) was interpreted as enhanced presence of suspended particles including zooplankton and consequent increased bioluminescence \cite{ANTARES:20}. These events were coincident with deep dense-water formation occurred in the Ligurian sub-basin, thus providing a possible explanation for these phenomena.

\subsection{Other observations on sea science\label{sec:ESSresults}}
The above observations, as mentioned in \S \ref{sec:biofouling}, reflect the fact that the deep sea hosts numerous pelagic organisms able to emit light. 
A sample of a 2.5-year long record of ANTARES optical data, jointly with synchronous hydrological records, was used for a detailed study of the light emission by deep-sea organisms \cite{ANTARES:31}. This represented the longest continuous time-series of deep-sea bioluminescence ever recorded. The data revealed several weeks long, seasonal bioluminescence blooms with light intensity up to two orders of magnitude higher than background values, which correlate to changes in the properties of deep waters. 
Such changes are triggered by the winter cooling and evaporation experienced by the upper ocean layer that leads to the formation and subsequent sinking of dense water through a process known as \textit{open-sea convection} (see references from 18 to 26 on \cite{ANTARES:31}). It episodically renews the deep water of the study area and conveys fresh organic matter that fuels the deep ecosystems. Luminous bacteria most likely are the main contributors to the observed deep-sea bioluminescence blooms.

The observations demonstrate a consistent and rapid connection between deep open-sea convection and bathypelagic biological activity, as expressed by bioluminescence. 
In a setting where dense water formation events are likely to decline under global warming scenarios enhancing ocean stratification, in situ observatories become essential as environmental sentinels for the monitoring and understanding of deep-sea ecosystem shifts.

\section{Results on particle physics  \label{sec:ppresults}}
Atmospheric muons and atmospheric neutrinos constitute by far the largest fraction of triggered events in neutrino telescopes, representing both an interesting physical signal and the irreducible background for cosmic neutrino study.
The ANTARES detector contributed to the characterization of the muon flux in the deep sea and to the measurement of the atmospheric $\nu_\mu$ and $\nu_e$ spectra in a wide energy range. 
Below 100 GeV, Earth-crossing atmospheric neutrinos are subject to oscillations through the “standard” mass-flavor mechanism. ANTARES contributed to the measurement of the oscillation parameters of the atmospheric sector, set  constraints to 3+1 models of oscillations including a sterile neutrino, and set limits on non-standard neutrino interactions. 
The detector was sufficiently versatile to contribute as well to some of the fundamental open questions of high energy physics: the search for relic massive particles in the cosmic radiation, and the search for dark matter candidates through indirect methods \cite{Heros:20}.

\subsection{Atmospheric muons and neutrinos\label{sec:atmunu}}

Atmospheric muons and neutrinos are produced by cosmic rays (CRs) interacting with atmospheric nuclei. Up to $\sim$100 TeV, muons and neutrinos are produced mainly by decays of charged pions and kaons in the cascade resulting from these interactions, and their spectra are related by the kinematics of the $\pi \to \mu\nu $ and $K \to\ \mu\nu$ decays. Additional lower-energy neutrinos are produced by muon decays. 
The flux of downward-going atmospheric muons exceeds the flux induced by atmospheric neutrino interactions by many orders of magnitude, as shown in Fig. \ref{fig:atmunu} for two underwater depths and two threshold energies for atmospheric neutrinos. 

\begin{SCfigure}
  \centering
\includegraphics[width=0.55\textwidth]%
    {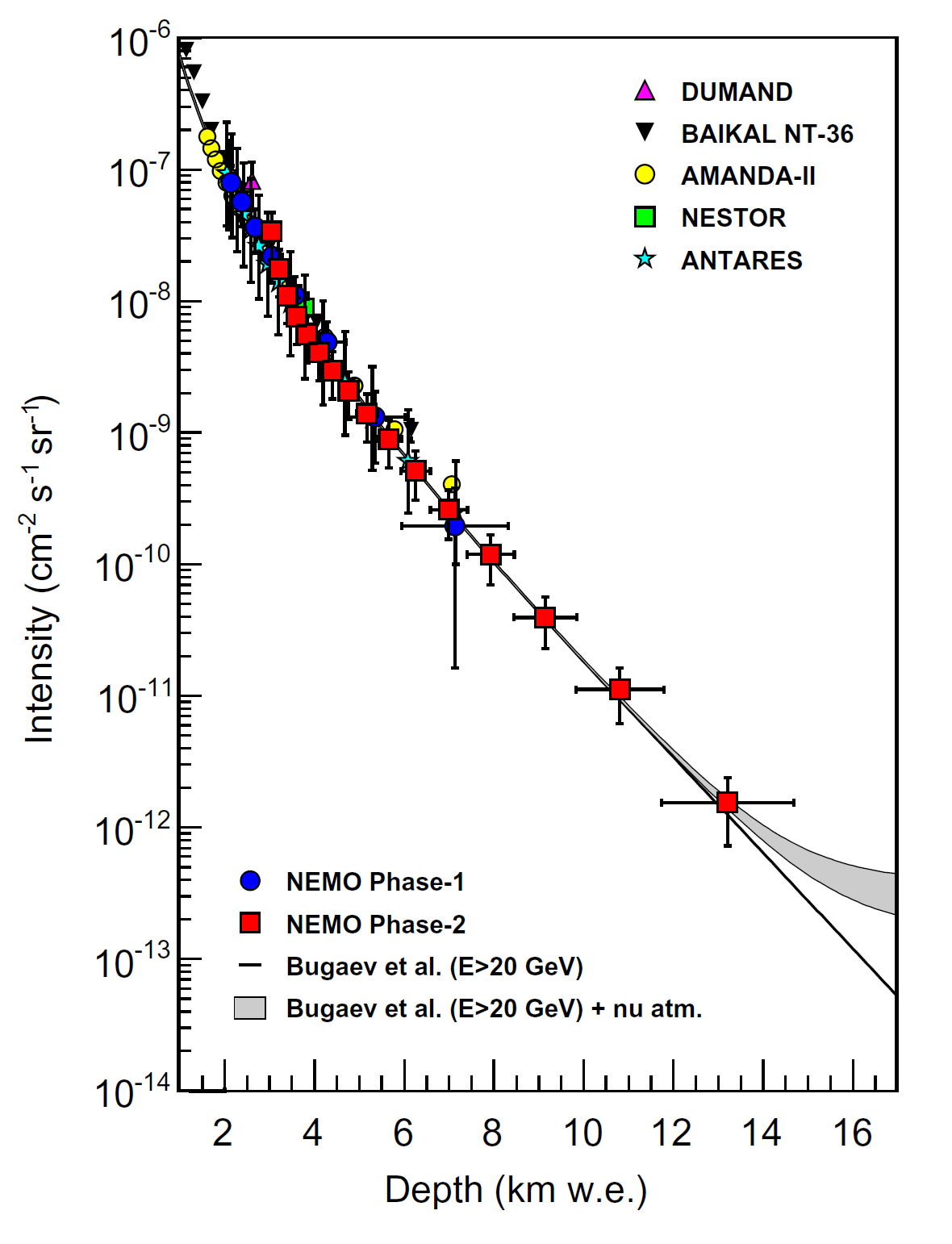}
  \caption{\small \textcolor[rgb]{0,0,0}{Vertical muon intensity versus depth measured in underwater/ice experiments. The solid line is the prediction of Bugaev et al. \cite{Bugaev:98}. The shaded area at large depths includes the contribution of muons induced by $\nu_\mu$ interactions. The ANTARES data are taken from \cite{ANTARES:13}. Refer for details to \cite{Nemo:15}.}}    \label{fig:DIR}
\end{SCfigure}

To be detectable at the depth of ANTARES, muons must have an energy larger than 1 TeV at sea levels; such energetic muons are induced by interacting CRs whose energy is at least one order of magnitude larger.
The atmospheric muon flux is strongly reduced when the amount of material (generally expressed in km of water equivalent, km.w.e.) increases, as shown in Fig. \ref{fig:DIR}. 
The vertical muon flux is determined as a function of the muon slant depth in water, i.e. from the angular distribution, computing the related Depth Intensity Relation (DIR). 
A muon, reaching the detector located at a depth $D$ from a zenith angle $\theta_z\lesssim 70^\circ$ propagates through a water slant $h= D/\cos\theta_z$, neglecting the curvature of the Earth.
\textcolor[rgb]{0,0,0}{The results from the ANTARES experiment reported in the compilation shown in  figure are taken from \cite{ANTARES:13}. }
These observations of penetrating muons relate to primary CR energies not reachable in  CR experiments carried outside the Earth atmosphere and to kinematic regions of hadronic interactions beyond those studied at accelerators or colliders and are used to constrain theoretical models, see for instance 
\cite{Fedy:muon}.

Each ANTARES search for a specific signal (i.e., cosmic or atmospheric neutrino-induced events, penetrating massive particles as  magnetic monopoles) defined analysis-dependent cuts to suppress the background due to atmospheric muons wrongly-reconstructed as upward-going.
The characterization of this background was always done with the  MUPAGE  \cite{MUPAGE:08} generator of muon bundles (see \S \ref{sec:generator}), 
whose  simulated distributions were always in reasonable agreement with those of the data after analysis cuts.
Finally, atmospheric muons were used for a real-time monitoring of the detector status \cite{ANTARES:45}, and for the evaluation of the systematic effects due to uncertainties on environmental and detector parameters \cite{ANTARES:13}, in particular related to the PMT efficiencies \cite{ANTARES:11}. 
The uncertainties assessed in this way were used in all physical studies mentioned in the following.

\begin{figure}[tbh]
  \centering
\includegraphics[width=0.67\textwidth]%
    {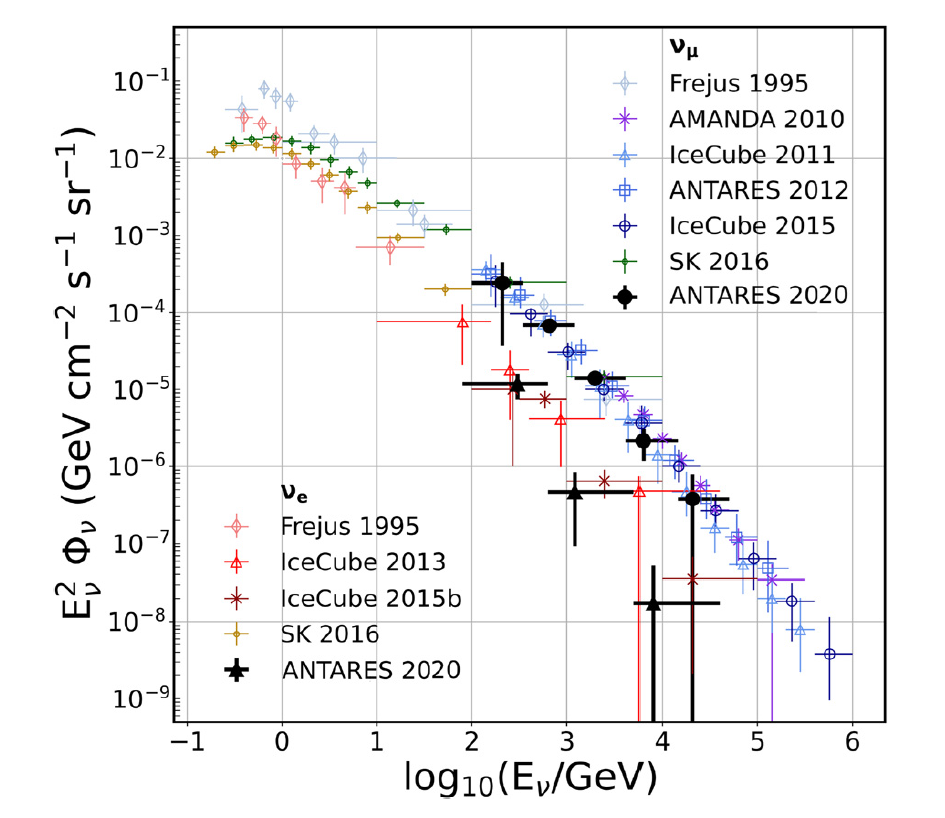}
  \caption{\small Measured energy spectra of the atmospheric $\nu_e$ and $\nu_\mu$ using shower-like and starting track events in the ANTARES detector (black dots) \cite{ANTARES:90}, and the $\nu_\mu$ flux measurement using through-going tracks (blue empty squares) \cite{ANTARES:34}. The vertical error bars include all statistical and systematic uncertainties. The plot includes the measurements by Frejus, AMANDA-II, IceCube, and Super-Kamiokande: refer to \cite{ANTARES:90} for details.}
    \label{fig:atmoneutrinos}
\end{figure}
The flux of {atmospheric neutrinos} from charged pion and kaon decays is usually referred to as the \textit{conventional atmospheric neutrino flux} and it is dominated by the $\nu_\mu$ flavor.
Above 1\,TeV and up~to $\approx$100\,TeV, the flux can be expressed with a simple power-law 
${\text{d}\Phi_\nu}/{\text{d}E }  \propto  E^{-\gamma_\nu}$ 
where $ \gamma_\nu\simeq \gamma +1$ and $\gamma\simeq 2.7$ corresponds to the measured spectral index for CRs below the \textit{knee} (i.e., below a few $10^{15}$ eV). 
In this energy range, the $\nu_e$ flux is expected to be one order of magnitude smaller than that of $\nu_\mu$.
Using the ANTARES data, the  $\nu_\mu + \bar{\nu}_\mu$ energy spectrum was measured using an unfolding method in the energy range 0.1--200 TeV \cite{ANTARES:34}. The observed spectral index  was $\gamma_\nu=3.58\pm 0.12$, compatible with the single power-law dependence and with the measurements of the IceCube collaboration.
With a larger detector lifetime, ANTARES performed in 2021 one of the very few measurements of the atmospheric $\nu_e + \bar{\nu}_e$ sample in the 100 GeV -- 50 TeV energy range. 
In both measurements a detailed estimation of systematic uncertainties, connected to water properties and optical module acceptance, was used.
In Fig. \ref{fig:atmoneutrinos} the compilation of all existing measurement of the atmospheric neutrino spectrum is reported.

A  process similar to the production of atmospheric neutrinos happens when CRs interact with the solar medium, resulting in the production of \textit{ solar atmospheric neutrinos}. These neutrinos represent an irreducible source of background for indirect searches for dark matter from the Sun, \S \ref{sec:DM}. 
The detection of these $\nu$'s would provide useful information on the composition of primary CRs as well as on the solar density.
ANTARES used an unbinned likelihood method to search for solar atmospheric neutrinos \cite{ANTARES:95}: no signal was found and an upper limit at 90\% confidence level (CL) equal to $7\times 10^{-11}$ [TeV$^{-1}$ cm$^{-2}$ s$^{-1}$] at 1 TeV was obtained.

\subsection{Neutrino oscillation and NSI studies\label{sec:nuosci}}

\begin{SCfigure}
  \centering
\includegraphics[width=0.65\textwidth]%
    {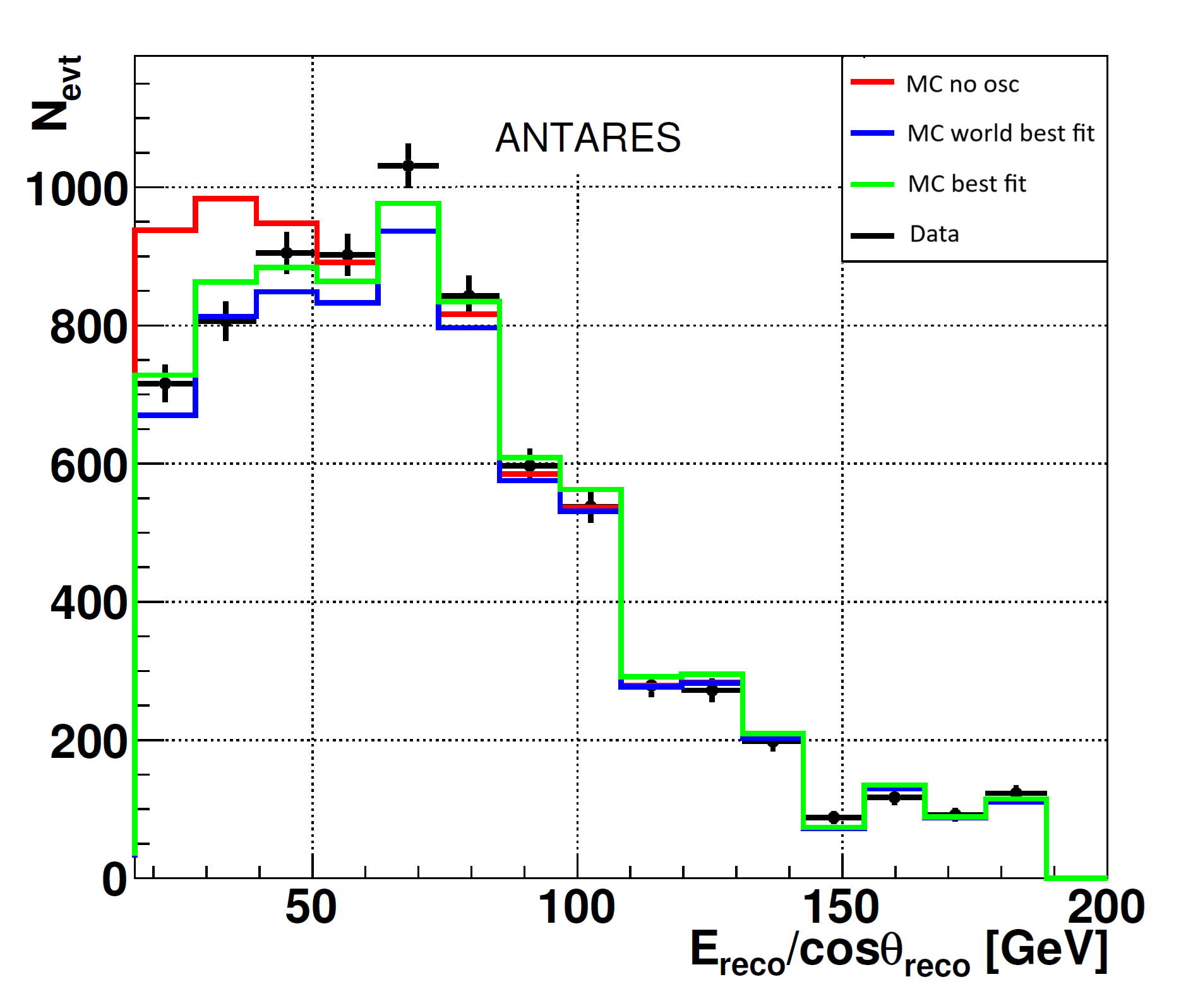}
  \caption{\small $E_{\rm reco}/\cos\theta_{\rm reco}$ distribution for ANTARES data (black), Monte Carlo predictions (MC) without oscillation (red), MC assuming the world's best-fit values (blue), and MC assuming best-fit values of the ANTARES analysis presented  in \cite{ANTARES:79} (green).}
\label{fig:STosci}
\end{SCfigure}

The flux of atmospheric neutrinos at energies $E_\nu< 100$  GeV is affected by neutrino oscillations.  
The ANTARES trigger conditions (\S \ref{sec:DAQ}) allowed  detecting muons induced by $\nu_\mu$ interactions above an energy threshold of $\sim$10--20 GeV, and to study the $\nu_\mu$ disappearance channel using the observed $E_\nu/L_\nu$ ratio.
Once the trajectory of an upward-going $\nu_\mu$-induced track was reconstructed, the distance $L_\nu$ between the neutrino production point (in the atmosphere, $\sim$20 km above the Earth surface) and the detector can be obtained from the measured  zenith angle $\theta_{\rm reco}$. 
The measured length of the muon inside the detector was used as an energy proxy assuming an energy loss of 0.25 GeV/m
\cite{ANTARES:27}. 
Fig. \ref{fig:STosci} shows the distribution of ANTARES data vs. predictions as a function of $E_{\rm reco}/\cos\theta_{\rm reco}$. 
The non-oscillation hypothesis was discarded with a significance of 4.6$\sigma$ \cite{ANTARES:79}, and the allowed values in the  $(\sin^2\theta_{23}, \Delta m^2_{32}$) parameter space were consistent with the world's best-fit values \cite{NUFIT:2020,BARIFIT:21}.

\begin{SCfigure}
  \centering
\includegraphics[width=0.65\textwidth]%
    {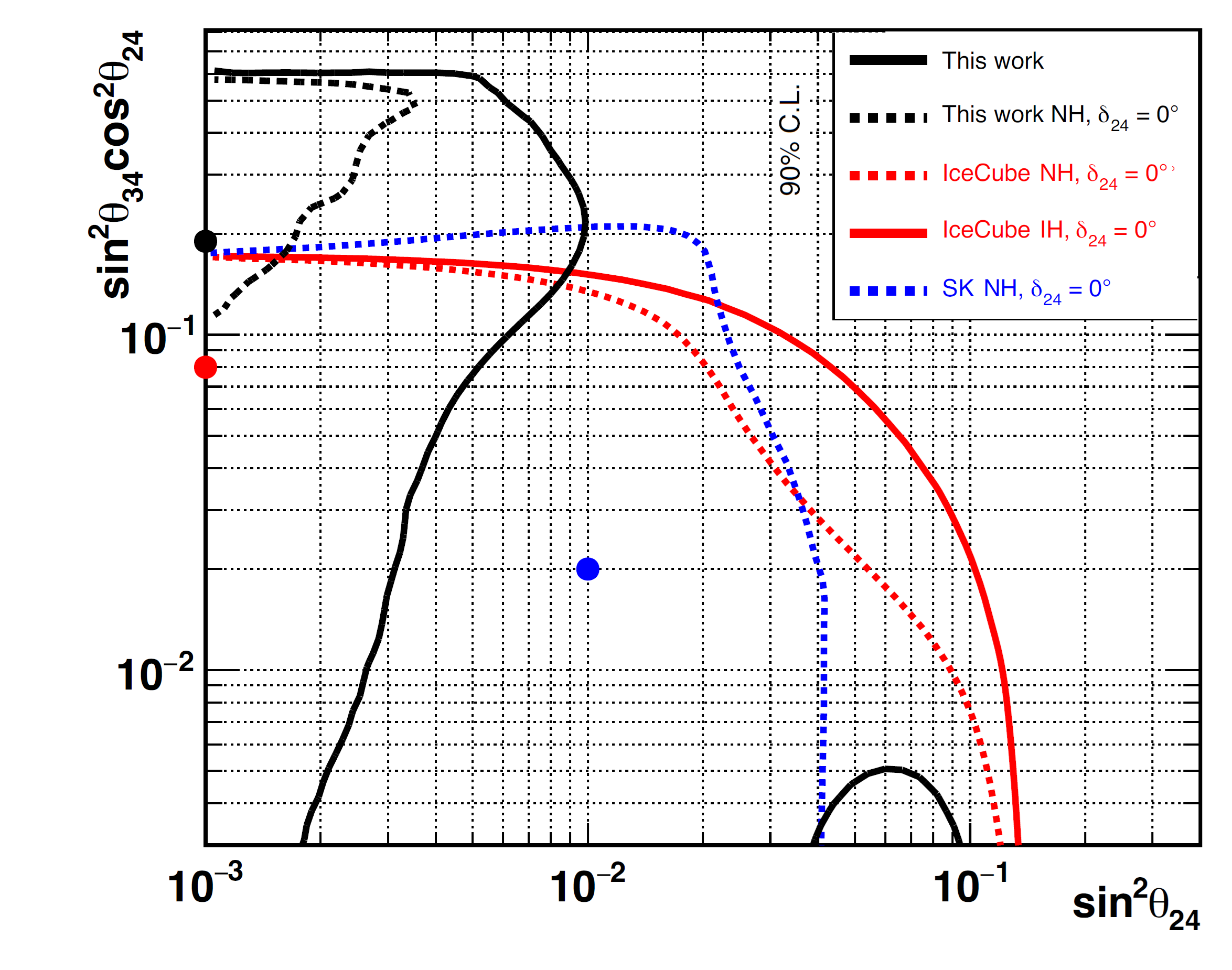}
  \caption{\small ANTARES 90\% CL limits  for the 3+1 neutrino model in the parameter plane of $|U_{\mu 4}|^2\equiv \sin^2\theta_{24}$ and $|U_{\tau 4}|^2\equiv \sin^2\theta_{34}\cos^2\theta_{24}$. Similar upper limits from the IceCube and Super-Kamiokande (SK) experiments are also reported. Refer to \cite{ANTARES:79} for details.}
\label{fig:NSI}
\end{SCfigure}

In a similar way, constraints on the 3+1 neutrino model, which foresees the existence of one sterile neutrino, can be inferred. Even though a sterile neutrino does not interact as active flavors do, its presence would modify the standard oscillation patterns, due to the fact that the $\nu$ mass eigenstates (e.g, that of eigenvalues $m_2$ or $m_3$) could oscillate into an additional fourth sterile state through mixing angles $\theta_{24}$ or $\theta_{34}$. 
Comparing data and Monte Carlo (MC) under these additional oscillation possibilities, exclusion contours were built. In Fig. \ref{fig:NSI}  the resulting ANTARES 90\% CL exclusion limits were computed on a 2D grid in the plane of the oscillation matrix elements depending on mixing angles $\theta_{24}$ and $\theta_{34}$ for $\Delta m^2_{41} > 0.5$ eV$^2$.

Non-standard interactions (NSIs) of $\nu$'s arise in many beyond Standard Model theories: they can alter atmospheric $\nu$ propagation in Earth through matter effects \cite{Wolf:78, MikeSmir:85}. NSIs are quantified through dimensionless constants $\epsilon_{\alpha\beta}$ ($\alpha, \beta=e, \mu,\tau$)  that appear in the four-fermion Lagrangian containing new interaction terms. ANTARES \cite{ANTARES:97}  used a log-likelihood ratio test on $\epsilon_{\mu\tau}$ and $\epsilon_{\tau\tau}-\epsilon_{\mu\mu}$, which provided no evidence of deviations from standard interactions. 
The derived constraint on $\epsilon_{\mu\tau}$ in the ${\mu-\tau}$ sector is among the most stringent to date for NSIs. 

\subsection{Dark matter searches\label{sec:DM}}

Neutrinos can be trackers of the annihilation of dark matter (DM) particles inside an astrophysical environment. A common hypothesis assumes that dark matter could be composed of (yet unobserved) weakly interacting massive particles (WIMPs), which naturally display an interaction strength of the same order as the weak interaction \cite{Bertone:08}. WIMPs might annihilate or decay producing {Standard Model} (SM) particles that can decay with a neutrino in the final state. 
Thus, indirect evidence for WIMPs can be sought in astrophysical regions which are dense enough to bind WIMPs gravitationally \cite{Bertone:05}. 

These regions must have a small angular size (as the Sun); the case of the Galactic Center is remarkable as is the unique extended and close-by over-dense region.
All indirect DM search strategies rely on assumptions on the annihilation channels for the secondary neutrino yields, which impact the shape of the signal that would appear in the detector \cite{Cirelli:11}. 

ANTARES, located in the Northern hemisphere, was able to perform a search looking towards the {Galactic Center} using upward-going neutrinos.
The expected differential flux of secondary neutrinos from DM self-annihilation in the Galactic Center is ~\cite{NeutrinoFluxDM}:
\begin{equation}
    \frac{\mathrm{d}\phi_{\mathrm{\nu}}}{\mathrm{d}E_{\mathrm{\nu}}} = \frac{1}{4\pi} \, \frac{\langle \sigma_A \upsilon \rangle}{2 \, m_{\mathrm{DM}}^2} \; \frac{\mathrm{d}N_{\mathrm{\nu}}}{\mathrm{d}E_{\mathrm{\nu}}} \; J \, ,
    \label{eq:sig_expectation}
\end{equation}
where $\langle\sigma_A\upsilon\rangle$ is the thermally-averaged self-annihilation cross section, $m_{\mathrm{DM}}$ is the mass of the DM particle and $\mathrm{d}N_{\mathrm{\nu}}/\mathrm{d}E_{\mathrm{\nu}}$ is the differential number of neutrinos per annihilating DM pair. 
The $J$-factor is defined as the integral over the solid angle, $\Delta \Omega$, of the squared dark matter density evaluated along the line of sight. It strongly depends on the theoretical Galactic halo model that is used. 
The density distribution of DM as a function of the distance $r$ to the Galactic Center is one of the most critical parameters, as different models exist that differ in a significant way. 
\begin{figure}
  \centering
\includegraphics[width=0.89\textwidth]%
    {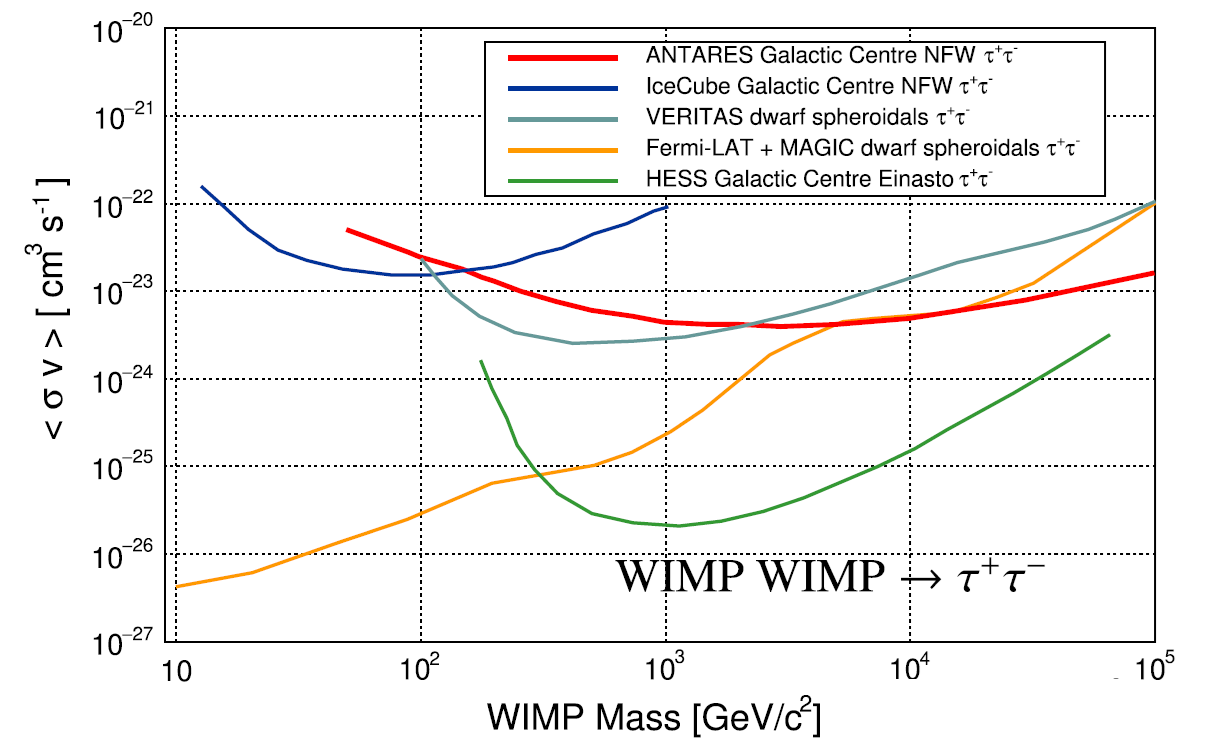}
  \caption{\small Limits from the Galactic Center on the thermally-averaged cross section for self-annihilation WIMP pairs set with ANTARES (red line),  IceCube and from $\gamma$-ray telescopes H.E.S.S., VERITAS and Fermi-LAT + MAGIC. All curves are for the $\tau^+\tau^-$ benchmark channel. Refer to \cite{ANTARES:84} for further details.}
\label{fig:DMGC}
\end{figure}

The ANTARES selection criteria for these analyses were defined to maximize the sensitivity to possible signals produced by the self-annihilation of WIMPs with respect to the atmospheric background.
The directional information arising from $\nu_\mu$'s yielding track events and the spectral features of annihilating DM pairs entered  into the unbinned likelihood method using 3170 days of data \cite{ANTARES:84}. This study updated a previous binned analysis of 1321 days of data \cite{ANTARES:43}
and one with an unbinned method using 2102 days of data \cite{ANTARES:56}.
The non-observation of DM was converted into limits on the velocity-averaged cross section for self-annihilation WIMP pairs, as shown in Fig. \ref{fig:DMGC} when the Navarro-Frenk-White \cite{NFW:96} halo profile is used.
Similar null results were obtained when combining ANTARES data with that of the IceCube collaboration \cite{ANTARES:86}.

A second interesting region candidate is the {Sun center}, which is not an expected source of neutrinos above a few tens of MeV.
In the Sun (containing both nuclei with odd number of nucleons and relatively heavy elements) WIMPs can interact with ordinary matter through spin-dependent and spin-independent interactions. In the former case, the WIMP coupling is through the spin of the target nucleon, while in the latter coupling is through the nucleus mass. 
The upper limits placed by ANTARES \cite{ANTARES:35,ANTARES:51} are competitive with those obtained by both direct and indirect  searches in the 100--1000 GeV/c$^2$ DM mass range, in particular for the case of spin-dependent WIMP-nuclei  scattering interactions. 

The possibility that neutrinos can be produced by DM annihilation in the center of the Earth~\cite{ANTARES:59} was also considered. The signature would be an enhancement of vertically upward going tracks.  No events were found and the obtained limits are particularly stringent for DM masses close to 56~GeV/c$^2$, the mass of iron nuclei, which dominates the Earth core.

An  hypothesis alternative to the previous capture and annihilation scenario is based on the idea that DM is \textit{secluded} from SM particles and that the annihilation is only possible through a metastable mediator ($\phi$), which subsequently decays into SM states \cite{secluded:08,secluded:09}. These models retain the thermal relic WIMP DM scenario while at the same time explain the positron-to-electron ratio observed by different space experiments, see \cite{REVDM:24} for a recent review.
In the secluded scenario, the presence of a mediator dramatically changes the annihilation signature of DM captured in the Sun.
If the mediators $\phi$ live long enough to escape the massive objects before decaying, they can decay into leptons near the Earth or produce fluxes of charged particles, $\gamma$-rays or neutrinos that could reach the Earth and be detected. 

The signature of leptons arising from $\phi$ decays may differ substantially from other DM models when decaying into $\mu^+\mu^-$,  $\tau^+\tau^-$, $b \bar{b}$, or $\nu \bar{\nu}$.
A search in the ANTARES data, based on an optimized selection of signal events from the direction of the Sun \cite{ANTARES:50} over the atmospheric background, was performed.
The result was consistent with the background-only hypothesis, and 90\% CL upper limits on WIMP-nucleon spin-dependent and spin-independent cross sections as a function of WIMP mass were set for different values of the mediator lifetimes.
A second analysis considered the case of the Galactic Center.  
This secluded scenario allows the theory to accommodate heavier DM particles than WIMPs \cite{Cirelli:19}. 
The ANTARES detector explored for the first time masses up to 6 PeV exploiting the performance of the detector at high energies, and velocity-averaged cross sections \cite{ANTARES:96} were compared with theoretical expectations for the maximal possible annihilation signals in different models.

\subsection{Stable massive particle searches\label{sec:exotic}}

Stable massive particles \cite{SpurioSMP:19}, e.g. magnetic monopoles, strange quark matter and supersymmetric particles, are not included in the Standard Model: they could be present in the cosmic radiation, and can be searched for in large detectors as the neutrino telescopes. In fact, stable massive particles are sufficiently long lived that they can be directly observed via strong and/or electromagnetic interactions in a detector rather than via their decay products. 
Their stability means that if they were produced at any time in the thermal history of the Universe, they would still be present as relic particles. Their motivation is usually connected with general considerations on cosmology and dark matter.
The searches of relic particles is also a fundamental aspect of many astroparticle physics experiments in space, on the Earth surface and underground/ice/water \cite{PatSpu:15}. 

\subsubsection{Magnetic monopoles}
Magnetic monopoles ($\cal M$) are particularly intriguing particles related also to the inner symmetries of electromagnetic interactions.
Their existence (first hypothesized by Paul Dirac in 1931) would restore the symmetry in Maxwell’s equations with respect to magnetic and electric fields \cite{Dirac:31}. 
The interesting aspect of the Dirac theory is that the existence of a free magnetic charge $g_D$ would explain the quantization of the electric charge, $e$. 
Dirac established the basic relation between $e$ and $g$ as:
\begin{equation}\label{eq:1.g}
\frac{eg}{c} =\frac{n\hbar}{2}  \longrightarrow g = n\cdot g_D = n\cdot \frac{1}{2} \frac{\hbar c}{e}  \sim~n\cdot \frac{137}{2} e \ ,
\end{equation}
where $n$ is an integer.

Magnetic monopoles could be produced in the early Universe \cite{Kibble:76} and, as the Universe expanded and cooled down, the energy of $\cal M$ decreased. 
After galaxy formation, $\cal M$ were re-accelerated by galactic magnetic fields, yielding an isotropic intergalactic flux of relatively high-energy objects.
For the typical values in our Galaxy, 
a $\cal M$ with the Dirac charge $g_D$ is relativistic up to $m_{\cal M}\simeq 10^{11}$ GeV/c$^2$. 
In models in which the cosmic magnetic field is strongly correlated with the large-scale structure of the Universe, or the monopole crosses several magnetic field domains over the lifetime of the Universe, they are relativistic up to  $m_{\cal M} \lesssim 10^{14}$ GeV/c$^2$ \cite{Ryu:98,Wick:03}.
An upper bound on the expected flux (called the Parker bound \cite{Parker:70}) was obtained by requiring the rate of 
energy gain of the accelerated $\cal M$s to be small compared to the timescale on which the galactic magnetic field can be regenerated. The Parker bound corresponds to a flux $<10^{-15}$ cm$^{-2}$ s$^{-1}$ sr$^{-1}$ for $m_{\cal M}< 10^{17}$ GeV/c$^2$.

A $\cal M$ induces Cherenkov radiation if its velocity is larger than the phase velocity, i.e.  $v=c/n\gtrsim 0.74$ in water, referred to as the Cherenkov limit.
The number of Cherenkov photons induced by a $\cal M$ with magnetic charge $g$  is described by the Frank-Tamm formula, eq. \ref{eq:FT}, with the replacement $e \to gn$ \cite{MMloss:65}.
For a $\cal M$ with the Dirac magnetic charge this would corresponds to a light yield $(g_D/e)^2\simeq 8200$ larger than that emitted by a muon at the same velocity.

Part of the $\cal M$ ionization energy loss is transferred to the medium in collisions large enough to knock the electrons out of their atomic orbits (referred to as $\delta$-rays), which could have enough kinetic energy to induce additional Cherenkov light. 
The production of $\delta$-rays is described by the differential cross section of Kasama, Yang and Goldhaber (KYG) \cite{MMKYG:77} or by the more conservative (in terms of photon yield) Mott cross section \cite{MMAhlen:76}.

\begin{figure}
  \centering
\includegraphics[width=0.85\textwidth]%
    {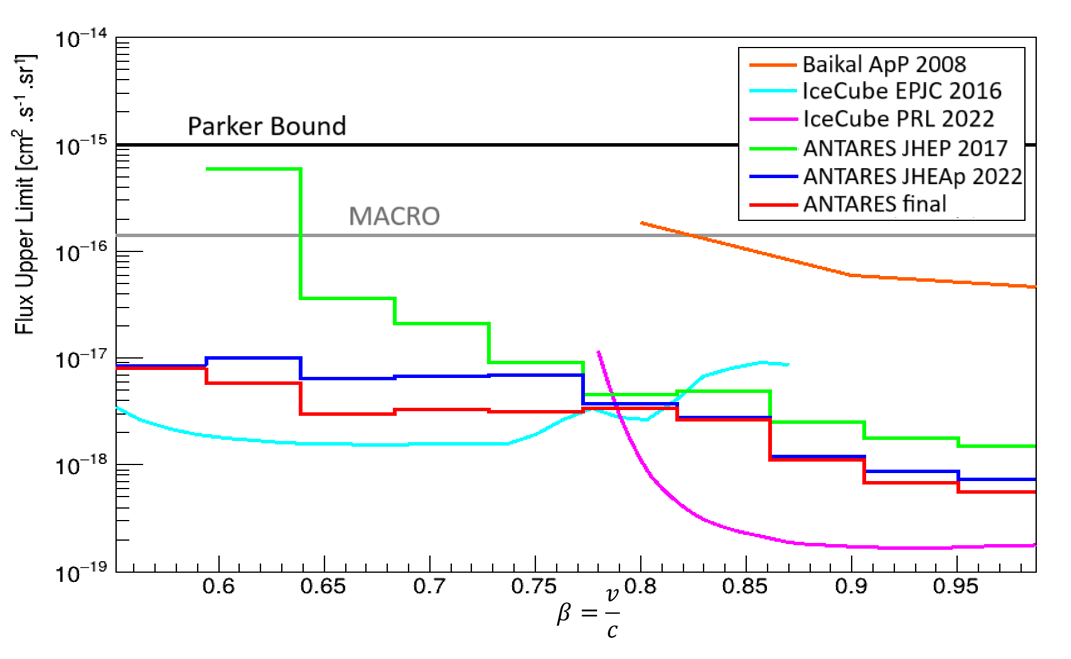}
  \caption{\small 90\% CL upper limit on the magnetic monopole flux  obtained using the whole ANTARES data set (red line) and in previous analyses (green line  \cite{ANTARES:63} and blue line \cite{ANTARES:94}). Upper limits from MACRO (gray line \cite{MACRO:02}), IceCube (cyan  \cite{MMIceCube:16} and magenta \cite{MMIceCube:22}), and Baikal (orange \cite{MMBaikal:08}) as well as the theoretical Parker bound (black \cite{Parker:70}) are reported. }
\label{fig:MM}
\end{figure}

An initial search for $\cal M$ in the ANTARES data \cite{ANTARES:63} was based on the Mott cross section, which starts to produce $\delta$-rays for the minimum monopole velocity of $v_{\cal M} \simeq 0.60c$. 
In order to remove the bulk of downward-going events from atmospheric origin, only upward-going candidates were searched for. 
In ANTARES data, no signal above the background expectation was observed, and upper limits on the $\cal M$ flux 
were set for velocities $0.6<v/c<0.99$,  $g=g_D$ and mass in the range $10^{10} \lesssim m_{\cal M}\lesssim 10^{14}$ GeV/c$^{2}$ (green histogram of Fig. \ref{fig:MM}).
The final analysis used the whole data set (updating that reported in \cite{ANTARES:94}) and the KYG cross section for $\delta$-rays production. Under this assumption,  the detector is sensitive to $\cal M$ with velocities $v_{\cal M}>0.55c$; no candidates were found and the upper limits given by the red histogram in Fig.  \ref{fig:MM} were obtained. The limit set by the MACRO experiment \cite{MACRO:02} refers also to downward-going candidates, surviving the 3000 meters of water equivalent of the Gran Sasso mountain overburden. Thus, their limit holds for ${\cal M}$ of lower mass (starting from $10^6$ GeV/c$^2$). 

\subsubsection{Strange Quark Matter}
In 1984 Witten formulated the hypothesis \cite{Witten:84} that strange quark matter composed of comparable amounts of $u, d$ and $s$ quarks might be the ground state of hadronic matter.
The term {nuclearites} (${\cal N}$) is used to design higher mass (mass number $A> 10^7$) objects. They are electrically neutral atom-like systems, as they would be expected to possess a electron cloud around the core. For $A > 10^{15}$, electrons would be largely contained within the bag of nuclear matter. 
Nuclearites are generally assumed to be bound to astrophysical objects, and with a speed determined by the virial theorem: in the case of ${\cal N}$ bound to the Milky Way, $v_{\cal N}\sim~10^{-3}c$.
The main energy loss mechanism for ${\cal N}$ passing through matter is elastic or quasi-elastic atomic collisions. Hence, they leave a distinct signal during the passage in a transparent medium such as water: the signal could be detected by using the light emission from their overheated path as a black-body radiation from an expanding cylindrical thermal shock wave \cite{DeRujula:84}. 

\begin{SCfigure}
  \centering
\includegraphics[width=0.7\textwidth]%
    {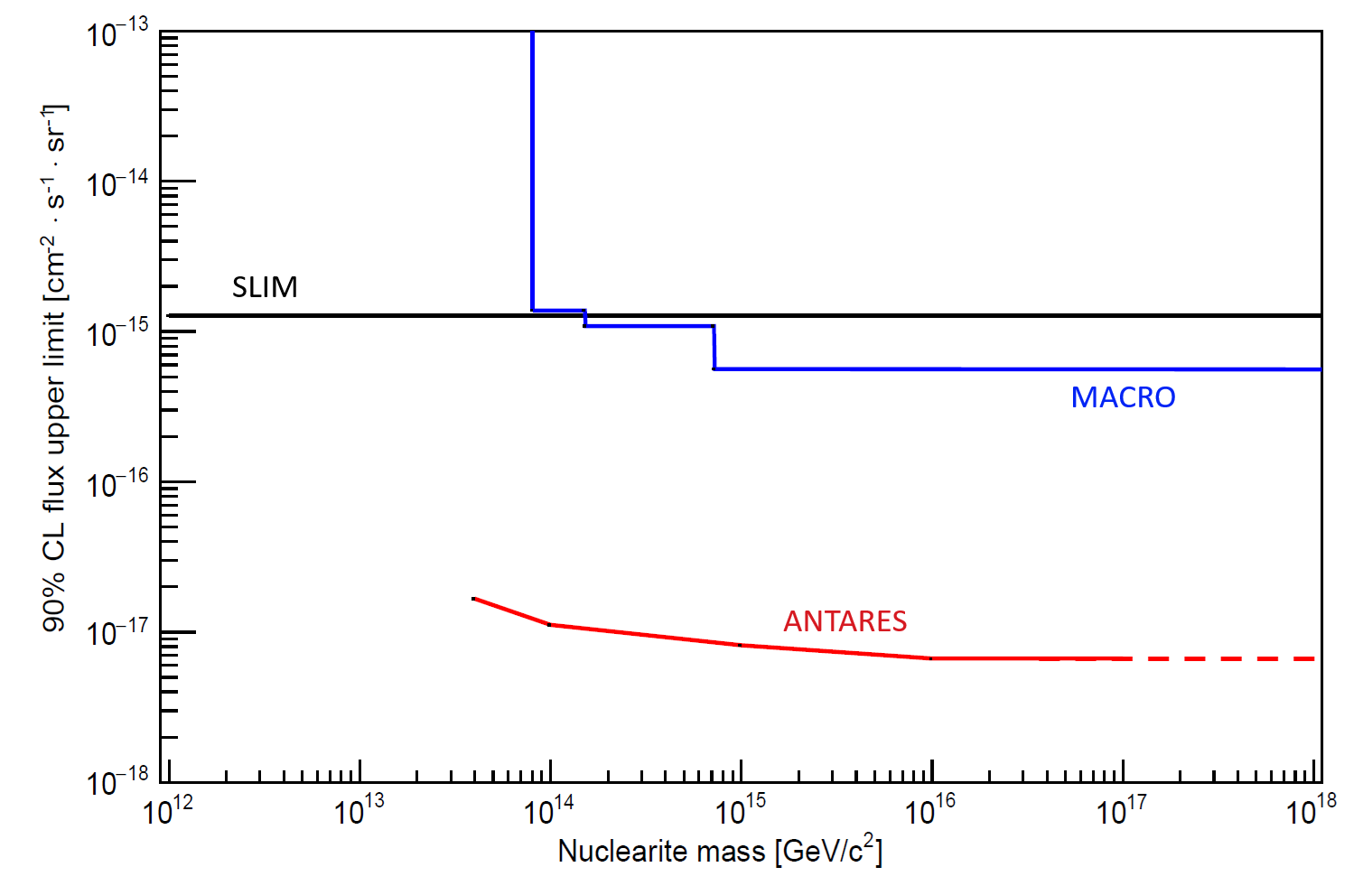}
  \caption{\small ANTARES 90\% CL upper limit (red line) on the flux of nuclearites with Galactic velocities ($v = 10^{-3}c$), using data collected between 2009 and 2017. Limits from MACRO \cite{MACROnuclea:99} and SLIM \cite{SLIMnuclea:08} experiments are also reported.}    
\label{fig:nuclearite}
\end{SCfigure}

The ANTARES collaboration searched for nuclearites \cite{ANTARES:99} by simulating the energy loss process during the passage through matter of these particles, and with a detailed description of the detector response and of the data acquisition conditions. 
A downward-going flux of cosmic ${\cal N}$ with Galactic velocities and different masses $m_{\cal N}$ was considered. 
The mass threshold for detecting these particles at the detector level is $m_{\cal N}\simeq 10^{13}$ GeV/c$^2$. 
Nuclearites with $m_{\cal N}\gtrsim  10^{20}$ GeV/c$^2$ are stopped before reaching the detector.
The 90\% CL upper limit on the ${\cal N}$ flux set by ANTARES is reported in Fig. \ref{fig:nuclearite}. 
The dashed interval for masses above $10^{17}$ GeV/c$^2$ represents the region where the signal would be similar or larger to that produced by the highest simulated nuclearite mass.
The MACRO limit (under a rock coverage in m.w.e. similar to that of ANTARES) is valid for the same mass range as ANTARES, but covers a wider velocity range ($4\times 10^{-5}<v/c<0.5$). 
The SLIM experiment, being at the Chacaltaya high altitude laboratory, can be reached by nuclearites with masses $m_{\cal N}\gtrsim 3\times  10^{10}$ GeV/c$^2$ at $v/c=10^{-3}$.

\section{Results on astrophysics\label{sec:apresults}}

Evidence of the existence of a high-energy diffuse flux of cosmic neutrinos has emerged in the last decade from several observations by the IceCube collaboration. 
A few individual source contributions have been so far identified: neutrinos from the blazar TXS 0506+056, neutrinos from the active galaxy NGC 1068, and neutrinos from the Galactic Plane. 
In addition, there have been set a number of interesting constraints on source classes which are believed to contribute significantly to the astrophysical diffuse flux.

A neutrino detector located in the Mediterranean Sea can monitor, with upward-going tracks, the Southern sky hemisphere including most of the Galactic Plane and its Center. Furthermore, the inherent optical properties of seawater compared to ice guarantee a better angular resolution for events arising from $\nu_\mu$ CC interactions.
The geographical location of the IceCube detector, on the other hand, offers a privileged  point of view for studying the Northern sky. 
This section describes the contribution of the ANTARES detector to these studies.

\subsection{Full-sky diffuse flux\label{sec:adiffuse}} 

Neutrinos created in the atmosphere by interactions of charged cosmic rays are indistinguishable from neutrinos created in cosmic sources: 
without focusing on directional searches, the only difference between background and signal is given by their energy spectra.
The first observation of an excess of high-energy astrophysical neutrinos over the expected background was reported by the IceCube collaboration using data collected from May 2010 to May 2013 \cite{IChese:13}, the  High Energy Starting Events (HESE). The latest update is in \cite{ICHESE:22}.
The excess was observed when focusing on events that had their reconstructed interaction vertex contained within the instrumented ice volume, using the edges of the IceCube detector as a veto. 
The excess was at neutrino energies above 60 TeV, arising primarily from the Southern hemisphere, with a sample mostly composed by shower events with poor angular resolution (${\sim}15^\circ$). 
This prevented an accurate localization of the parent neutrino's direction in the sky.
Events above the PeV energy were also observed \cite{ICpev:13}.

A second sample of events from the IceCube detector with a significant contribution of cosmic neutrinos consisted of upward-going muon tracks, stemming from CC interactions of muon neutrinos \cite{ICpasmu:16}.
The field of view for these events is restricted to the Northern Sky (NS) hemisphere and their energy  ranges from 15 TeV to 5 PeV (NS tracks, updated in \cite{ICpasmu:22}).
Additional event samples detected with the IceCube detector showed a significant contribution of cosmic neutrinos:
a  sample  of cascades \cite{ICcasc:20}, dominated ($\sim$90\%) by cosmic $\nu_e$ and $\nu_\tau$ interactions in the  energy range from 16 TeV to 2.6 PeV; a sample of events with a contained vertex and exiting track induced by $\nu_\mu$'s undergoing a CC interaction (ESTES) in the 3--550 TeV energy range \cite{ICStarting:24}.

\begin{figure}[tbh]
  \centering
\includegraphics[width=0.8\textwidth]%
    {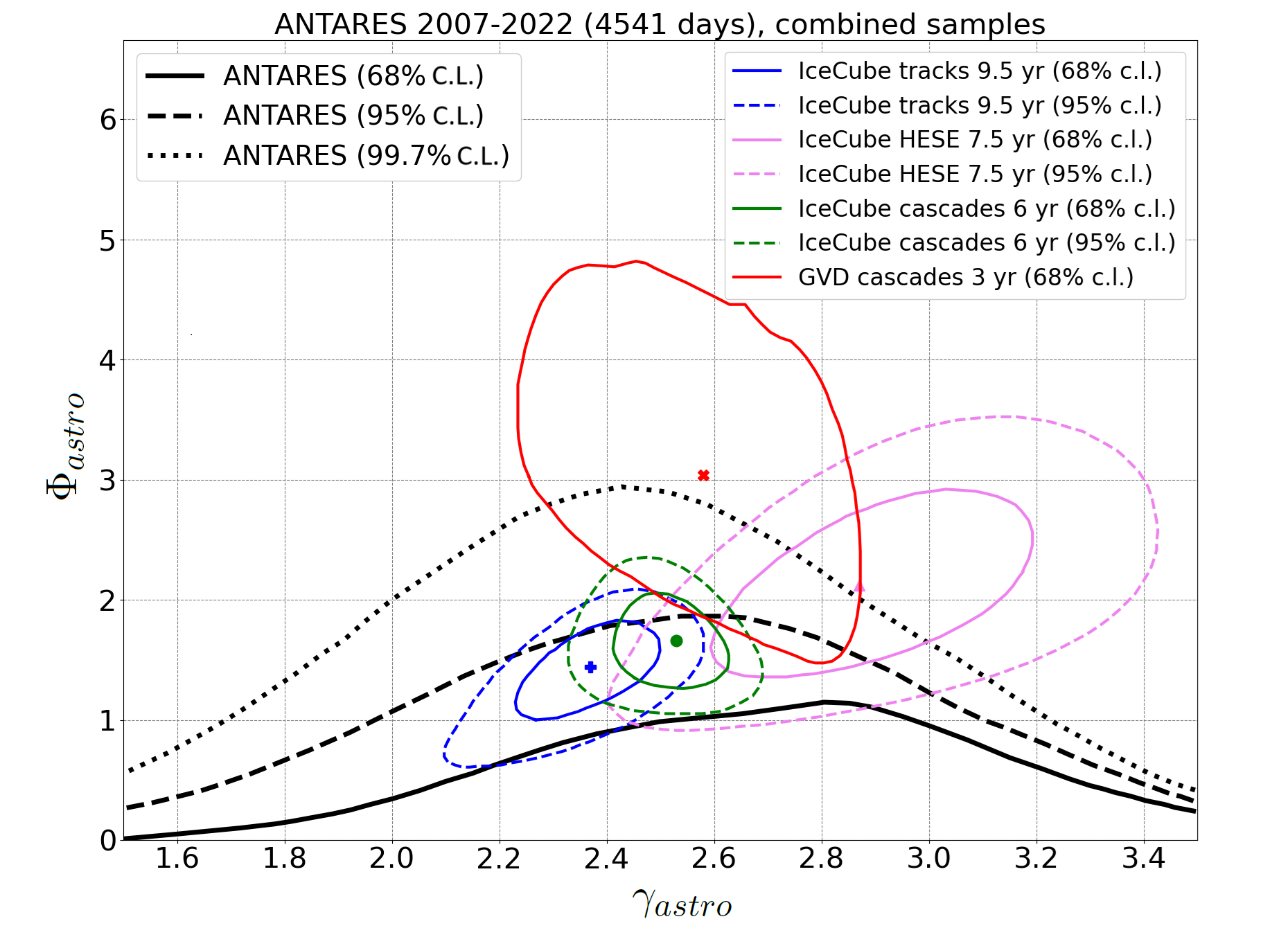}
  \caption{\small Space of the $(\gamma_{astro}, \Phi_{astro})$ parameters defined in eq. \ref{eq:fit}.
  Contours at 68\% (solid) and 95\% (dashed) CL from IceCube analyses (HESE \cite{IChese:13} in pink, tracks \cite{ICpasmu:22} in blue, cascades \cite{ICcasc:20} in green) compared to the 68\% (solid), 95\% (dashed), and 99.7\% (dotted) posterior probability credible areas obtained in the combined analysis of the  ANTARES samples (black lines). The IceCube best-fit points are shown with symbols. The Baikal-GVD 68\% CL contour and best-fit point \cite{GVDdiff:23} are also shown in red. Adapted from \cite{ANTARES:106}.}
\label{fig:diffuse}
\end{figure}

In all the mentioned samples, a single power law flux 
\begin{equation}\label{eq:fit}
\frac{d \Phi_{1\nu}}{dE}= {\Phi_{astro}}\cdot 10^{-18} 
\biggl( \frac{E}{100 \textrm{ TeV}} \biggr)^{-\gamma_{astro}} \textrm{ GeV}^{-1}\textrm{ cm}^{-2}
\textrm{ s}^{-1}\textrm{ sr}^{-1}  
\end{equation}
was assumed. The  energy range reported above for each sample was computed under this assumption. 
A summary of the IceCube results is shown in Fig. 19 of \cite{ICStarting:24}. Fig. \ref{fig:diffuse} shows the 68\% and 95\% CL regions for some of the mentioned  IceCube samples.
The allowed parameter space presents some 68\% CL non-overlapping regions.
The  diffuse flux from unresolved sources observed by the IceCube collaboration can originate from faint extragalactic sources, by interactions during hadron propagation, and can include a possible contribution from our Galaxy. 

ANTARES, embedded in a different detection medium, tried to measure the cosmic neutrino diffuse flux exploiting its different sky coverage with  upward-going events \cite{ANTARES:15}.
Using track-like and shower-like events in 9 years of data \cite{ANTARES:71}, ANTARES found a mild excess of high-energy events over the expected atmospheric background in both samples. The fit to eq. \ref{eq:fit} yielded best-fit values of $(\Phi_{astro},\gamma_{astro})=(1.7\pm 1.0, 2.4\pm 0.5)$.  The significance of the excesses was estimated to be $\sim$1.6$\sigma$ and the null-cosmic hypothesis rejected at 85\% CL.
 
\begin{figure}[tbh]
  \centering
\includegraphics[width=0.9\textwidth]%
    {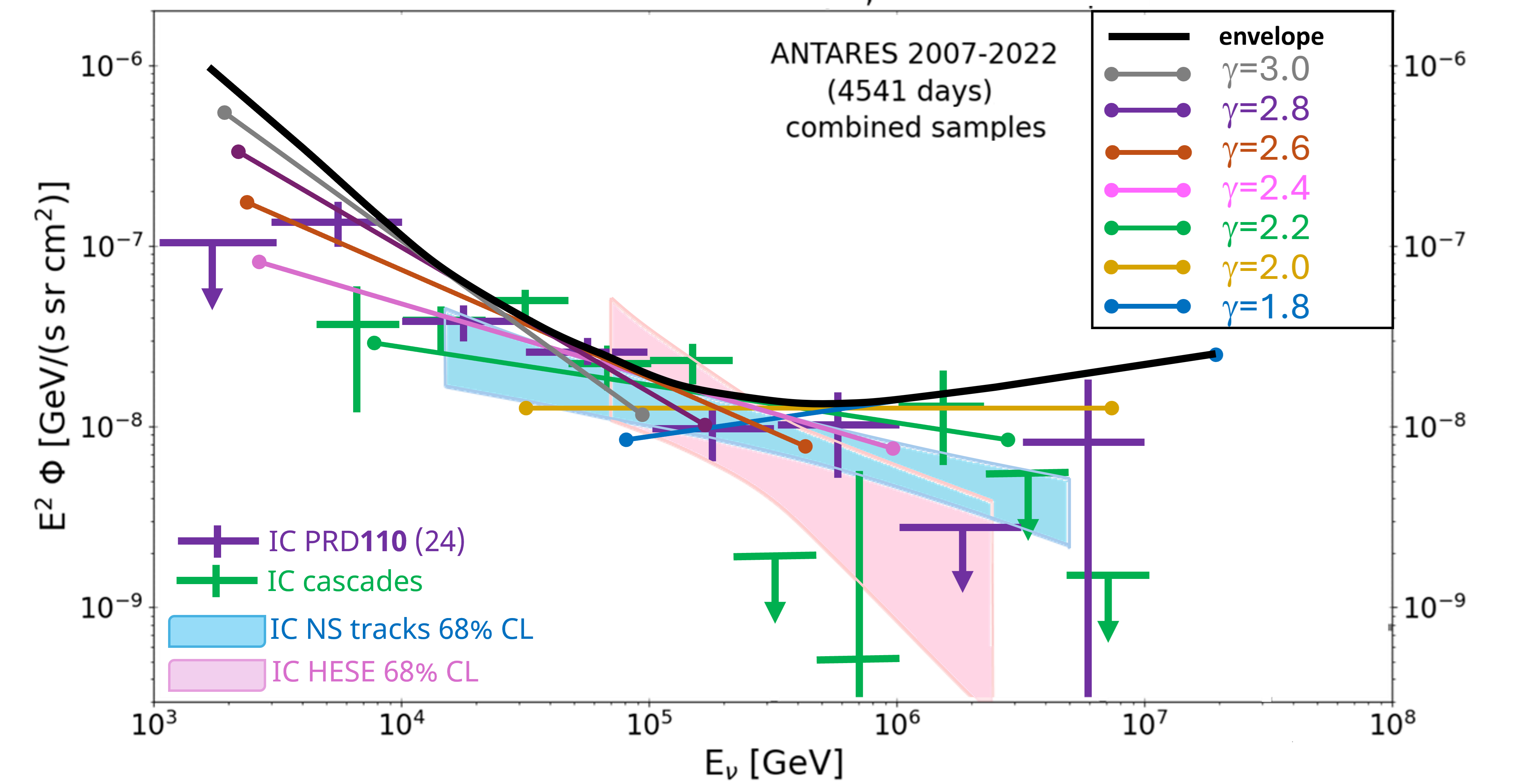}
  \caption{\small  The ANTARES 15-years 95\% probability upper limits for different spectral indexes $\gamma_{astro}=3.2, 3.0,\dots , 1.8$ (colored lines in the legend) are reported in the figure. The envelope of the limits (black) is taken as the least restrictive limit at every energy.
	The shaded areas represent the 68\% CL intervals for the measurements obtained with the IceCube HESE sample \cite{ICHESE:22} in pink and the IceCube track sample \cite{ICpasmu:22} in blue. The green points correspond to the cascade sample \cite{ICcasc:20}. The results from the $E^{-2}$ segmented fit of the IceCube combined samples \cite{ICStarting:24} are also shown in violet. }
\label{fig:diflim}
\end{figure}

The final ANTARES analysis \cite{ANTARES:106} using all the available data sample (4541 days of livetime) did not provide a statistically significant observation of the cosmic diffuse flux. 
The energy range of validity of the ANTARES data fit to equation \ref{eq:fit} for different values of $\gamma_{astro}$ was estimated.
For soft spectra, the sensitivity extends to the TeV region, below what has been obtained with IceCube data.
As already pointed out by the IceCube collaboration, the hypothesis of a single unbroken power-law spectrum over the  energy range from TeV to multi-PeV may not be valid. The combined analysis of different IceCube data samples \cite{ICStarting:24}, indeed, shows some preference for a spectral break in the 10--30 TeV interval, were also some tensions in the energy spectra from different samples is present, as shown in Fig. \ref{fig:diflim}.
In the plot, different IceCube results are compared to the ANTARES 95\% probability upper limits on the cosmic flux normalization
obtained for different spectral indexes $\gamma_{astro}$ in the energy range of validity. The envelope of the ANTARES limits is also shown as a black curve, considering for every energy the least restrictive available limit. 

The result was  converted into 68\%, 95\% and 99.7\% CL limits on the properties of the cosmic neutrino spectrum, as shown in Fig. \ref{fig:diffuse}. 
It should be remembered that graphs similar to the one in Fig. \ref{fig:diffuse}  only convey a limited amount of information. 
Each analysis reported  is most sensitive in a well defined energy range; each sample is dominated by events arising from different regions of the sky; different neutrino flavors and interaction channels contribute differently in each.

\subsection{Diffuse flux from Galactic regions\label{sec:agaldiffuse}} 
Charged cosmic rays (CRs) accelerated within our Galaxy propagate through complex trajectories due to the presence of the Galactic magnetic field. During their propagation, interactions with interstellar gas produce pions and heavier mesons that decay into $\gamma$-rays (from, e.g., $\pi^0$) or neutrinos (from, e.g., $\pi^\pm$). 
The  $\gamma$-ray flux with $E_\gamma\gtrsim$ 1 GeV arising from the Milky Way was measured by the  LAT experiment on the FERMI satellite (Fermi-LAT) \cite{LATdifg:12} and at $E_\gamma\gtrsim 1$ TeV by the HAWC \cite{HAWC:24} and LHAASO \cite{lhaaso:24} observatories.

Although recent progress has been made by IceCube \cite{ICgalplane:23}, our understanding of the Galactic neutrino flux remains limited. Critical aspects such as the absolute flux, spectral shape, and spatial origin lack stringent constraints. Moreover, no high-energy neutrino sources within the Galaxy have been conclusively identified, and the processes responsible for the hardening of the CR spectrum toward the Galactic Center observed in $\gamma$-rays remain poorly understood.
The central Galactic region is visible in IceCube only with downward-going neutrinos. This region can be investigated with upward-going neutrinos in the ANTARES detector, in particular with the  $\nu_\mu$ CC sample.

\begin{figure}
  \centering
\includegraphics[width=0.8\textwidth]%
    {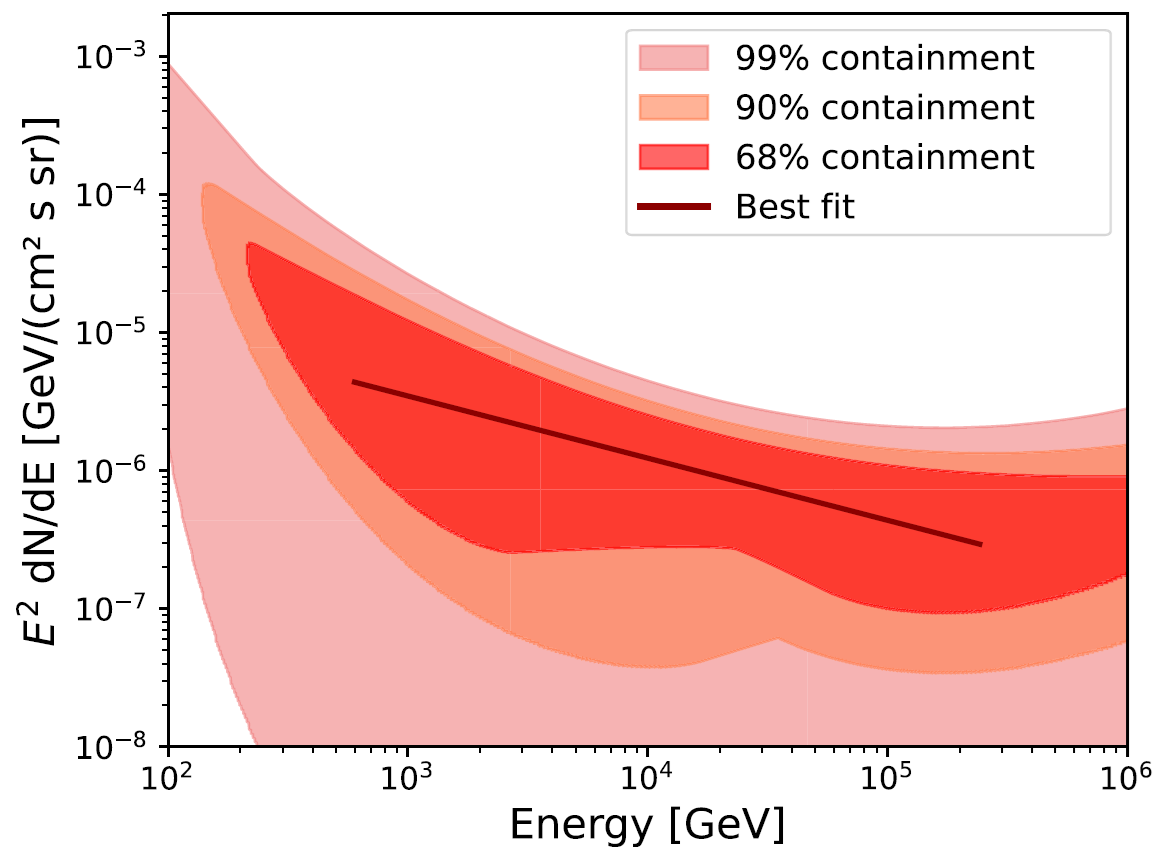}
  \caption{\small ANTARES constraints on the per-flavor neutrino energy-flux $E_\nu^2 \Phi(E_\nu)$ in the Galactic region $ |\ell|< 30^\circ$ and  $|b|<2^\circ$  as a function of neutrino energy. The red-shaded bands show the envelopes of the 68\%, 90\% and 99\% probabilities and the dark red line represents the best-fit flux. The endpoints on the $x$-axis illustrate the central energy ranges where 90\% of the considered neutrino signal is located for the various power-law spectra probed in the search.}    
\label{fig:galfluxof}
\end{figure}

The ANTARES collaboration initially searched for $\nu_\mu$ arising from a Galactic region with Galactic longitude $ |\ell|< 40^\circ$ and latitude $|b|<3^\circ$.
The expected background in the search region was estimated using a method employing off-zone regions with similar sky coverage \cite{ANTARES:52}. 
No excess of events was observed and model-independent upper limits were set.
A second study including both track and shower events with a larger data set  \cite{ANTARES:102} was done with a similar on/off method in the region $ |\ell|< 30^\circ$ and  $|b|<2^\circ$.
The energy distribution of detected events in the signal region was found inconsistent with the background expectation at 96\% CL.
The result reported in Fig. \ref{fig:galfluxof} was confirmed when the whole available data set was analyzed.
The mild excess in the on region is consistent with a neutrino flux with a power law of spectral index $\gamma=2.45^{+0.22}_{-0.34}$ and a flux normalization $\Phi=4.0^{+2.7}_{-2.0}\times 10^{-16}$ GeV$^{-1}$cm$^{-2}$s$^{-1}$sr$^{-1}$ at 40 TeV reference energy. 

The observations of $\gamma$-rays of Galactic origin have been used to fit different models of neutrino production from transport of CRs in the Galaxy. The model derived from the data in \cite{LATdifg:12} predicts a single power-law flux $\propto E_\nu^{-2.7} $ over the full sky. 
This template (called $\pi^0$ model) reproduces roughly the behavior of the diffuse galactic $\gamma$-ray flux up to a few hundreds of GeV. However, it fails to predict the hardening of CR emission in the Galactic Center derived from  observations at higher energies, which has prompted the development of several phenomenological models.
These models consider a non-isotropic transport of CRs, theoretically expected to originate from a higher level of turbulence near the Galactic Center, with a diffusion coefficient whose energy-dependence varies as a function of the galactocentric radius, as for instance in \cite{KRAg:15,KRAg:23}.
Alternative models incorporate a CR diffuse component as well as the possible neutrino emission by Galactic unresolved sources, as for instance the one of \cite{Vecchiotti:23}.

The first ANTARES study based on templates \cite{ANTARES:64} was based on the KRA$_\gamma$ models \cite{KRAg:15} and used 9 yr of data, considering all-flavor neutrino interactions. No excess of events was observed, and upper limits on the neutrino flux of the order of 1.1--1.2 times the predictions were derived.
The upper limit just above the model motivated the joined analysis with 7 yr of IceCube track data \cite{ANTARES:75}, but again without a statistically significant signal ($p$-values $\ge$0.02).
Later, the development of deep learning techniques in the IceCube collaboration enabled the possibility to identify a larger number of neutrino interactions in 10 yr of data. 
\textcolor[rgb]{0,0,0}{A neutrino emission from the Galactic Plane with pre-trial $p$-value (significance) calculated with respect to the background-only hypothesis of $1.26\times 10^{-6} (4.71\sigma)$ for the  $\pi^0$ diffuse template and $6.13\times 10^{-6} (4.37\sigma)$ for the KRA$^5_\gamma$ model was reported.
The trial-corrected $p$-value for the most significant template results in a significance of 4.48$\sigma$} \cite{ICgalplane:23}.
The signal was consistent with modeled diffuse emission from the Galactic Plane, but could also arise from a population of unresolved point sources.

\begin{figure}   \centering \includegraphics[width=0.99\textwidth]%
    {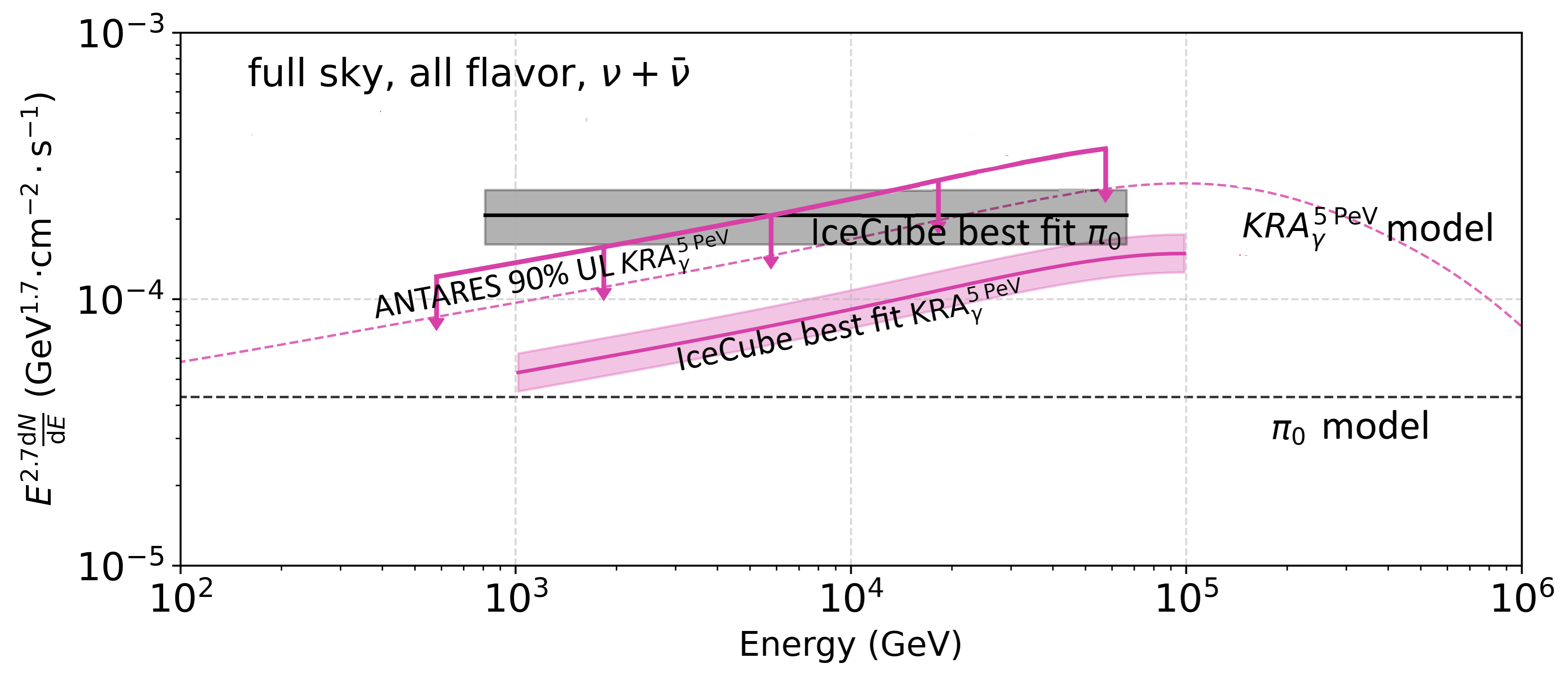}
  \caption{\small Comparison of the ANTARES upper limits for $\nu+ {\bar \nu}$ produced in our Galaxy with IceCube best fits for the $\pi^0$ model (black) and KRA$_\gamma^5$ model (pink).} 
	\label{fig:GalFlux}
\end{figure}

The final ANTARES analysis used the same all-flavor neutrino data set defined in \cite{ANTARES:106}. A maximum likelihood ratio method was built to probe  various Galactic emission model templates \cite{LATdifg:12,KRAg:15,KRAg:23,Vecchiotti:23,CRINGE}. This method was notably employed to evaluate the compatibility of these models with the observed spatial and energy distributions of neutrino events. 
Although the results do not yield stringent constraints on the tested models, upper limits on the diffuse neutrino flux were derived, which are compatible with the IceCube observations, as shown in Fig. \ref{fig:GalFlux}.
The growing KM3NeT detector will soon be able to significantly help disentangling the different hypotheses by observing $\nu_\mu$ CC interactions, due to its larger acceptance and higher angular resolution.

A second extended region that was investigated is the large emission halo referred to as \textit{Fermi Bubbles} \cite{FBubbles:10}. This region consists of two extended structures above and below the Galactic Center emitting $\gamma$-rays and discovered by analyzing Fermi-LAT data.
If hadronic processes are responsible for the $\gamma$-ray excess,  emission of high-energy neutrinos and $\gamma$-rays with similar fluxes is expected.    Data collected from 2008 to 2011 were used: no statistically significant excess of events was observed, and therefore upper limits on the neutrino flux in the TeV range  were obtained \cite{ANTARES:36}.

\subsection{Searches for steady neutrino sources in the Southern sky\label{sec:aps}}
Due to its geographical location, its sensitivity for $E_\nu< 10$ TeV, and the inherent optical properties of seawater, the ANTARES telescope (although much smaller in size than IceCube) provided competitive limits on the presence of point-like neutrino sources in the Southern sky.
In these studies, a model describing neutrino emissions following a simple unbroken power-law is considered:
\begin{equation}\label{eq:fitps}
\frac{d \phi_{\nu}}{dE}= {\phi_{\rm ps}}\cdot  
\biggl( \frac{E}{1 \textrm{ GeV}} \biggr)^{-\gamma_{\rm ps}} \textrm{ GeV}^{-1}\textrm{ cm}^{-2}\textrm{ s}^{-1} 
\end{equation} 
where ${\phi_{\rm ps}}$  is the flux normalization factor at Earth at the energy of 1 GeV, assuming flavor equipartition.
The spectral features of the flux depends solely on the spectral index $\gamma_{\rm ps}$.

The advantageous field of view of the ANTARES detector for the Southern sky was evident with few years of data and using the track sample only \cite{ANTARES:21,ANTARES:28}. 
This motivated to use the ANTARES sensitivity for neutrinos in the TeV range and the large size of the IceCube detector to improve the possibility to observe neutrino sources in the Southern sky in two joint studies.
The first one combined data recorded by ANTARES from 2007 to 2012, and by IceCube from 2008 to 2011 \cite{ANTARES:48}.
The second \cite{ANTARES:83} used nine years of ANTARES track-like and shower-like neutrino candidates pointing in the direction of the Southern sky \cite{ANTARES:38}, combined with seven years of throughgoing track-like IceCube events \cite{ICps7:17}. 
No significant evidence for cosmic neutrino sources was found, improving the upper limits by a factor of two compared to both individual analyses. 
\begin{figure}[tbh]
  \centering
\includegraphics[width=0.9\textwidth]%
    {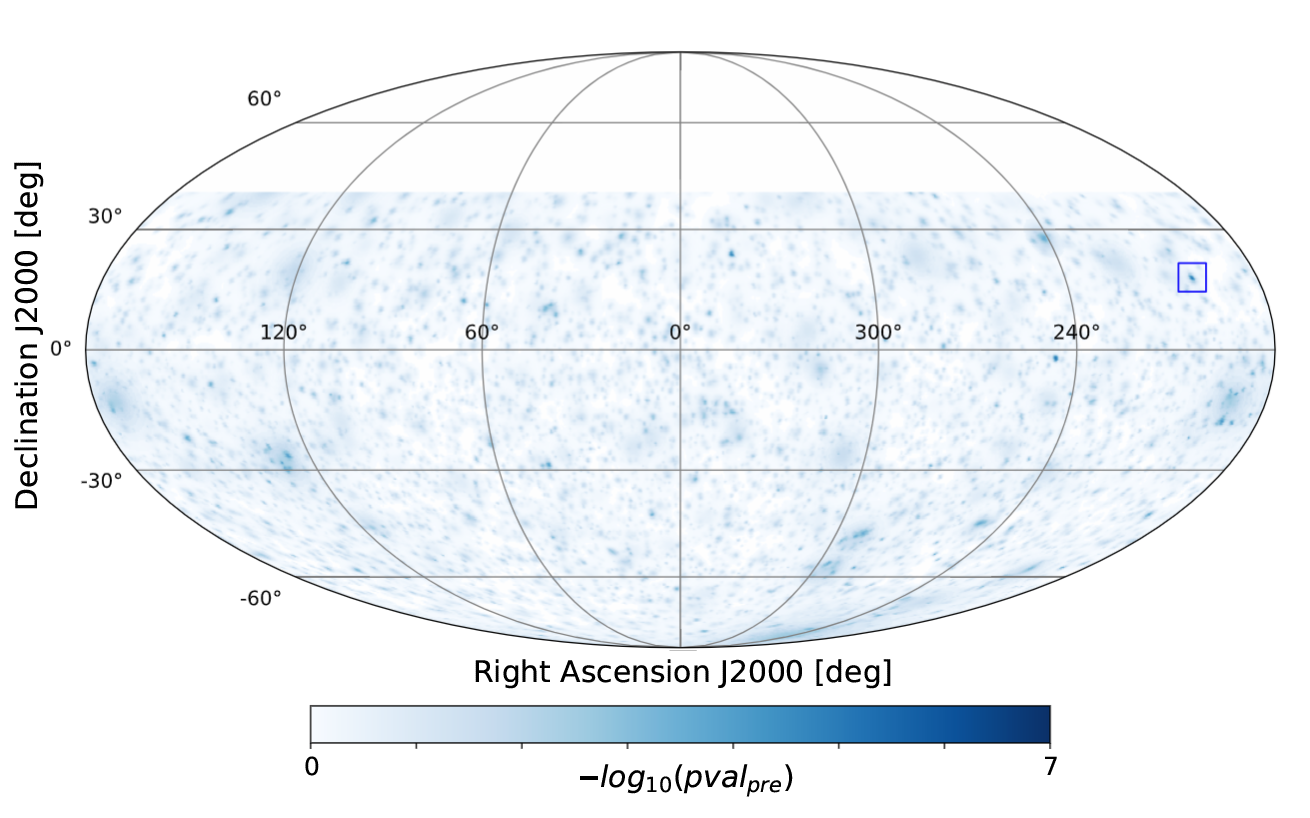}
  \caption{\small Sky map in equatorial coordinates of pre-trial $p$-values of the ANTARES visible sky. The position of the hotspot is marked with a blue box. }
\label{fig:psFS}
\end{figure}

The last ANTARES published results \cite{ANTARES:65} combined both the track and shower channels to search for cosmic neutrino sources.
 In the following, we refer to the update using the full data set \cite{Salves:25}.
 
In the \textit{full sky search} approach, the visible sky was explored looking for an excess of events regardless of any hypothesis on the position of the source, scanning regions of $0.11^\circ\times 0.11^\circ$ in size \cite{HEALPix:05}. 
The significance of the observation at each location is shown in terms of $p$-values (a value of $p$ close to 1 is compatible with the background hypothesis) in Fig. \ref{fig:psFS}. 
The smallest $p$-value (\textit{hotspot}) found corresponds to $3.2\times 10^{-6}$ (4.5$\sigma$) at equatorial coordinates (RA, $\delta$) = (200.5$^\circ$,17.7$^\circ$), indicated in Fig. \ref{fig:psFS} with a blue box. Since more than $2\times 10^{6}$ directions were scanned, corrections due to the “look-elsewhere-effect” were applied, making the observation 39\% compatible with background. 
No known source was found near the hotspot, being the closest one the radio blazar J1318+1807 at  $\sim~1^\circ$ distance. 

The above survey was complemented with a search over a predefined list of 169 potential cosmic neutrino sources. 
By investigating only a preselected list of promising candidates, the trial factor penalization arising from inspecting a huge number of sky locations is drastically reduced.
In the selected candidate list, 14 were assumed to have a spatially extended emission profile: in this case, the source emission was parametrized with a symmetrical 2D Gaussian function, with variance $\sigma_{\rm ext}$ given by the distance at which 68\% of the source is contained. 
Neutrino emission following eq. \ref{eq:fitps} assuming spectral index $\gamma_{\rm ps}$=2.0 (2.5)  was looked for at the direction of each of the sources.
For the seven most significant objects, the name, equatorial coordinates, the variance $\sigma_{\rm ext}$, the results found in the search in terms of best-fit number of signal events, the corresponding pre-trial $p$-value, and the 90\% CL upper limits on the all-flavor flux normalization factor ${\phi_{\rm ps}}$ are reported in Tab. \ref{tab:ps}. 
\begin{table}
    \centering
    \begin{tabular}{lcccccc} \hline
Name         & $\delta [^\circ]$ & RA$[^\circ]$ & $\sigma_{\rm ext}[^\circ]$ &$\hat{\mu}_{\rm sig}$ & $p$-value  & ${\phi^{90\%}_{\rm ps}}$\\ \hline
Vela X    				 & -45.60 & 128.75 & 0.58 &2.76 &  0.027 & 1.0\\
Galactic Centre    & -29.01 & 266.43 & 0 & 2.06 &  0.017 & 1.2\\
J0609-1542         & -20.12 & 287.79 & 0 &1.22 & 0.0073 & 1.4\\
3C403      			   &  2.51	& 298.07 & 0 &2.47 & 0.00048 & 2.0\\
TXS 0506+056       &  5.7 	&  77.35 & 0 &2.23 & 0.0075 & 1.6\\
J0242+1101         &  11.02	&  40.6 &  0 &3.67	&  0.0074& 1.6\\
MG3 J225517+2409   &  24.19 &  343.82& 0 &3.97 & 0.00024 & 2.3\\ \hline
    \end{tabular}
    \caption{\small {List of the seven candidates in the predefined list of 169 potential sources with the most significant fitted signal $\hat{\mu}_{\rm sig}$ in terms of $p$-values.	
The list is ordered from lower to higher declination $\delta$.	When $\sigma_{\rm ext}=0$ the source is assumed to be point-like. The 90\% CL upper limits on the all-flavor flux normalization factor ${\phi_{\rm ps}}$ is computed for ${\gamma_{\rm ps}}=2.0$ as from  eq. \ref{eq:fitps} in units of $ 10^{-8} \textrm{ GeV}^{-1}\textrm{ cm}^{-2}\textrm{ s}^{-1} $.}}
    \label{tab:ps}
\end{table}

\begin{figure}[tbh]
  \centering
\includegraphics[width=0.89\textwidth]%
    {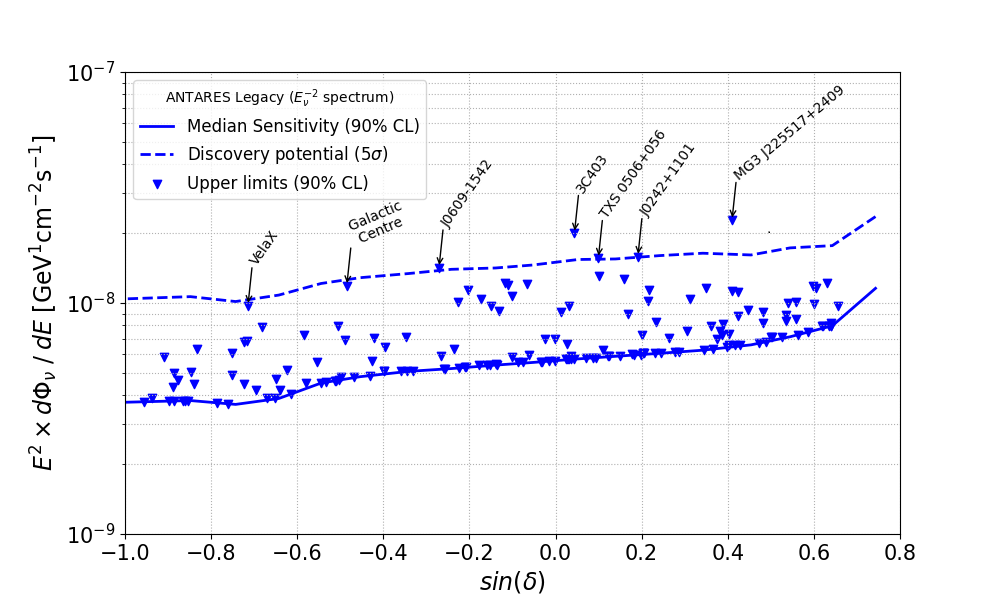}
  \caption{\small  90\% CL upper limits (blue points) on the one-flavor neutrino flux for the 169 potential sources vs. the sinus of the declination $\delta$.  The solid line indicates the 90\% CL median sensitivity, while the dashed line the $5\sigma$ discovery potential assuming a $E_\nu^{-2.0}$ energy spectrum. Refer to \cite{ANTARES:65} for the definition of sensitivity and discovery potential.}
\label{fig:ps}
\end{figure}

Blazar MG3 J225517+2409 was the most significant object found, with a local $p$-value of  $2.4\times 10^{-4}$ for  ${\gamma_{\rm ps}}=2.0$ ($6.4\times 10^{-5}$ for  ${\gamma_{\rm ps}}=2.5$).
After correcting for the number of trials, this observation only deviates 2.0$\sigma$ from the background-only hypothesis, and 90\% CL upper limits were set on the flux normalization.  
A high energy IceCube $\nu_\mu$ event, $\sim~1.1^\circ$ away from the object, was recorded when the blazar was in a flaring state 
(see also \S \ref{sec:otc}.) 
The sources with a pre-trial significance over 2$\sigma$ reported in Tab. \ref{tab:ps} were indicated with arrows in Fig. \ref{fig:ps}, together with the 90\% average upper limit  and discovery flux for the spectral index  $\gamma_{\rm ps}$=2.0. 
In this list, the pulsar wind nebulae Vela X is the only source assumed as extended, with $\sigma_{\rm ext}=0.58^\circ$.
The solid line indicates the 90\% CL median sensitivity and the dashed line the $5\sigma$ discovery potential assuming a $E_\nu^{-2}$ energy spectrum. 
As a convention in the field, the maximum between the sensitivity and the upper limit for the particular location of the source is reported in the figure.

The active galactic nuclei NGC 1068 was also considered in the list. The results of the IceCube collaboration \cite{ICngc:22} point toward a very soft spectrum,  $\gamma_{\rm ps}$=3.2, for this source. For this reason, a dedicated ANTARES search assuming $\gamma_{\rm ps}$=3.0 was performed, obtaining a pre-trial $p$-value of 0.22. The derived 90\% CL upper limits was compatible with the IceCube observation.

\subsection{The TXS 0506+056 source\label{sec:apstx}}
In 2017, an extremely high-energy track event (IC170922A) observed by the IceCube detector was reported in a GCN circular. Its arrival direction was consistent with the location of the known $\gamma$-ray blazar TXS 0506+056, observed to be in a flaring state by  Fermi-LAT. 
An extensive multi-wavelength campaign followed, ranging from radio frequencies to $\gamma$-rays, and IC170922A represented the first observation of a neutrino event in spatial coincidence with a $\gamma$-ray emitting blazar during an active phase \cite{ICtxsMM:18}.
\textcolor[rgb]{0,0,0}{The MAGIC $\gamma$-ray telescopes monitored the blazar TXS 0506+056 in the very-high-energy band for $\sim$41 hours from 1.3 to 40.4 days after the neutrino detection \cite{MAGIC:18}. The MASTER optical robotic telescope found the blazar to be in the off-state one minute after the observation of IC170922A and then switched to the on-state no later than two hours after the event \cite{MASTER:20}.}
Prompted by this association, the IceCube collaboration searched for potential other neutrino observations from this direction in their data (9.5 yr) and an excess of $13\pm 5$ neutrino candidates over background was found in the period between September 2014 and March 2015, with a statistical significance of $\sim~3.5\sigma$, independent of the 2017 flaring episode \cite{ICtxs:18}. 
However, during 2014--2015, the $\gamma$-ray flux of the source was one order of magnitude lower than its value during the 2017 bursting period and without time variability. 
These two observations from the direction of TXS 0506+056 pose challenging questions about the theoretical interpretation of the emission mechanism.
More multimessenger observations are then needed to identify the sources and to better model the astrophysical processes  creating neutrinos.

\begin{figure}[tbh]
  \centering
\includegraphics[width=0.99\textwidth]%
    {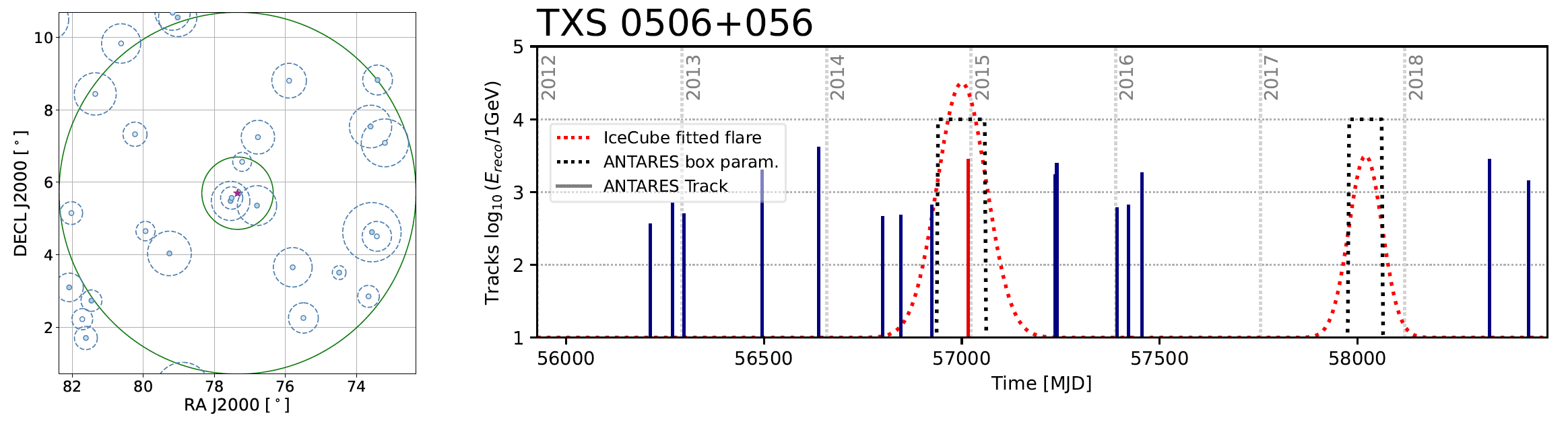}
  \caption{\small {Left: Distribution of the ANTARES neutrino candidates (all  track-like) close to TXS~0506+056 \cite{Salves:25}. The inner (outer) solid green line depicts the $1^\circ$ ($5^\circ$) distance from the blazar position. The dashed circles around the events indicate the angular error estimate. 	Right: Arrival time of neutrino candidates found within $5^\circ$ of the direction of the source. The IceCube neutrino-flare time profile is shown together with the characterization used for the analysis, both in arbitrary units. The height of the line is proportional to the muon reconstructed energy  \cite{AlvesGarre:20234B}}.}
\label{fig:TXS}
\end{figure}
The ANTARES collaboration  contributed providing additional experimental information \cite{ANTARES:72}. At neutrino energies $\lesssim 100$ TeV, the detector had competitive sensitivity with respect to IceCube. In fact, the reconstructed position in the sky of the IC170922A neutrino event corresponded, at the location of the ANTARES detector, to a direction 14.2$^\circ$ below the horizon. A possible neutrino candidate would thus be detected as an upward-going event. 
The data analysis showed no upward-going $\nu_\mu$ candidate event  within 3$^\circ$ around the IC170922A direction and within $\pm 1$ hour centered on the event time. 
However, an event was found in the ANTARES data coincident with the 2014--2015 neutrino flare at a pre-trial significance of $2.0\sigma$~\cite{AlvesGarre:20234B}, see the right panel of Fig. \ref{fig:TXS}. 
Apart from the transient emission hypothesis, TXS~0506+056 has also been monitored for steady neutrino emissions. The first time-integrated study with data covering the 2007--2017 period fitted 1.03 signal events with a pre-trial $p$-value of 3.4\% ($1.9\sigma$). Since then, as more data was collected and analyzed, its significance has remained stable, resulting in a total of 2.23 fitted signal events with a pre-trial significance of $2.4\sigma$ with the complete data set. 
The neutrino candidates around TXS~0506+056 during the whole ANTARES livetime are represented in the left panel of Fig. \ref{fig:TXS}.

\subsection{Stacked source analyses\label{sec:astack}}
Neutrinos of cosmic origin, with an angular distribution consistent with isotropy and suggesting a predominantly extragalactic origin, have been firmly observed. However, the sources of this diffuse flux remain inconclusively identified.
To this end, different strategies are adopted by the running neutrino telescopes in addition to the above-mentioned searches for clustering of neutrino events from given directions in the sky.
These include stacked cross-correlation of the observed neutrino  angular distribution with known catalogs of objects and searches for temporal and/or spatial correlations with transients observed with electromagnetic emission.

\subsubsection{Gamma Ray Bursts\label{sec:aGRB}}
Gamma Ray Bursts (GRBs) are among the most energetic phenomena in the Universe \cite{Piran:05}.
During their \textit{prompt phase}, copious amounts of keV--MeV (and sometimes GeV) photons have been observed.
GRBs have been empirically classified on the basis of the duration and the spectrum of the prompt phase as \textit{long} (if soft spectrum and longer than 2 s) or \textit{short} (if hard spectrum and shorter than 2 s). Long GRBs have been associated with supernovae, while short GRBs have been long believed to originate in the merger of compact objects, as corroborated by the observation of gravitational waves from GRB/GW 170817A, \S \ref{sec:GW}.
For a long time, GRBs have been proposed as plausible candidates for extragalactic CR acceleration sources \cite{Vietri:95,Waxman:95}.

On a theoretical basis, the interactions of these accelerated CRs with local environment radiation and/or matter during the different phases of the GRB would result in neutrino production: PeV neutrinos during the prompt phase \cite{WB:97,ZK:13}; TeV neutrinos in the precursor phase \cite{RMW:03}; EeV neutrinos during the afterglow \cite{WB:20}.
Models of the astrophysical processes of GRBs can also explain the extragalactic neutrino flux.
Due to the transient nature of the GRBs and the variety of their characteristics, it is essential to continuously monitor the whole sky in order to maximize the probability to observe a neutrino signal. 
During their  livetime, ANTARES and IceCube have each observed at least half of the sky and were sensitive to neutrino fluxes down to $E_\nu  \sim~100$ GeV.
Thanks to the expected time correlation between the emission of neutrinos and $\gamma$-rays from a GRB, backgrounds in searches for neutrinos originating from GRBs can be drastically reduced.

\begin{figure}[tbh]
  \centering
\includegraphics[width=0.8\textwidth]%
    {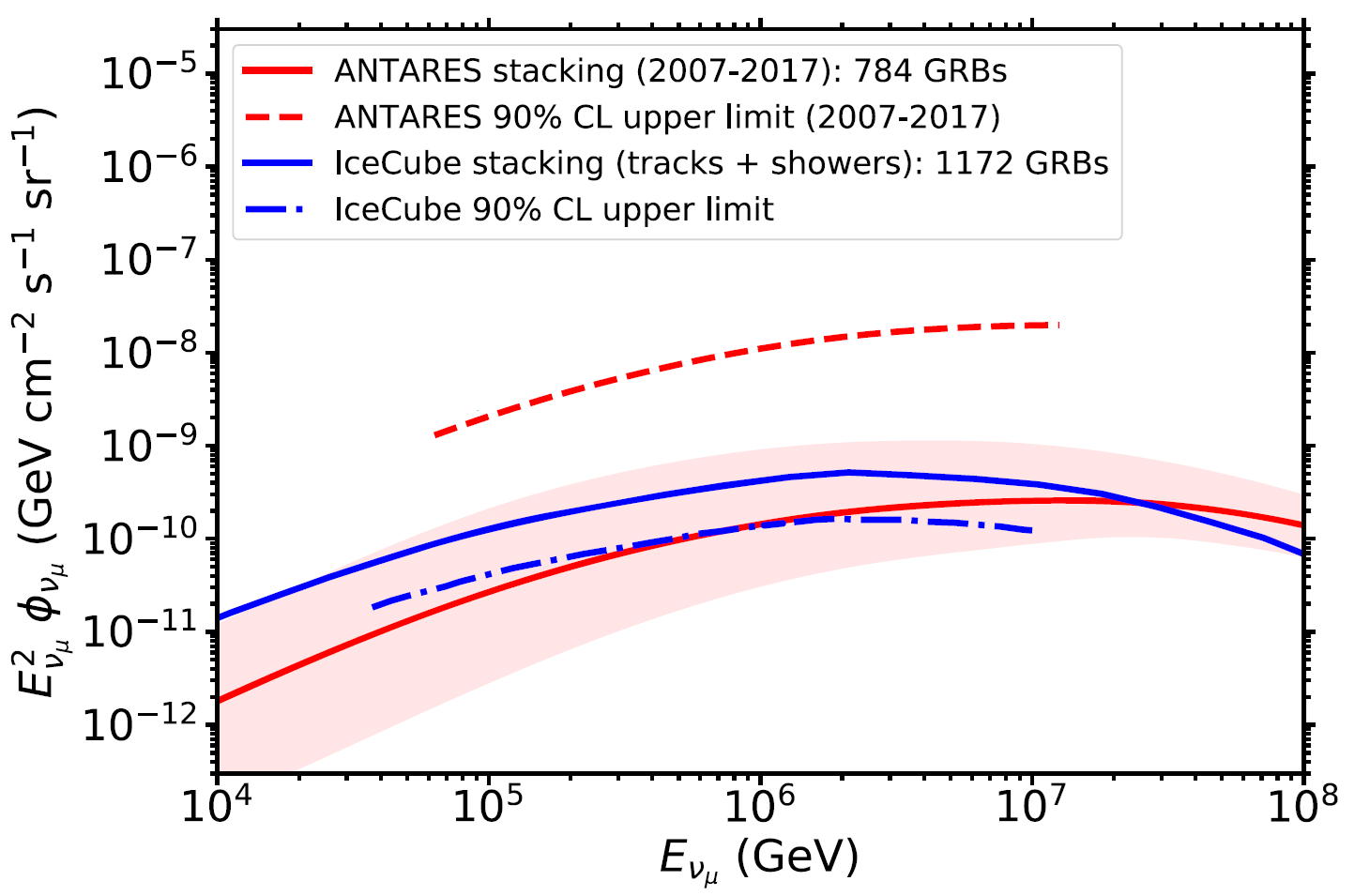}
  \caption{\small  Solid red (blue) curve: quasi-diffuse neutrino flux expected from the 784 GRBs in the ANTARES sample (1172 GRBs in IceCube). The shaded region indicates the error band, obtained from the sum of the individual maximum and minimum fluences for each GRB in the ANTARES sample. The red dashed line (dash–dotted blue line) represents the 90\% CL upper limit  with respect to the above expectation in ANTARES (IceCube) \cite{ANTARES:88}. 
	}.
\label{fig:GRB}
\end{figure}

The ANTARES collaboration produced different studies: the first refers to 296 GRBs detected from 2007 to 2011. Neutrino candidates were searched for in a time window  optimized for each GRB, whose average value was $\sim$80 s \cite{ANTARES:33}.
A search for high-energy neutrino emission outside the prompt-emission time window using a stacking approach of the time delays between reported GRB alerts and spatially coincident $\nu_\mu$ signatures was also performed \cite{ANTARES:53}.
Then, two dedicated studies were done for specific sets of observations.
In the first \cite{ANTARES:57}, four bright GRBs observed between 2008 and 2013 were investigated according to two scenarios of the fireball model: the internal shock scenario, leading to the production of neutrinos with  $E_\nu>100$ TeV, and the photospheric scenario, characterized by a low-energy component in neutrino spectra.
The second study \cite{ANTARES:91} refers to the first sub-TeV $\gamma$-ray burst detections by Imaging Atmospheric Cherenkov Telescopes: GRB 190114C (detected by MAGIC); GRB 180720B and GRB 190829A (observed by H.E.S.S.).  The search covered both the prompt and afterglow phases. 
The final ANTARES analysis \cite{ANTARES:88} considered the cumulative emission from the entire studied sample of 784 long GRBs occurred between 2007 and 2017. For each GRB, the expected neutrino flux was calculated within the framework of the internal shock model \cite{Piran:05}: the impact of the lack of knowledge of source redshifts and other intrinsic parameters of the emission mechanism was quantified in terms of uncertainties on neutrino flux expectations. 
The solid red curve in Fig. \ref{fig:GRB} corresponds to the quasi-diffuse neutrino flux expected from the 784 GRBs in the studied ANTARES sample.
This quantity represents the diffuse neutrino energy flux derived from the cumulative fluence from all the observed GRBs in the samples, as discussed in \cite{ANTARES:88}. The shaded region indicates the error band, obtained from the sum of the individual maximum and minimum fluences for each GRB. Similarly, the solid blue line represents the quasi-diffuse neutrino flux expected for the 1172 GRBs in IceCube  \cite{ICgrb:17}.
The dashed lines represents the 90\% CL upper limit  with respect to the above expectation in ANTARES (red dashed line) and IceCube (dash–dotted blue line). 

The IceCube updated recently the study to 2209  GRBs \cite{ICgrb:22} that were investigated for neutrino correlations from the precursor, prompt, and afterglow emission regions in a comprehensive manner with different strategies. 
Another analysis \cite{ICgrb:24} searched for neutrinos in the 10--1000 GeV range from 2268 GRBs over 8 years of IceCube-DeepCore data.

All ANTARES and IceCube analyses reported observations consistent with background only and, to date, there is no evidence of neutrino emission from GRBs.
The conclusion based on the ANTARES results is that, within standard assumptions of energy partition among accelerated hadrons, leptons, and magnetic fields, GRBs are not the main sources of the astrophysical neutrino flux, possibly contributing less than 10\% at energies around 100 TeV. 
IceCube observations are consistent with the hypothesis that GRBs, during the prompt phase, cannot be responsible for more than 1\% of the extragalactic neutrino flux \cite{ICgrb:22}. 

The brightest GRB ever observed, GRB 221009A, occurred when ANTARES was already turned off. 
GRB 221009A was studied in neutrinos from MeV to PeV by IceCube \cite{ICgrb22:23} and by  KM3NeT \cite{KM3grb:24}  and no evidence for neutrino emission was found. 

\subsubsection{Fast Radio Bursts}
Fast radio bursts (FRBs) are transient radio pulses serendipitously discovered by radio telescopes \cite{FRBpetro:19}. 
While FRBs bear similarities to pulsar pulses, their large dispersive delays imply origins far beyond the Milky Way.
Hence, these sources are intrinsically many orders of magnitude more luminous than pulsars.
The duration of a FRB ranges from a fraction of a millisecond, for an ultra-fast radio burst, to a few seconds, and their underlying mechanism is not understood yet.
Since most FRBs appear to occur only once, they are likely cataclysmic in nature, whereas the few sources that repeat clearly indicate the presence of a longer-lived central engine.
The all-sky event rate is high: there is a detectable FRB roughly once every minute occurring somewhere in the sky. 
Only a marginal fraction of FRB sources have been discovered in the last decade because the small fields of view of current operating radio telescopes. 

Highlights from these discoveries have included also detections of  FRBs with measured polarization profiles. For example, FRB 150215 has been observed close to the Galactic Plane with the Parkes radio telescope. 
This burst was detected in real time with its polarization recorded and followed up across multiple wavelengths and probes, including TeV $\gamma$-ray  by the H.E.S.S. telescope and neutrinos from ANTARES data, thus leading to the first constraints on neutrino flux from an FRB  \cite{ANTARES:60}.
The increasing automation of analysis methods allowed near-real-time triggering of multi-wavelength instruments to look for afterglows.
The follow-up for FRBs 151230 and 160102 was undertaken from X-ray to radio wavelengths including searches for associated neutrino emission with  ANTARES \cite{ANTARES:69}. 

To date, many models have been proposed to explain FRBs, but neither the progenitors nor the radiative and the particle acceleration processes at work have been clearly identified.
The ANTARES collaboration assessed whether hadronic processes may occur in the vicinity of the FRB source. If they do, FRBs may contribute to the high-energy CR and neutrino fluxes. A search for these hadronic signatures was carried out looking for TeV--PeV neutrinos that are spatially and temporally coincident with the detected FRBs in the period 2013--2017 in the field of view of the ANTARES detector \cite{ANTARES:76}. 
No coincident neutrino candidates were observed and upper limits on the per-burst neutrino fluence and on the energy released were derived using a power-law spectrum for the possible incoming neutrino flux.

\subsubsection{Blazars}
Radio-loud active galactic nuclei with jets pointing almost directly towards the observer, referred to as \textit{blazars}, have long been considered promising neutrino sources \cite{GHS:95}. 
Blazars generally show high time variability in their light curves at different wavelengths and on various timescales. The all-sky monitor Fermi-LAT probes the variability of gamma-ray bright blazars in the sky on timescales of hours to months \cite{LATgr:10}. 
Assuming hadronic models, for optically thin sources a strong correlation between the $\gamma$-ray and the neutrino fluxes is expected. 
The search for neutrino candidates in neutrino detectors correlated with blazars is especially facilitated if a narrow time window is selected.
A first ANTARES analysis \cite{ANTARES:25} used an unbinned method in data collected in 2008.
A second study applied a time-dependent analysis to a selection of  blazars emitting in $\gamma$-rays and observed by  Fermi-LAT  or by TeV Cherenkov telescopes. Data collected from 2008 to 2012 were analyzed \cite{ANTARES:44} and the results were compatible with background fluctuations. Upper limits on the neutrino fluence were produced and compared to the measured $\gamma$-ray spectral energy distribution.

The above searches focused on blazars observed in the GeV--TeV  $\gamma$-ray energy range. 
However, it was hypothesized that $\gamma$-ray emission and flares may not be tightly correlated with neutrino sources; in optically thick sources hadronic $\gamma$-rays are produced together with neutrinos but quickly cascade down in energy.
Synchrotron emission from blazar jets, detected on Earth as radio emission, could likely be a better tracer of relativistic beaming and activity happening close to the jet origin.

\begin{figure}
  \centering
\includegraphics[width=0.9\textwidth]%
    {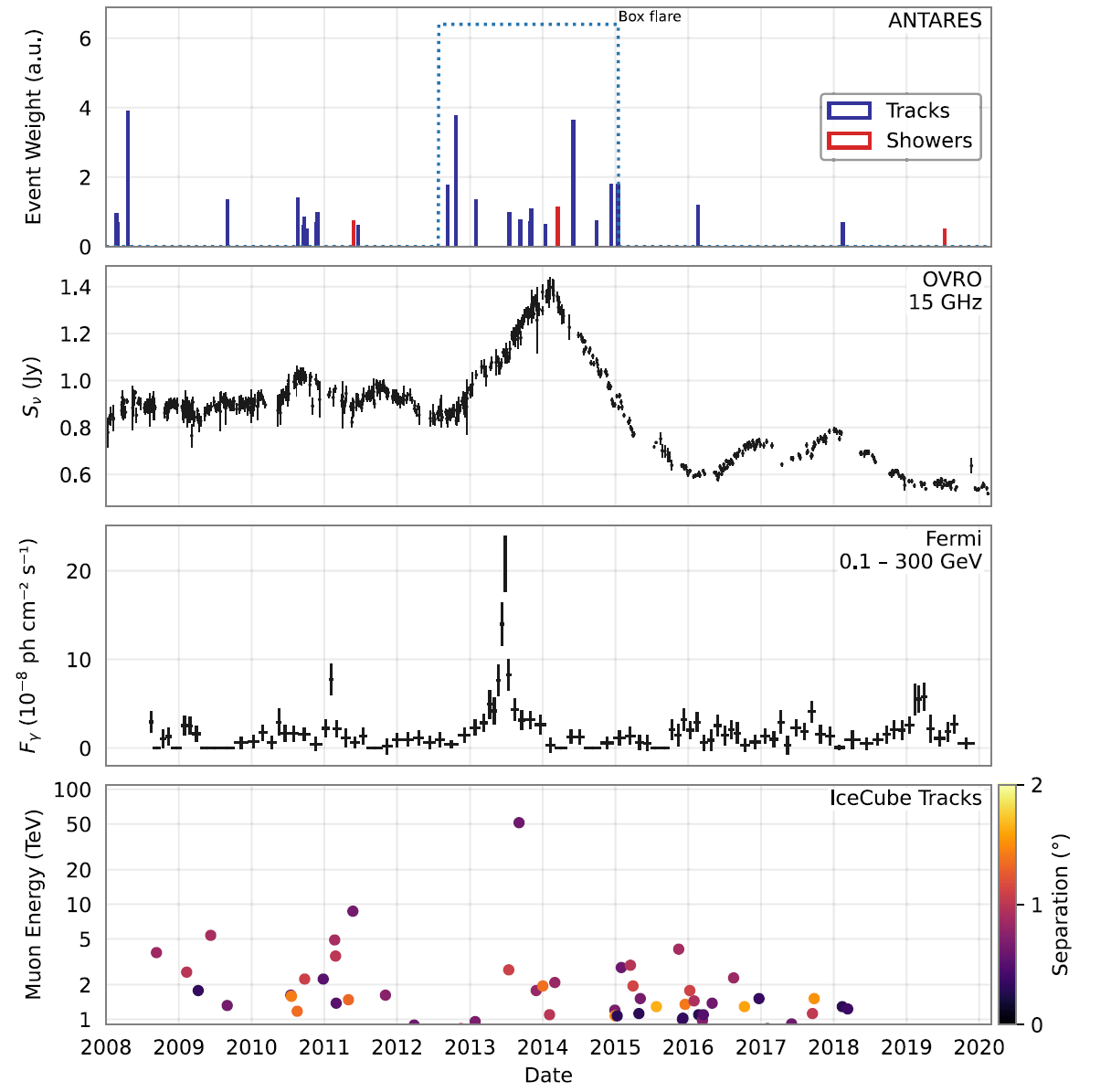}
  \caption{\small Multimessenger observations from the direction of the blazar J0242+1101 as a function of time since 2008. Top panel: weighted time distribution of the ANTARES track-like (shower-like) events within 5$^\circ$ (10$^\circ$) from the blazar. The box profile represents the best-fit neutrino flare identified in the ANTARES analysis. Second panel: OVRO radio light curve. Third panel: adaptive binned $\gamma$-ray light curve derived from Fermi-LAT data. Bottom panel: weighted time distribution of the IceCube track-like events closer to J0242+1101 than their 50\% angular uncertainty. 	The color scale indicates the event angular distance from the source. Refer to \cite{ANTARES:105} for further details. }
\label{fig:j02}
\end{figure}

A subsequent study, based on 13 years of data, analyzed a statistically representative sample of blazars selected for their bright radio emission \cite{ANTARES:105}.  
The hypothesis of a neutrino–blazar directional correlation was tested by pair counting and a complementary likelihood-based approach. 
The resulting post-trial $p$-value was $p=3.0$\%. Additionally, a time-dependent analysis was performed to search for temporal clustering of neutrino candidates.
None of the investigated sources alone reached a significant flare detection level. However, the presence of 18 sources with a pre-trial significance above 3$\sigma$ indicated  $p = 1.4\%$ detection of a time-variable neutrino flux. 
An a posteriori analysis revealed an intriguing temporal coincidence between neutrino, radio, and $\gamma$-ray flares from the J0242+1101 blazar, with a significance level of $p = 0.5\%$, see Fig. \ref{fig:j02}.
Altogether, the results of this study suggested a possible connection between neutrino candidates detected by the ANTARES telescope and radio-bright blazars.

The possibility to exploit gravitational lensing effects to improve the sensitivity of neutrino telescopes to the intrinsic neutrino emission of distant blazar populations was also studied in \cite{ANTARES:41}. 
The case of four distant and gravitationally lensed Flat-Spectrum Radio Quasars (FSRQ) was considered by estimating their magnification factor assuming a singular isothermal profile for the lens.  The strongest constraint was obtained from the lensed quasar B0218+357, providing a limit on the total neutrino luminosity of this source of $1.1\times 10^{-46}$ erg s$^{-1}$, about one order of magnitude lower than those already obtained in the standard  searches with non-lensed FSRQ.

\subsubsection{Other transients and catalogs\label{sec:otc}}
The large number of trials associated with the full sky searches implies that only very bright objects can be detected with $>5\sigma$ significance.
Searching in a short time window corresponding to flaring periods or among a limited set of directions provided by a catalog of sources is an efficient way to limit the trial factors and enable population studies.

For example, a time-dependent search \cite{ANTARES:58} using 2008--2012 data correlated ANTARES neutrino candidates with 33 X-ray binaries observed by satellites (RXTE/ASM, MAXI, and Swift/BAT).
The time window was restricted to periods when the sources exhibited high  flaring activities or during hardness transition states.
In addition, a search for neutrino emission from a sample of six microquasars, based on 2007--2010 data, was presented in \cite{ANTARES:40}.
By means of appropriate time cuts, the neutrino search was restricted to the periods when the acceleration of relativistic jets was taking place. 
The time interval selections were based on information from the X-ray telescopes RXTE/ASM and Swift/BAT, and, in one case, from Fermi-LAT.
No significant excess above the background was found  and the upper limits on the derived neutrino flux constrain the  jet parameters for some astrophysical models.

Concerning catalogs, a likelihood stacking method was used \cite{ANTARES:92} to search for a global excess of upward-going $\nu_\mu$ in correlation with: 
(a) a subsample of 1,420 Fermi-LAT blazars observed between 1--100 GeV \cite{LATcat15}; 
(b) a LAT catalog with 64 star-forming galaxies \cite{LATsf:12}; 
(c) a sample with the brightest and most accretion-efficient radio galaxies in the local sky,  identified with a double lobed radio morphology \cite{Bassani:16}; 
(d) a population of 15 active galactic nuclei (AGN) with jet dust-obscured \cite{Maggi:16}; and 
(e) the public sample of 56 high-energy track events from the IceCube experiment (extending a previous analysis \cite{ANTARES:80}).
None of the tested sources showed a significant association with the ANTARES neutrino sample. 
The smallest $p$-value  was obtained for the catalog of radio galaxies, with a pre(post)-trial $p$-value equivalent to a 2.8(1.6)$\sigma$ excess. 
Among all the tested individual sources, two objects exhibited a pre-trial significance of $\sim~3.8 \, \sigma$:  the blazar MG3 J225517+2409, and the radio galaxy 3C 403, with five and two ANTARES tracks, respectively,  located within $1^ \circ$ of each source.
An a posteriori significance of 1.9$\sigma$ for the combination of ANTARES and IceCube data with blazar MG3 J225517+2409 was estimated.

A dedicated study using 7.3 yr of ANTARES neutrino candidates and the whole Fermi-LAT $\gamma$-ray catalog to search for $\nu+\gamma$ transient sources or source populations within a time interval $\Delta t<1000$ s was finally performed, without any evidence for correlation \cite{ANTARES:81}.

\section{Contribution to multi-messenger astrophysics\label{sec:MM}} 

\subsection{Follow-up of ANTARES neutrino alerts}
High-energy neutrinos could be produced in the interaction of charged CRs with matter or radiation surrounding astrophysical sources. To look for transient sources possibly associated with neutrino emission, a follow-up program of neutrino alerts was operating within the ANTARES collaboration since 2009 \cite{ANTARES:19}. 
The alert system, dubbed TAToO (\textit{Telescopes-Antares Target of Opportunity}) sent alerts to partners operating classical electromagnetic telescopes. 
This approach does not require an a priori hypothesis on the nature of the underlying neutrino source. 
It relies only on the hypothesis that these astrophysical phenomena produce high-energy neutrino and electromagnetic radiation over a broad energy range. 
In particular, the system was mainly aimed to target very fast transient sources such as GRBs, transient sources as core-collapse supernovae (CCSNe), and long-term variable sources as flares of active galactic nuclei (AGN). Similar alert programs are in operation in IceCube \cite{ICrealtime:17} and in Baikal-GVD \cite{GVDale:23}. 

The TAToO program had four online neutrino trigger criteria \cite{ANTARES:46,ANTARES:107}:
1) \textit{High energy trigger } (HE) and  2) \textit{Very high energy trigger} (VHE): the detection of a single  high-energy neutrino with an energy $>5$ TeV (HE, with rate 1/month) and $>30$ TeV (VHE, with a typical rate of 3–5 events per year).
3) \textit{Directional}: the detection of a single neutrino for which the direction points toward ($\le 0.4^\circ$) a local galaxy ($\le 20$ Mpc), with a typical rate of one per month.
4) \textit{Doublet trigger}: the detection of at least two neutrinos coming from close directions ($\le 3^\circ$) within a predefined time window (15 min). No doublet trigger was ever issued.

The TAToO program triggered robotic optical telescopes (MASTER, TA\-ROT, ROTSE and the SVOM ground based telescopes) immediately after the detection of any relevant neutrino candidate.
For each alert, the optical observation strategy was composed of a follow-up within 24 hours after the neutrino detection, to search for fast transient sources such as GRB afterglows, complemented by several observations during the two following months, to detect for example the rising light curves of CCSNe or the flare of an AGN. 

A subset of VHE events were followed by the Swift satellite \cite{Swift:04} with its X-ray Telescope, which provided a unique opportunity to observe X-ray counterparts on account of its large field of view (FoV) and its very prompt and flexible scheduling processes.
Also the INTEGRAL satellite \cite{Integral:03}, thanks to its wide FoV, was able to derive a constraining upper  limit on impulsive $\gamma$-ray flux for every received trigger, even if the ANTARES alert position was not at the center.
A few alerts also triggered observations of the H.E.S.S. imaging atmospheric Cherenkov telescope.

A particular case was that of the radio follow-up of the Murchison Widefield Array (MWA), which studied two \textit{directional} trigger events.
MWA used real time data, archival data, and data collected up to a year after the neutrino triggers at frequencies between 118 and 182 MHz, to search for transient or strongly variable radio sources that are consistent with the neutrino positions. No  counterpart was detected \cite{ANTARES:47}.

In September 2015, ANTARES issued a VHE neutrino alert and during the follow-up, a potential transient counterpart was identified both by Swift and MASTER. A multi-wavelength follow-up campaign  allowed to identify the nature of this source and  proved its casual association with the neutrino \cite{ANTARES:107}. 
No other optical and X-ray counterpart was significantly associated with an ANTARES alert for a candidate neutrino. Constraints on transient neutrino emission were set in all the mentioned analyses. The return of experience is particularly important for the design of the alert system of KM3NeT \cite{KM3NeT:LoI}, the next generation neutrino telescope in the Mediterranean Sea.

\subsection{Real time response to external alerts}
Neutrino telescopes are well-designed to detect neutrinos emitted by transient astrophysical phenomena by constantly monitoring a very large portion of the sky. 
In the ANTARES online analysis framework, a dedicated real-time pipeline \cite{ANTARES:103} was developed to look for neutrino events in both temporal and spatial coincidence with transient events announced by public alerts distributed through the GCN system (\S \ref{sec:DAQ}) or by private alerts.
For interesting cases, the most optimized offline analyses, using the calibrated knowledge of the detector, were  performed to improve the online search.

An alert message is sent publicly via the GCN within a few tens of seconds once a GRB is detected by a $\gamma$-ray satellite. The ANTARES online system reacted in real time to bursts arising with directions below the detector horizon. 
In this case, a dedicated search  for neutrino-induced muons within a time window $[-250; +750]$ s around the detection time and in a cone centered on the GRB position was performed. No events were found and the results for neutrino correlation with GRBs (also after more refined off-line analyses) are described in \S \ref{sec:aGRB}.

Since 2016, the IceCube collaboration has been sending public triggers for interesting neutrino candidates. The events are received by the \textit{Astrophysical Multi-messenger Observatory Network} (AMON \cite{MMAMON:13}) and distributed to the community via an alert of the GCN.
ANTARES followed 37 of these alerts (out of 115 received) with a position on the sky below its horizon.  

Given the importance of some of the IceCube alerts, dedicated offline analyses were performed for IC170922A (see \S \ref{sec:apstx}), and IC191001A/ IC200530A.
The IC191001A event was followed, after a few hours, by the observation of the tidal disruption event (TDE) AT2019dsg,  by the \textit{Zwicky Transient Facility} (ZTF), and the TDE was indicated as a likely counterpart of the IceCube event. 
Motivated by this possible association, the follow-up campaign of the IceCube alerts by ZTF suggested a second TDE, AT2019fdr, as a promising counterpart of the IceCube candidate IC200530A. These intriguing associations were followed-up by searching for neutrinos in the ANTARES detector from the directions of AT2019dsg and AT2019fdr using a time-integrated approach. No significant evidence for space clustering of neutrino candidates close to both directions, was found \cite{ANTARES:93}.

Since mid 2019,  the HAWC ground-based $\gamma$-ray observatory   \cite{HAWC:17} has been issuing alerts of TeV transients lasting from 0.2 s to 100 s, mainly targeting for  GRBs. As ANTARES, also HAWC is able to monitor half the sky with a high duty cycle. The quest for TeV gamma rays produced by transient astrophysical sources is particularly interesting for high-energy neutrino telescopes. The alerts are channeled via the AMON framework and then distributed by the GCN. ANTARES followed 7 out of 22 HAWC alerts (up to the end of data taking) without any coincident neutrino. 
A dedicated search for coincidences in HAWC and ANTARES events that were below the threshold for sending public alerts in each individual detector was additionally performed for data  collected between 2015 and 2020 \cite{ANTARES:100}. 
During this time period, three coincident events with an estimated false-alarm rate of $<1$ coincidence per year were found, a number which is consistent with background expectations.

The results of the follow-up performed by the ANTARES telescope between January 2014 and February 2022, which corresponds to the end of the data taking, are summarized in \cite{ANTARES:103}. The searches triggered by ANTARES alerts are described in \cite{ANTARES:107}.
Although the online analysis found no coincidences, this effort has highlighted the multi-messenger program of the ANTARES neutrino telescope to a wide community and provided constraints on neutrino fluence from different astrophysical objects.

\subsection{Neutrinos and  UHECRs} 
The origin of ultra-high-energy cosmic rays (UHECRs) is an unsolved question of high-energy astrophysics \cite{MS:18}. 
UHECRs have energies $E>10^{18}$ eV (1 EeV), are not confined by galactic magnetic fields, and are likely of extragalactic origin. 
This assumption is confirmed by the Pierre Auger Observatory report of large-scale anisotropies in the arrival directions of UHECRs above 8 EeV with the excess flux directed from outside of our Galaxy \cite{PAOani:17}.

The Universe is opaque to $\gamma$-rays whenever the energy-dependent photon mean free path is smaller than the distance of the source.
The dominant process for the absorption of very high-energy photons of energy $E$ produced by astrophysical sources is pair-creation 
\begin{equation}\label{eq:9-ebl1}
{\mathop{\gamma}\nolimits_{E}} + {\mathop{\gamma}\nolimits_{\epsilon}}
\rightarrow e^{+}e^{-}
\end{equation}
on low-energy extragalactic background photons of energy $\epsilon$.
These photons extend from the Cosmic Microwave Background (CMB) to the near-ultraviolet (UV) wavelengths. 
As a result, galactic sources of photons with energy of $\gtrsim$100\,TeV  start being attenuated by the presence of the background photons. Photons from the nearby Universe (below some tens of Mpc) start to be attenuated above $\sim$10\,TeV by the presence of the CMB. The flux of $\gamma$-rays above 1 TeV is totally suppressed for distances larger than $\sim$1 Gpc.
On the contrary, the low-interacting neutrinos can propagate through the Universe without being significantly absorbed. 
They are expected to be produced at the acceleration site or during propagation, and are thus direct tracers of hadronic interactions of CRs as they are not deflected by magnetic fields. 

The direction of charged cosmic rays is modified by the presence of magnetic fields on their way from the source to Earth, a process that alters the possibility of association with the sources. 
The deflection of CRs under the presence of magnetic fields increases with increasing atomic number $Z$, which is not well determined at energies  above $\sim~10^{15}$ eV. 
In fact, the observable quantities on \textit{indirect} detection technique of extensive air shower array experiments depend marginally on the mass  $A$ or atomic $Z$ numbers of the incoming CR.
Nevertheless, UHECRs at the highest energies are deflected the least due to their extremely high magnetic rigidity \cite{UHECRs:19}, which makes them the most suitable for directional correlation searches.

One possible approach to disclosing the extragalactic accelerators is to correlate UHECRs with high-energy neutrinos. 
A joint effort between the ANTARES-IceCube-Pierre Auger-Telescope Array collaborations \cite{ANTARES:98} produced three different approaches for correlating the possible arrival directions of neutrinos with the arrival directions of UHECRs. 
Cosmic rays with energies above $\sim$50 EeV were selected in the Pierre Auger Observatory and in the Telescope Array data sets. For both the UHECR and neutrino data sets, the combination of the observatories provided full sky coverage.

The first analysis used  the measured UHECR directions, as well as basic magnetic deflection estimates, to identify regions where the high-statistics sample of $\nu_\mu$ CC interaction events can cluster. 
The neutrino samples used by IceCube and ANTARES are described in \cite{ICpev:13} and \cite{ANTARES:72}, respectively.
The second analysis searched for an excess of UHECRs in the direction of the highest-energy neutrinos. 
Finally, the third analysis searched for an excess of pairs of UHECRs and highest-energy selected neutrinos on different angular scales. 
All analyses found a result compatible with the assumed background hypotheses of either an isotropic neutrino flux or an isotropic UHECR flux.

\subsection{Neutrinos and gravitational waves\label{sec:GW}} 
\begin{SCfigure}
  \centering
\includegraphics[width=0.6\textwidth]%
    {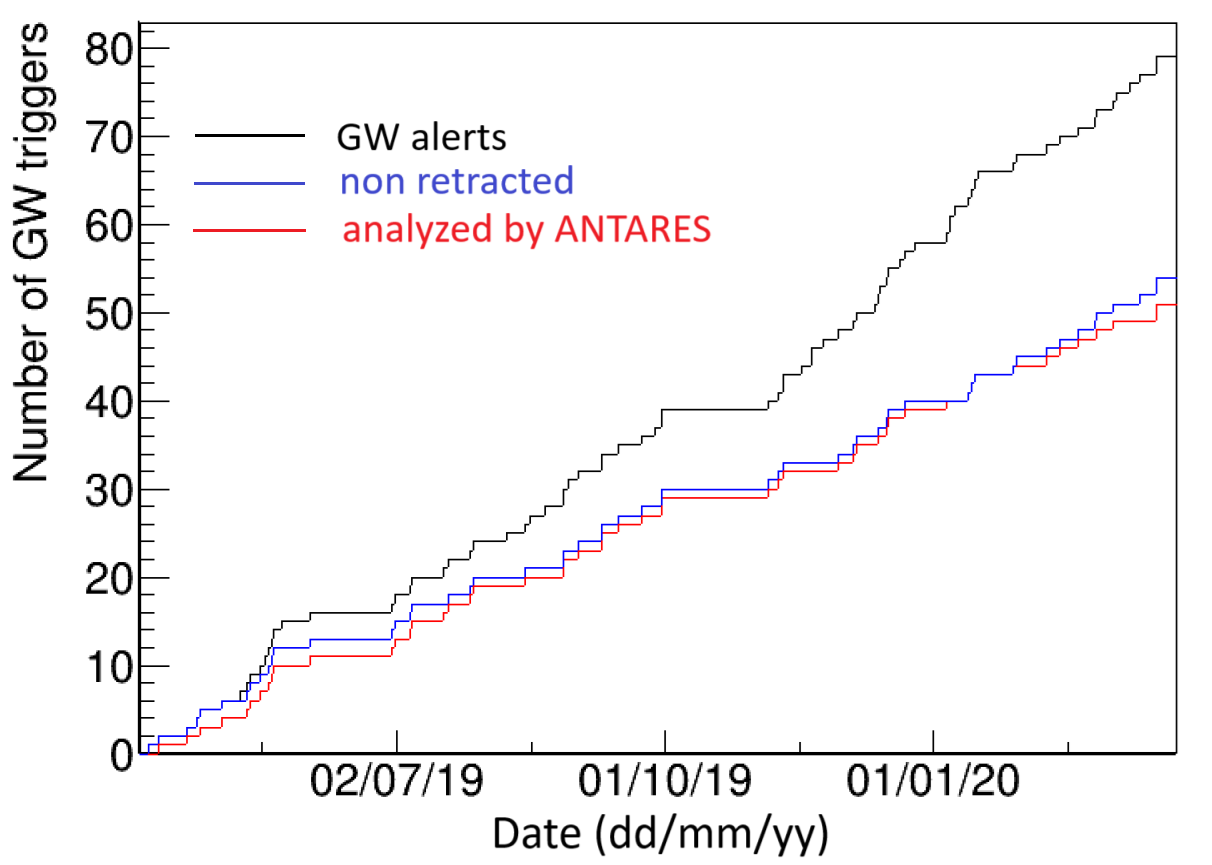}
  \caption{\small {Cumulative number of the GW candidates detected during the O3 run period as a function of the date \cite{ANTARES:103}. The black histogram refers to all GW triggers; the blue one to not terrestrial, nor retracted at the time of the analysis; the red histogram those analyzed by ANTARES.}}
\label{fig:GWt}
\end{SCfigure}

The  observation of gravitational waves (GWs) was initiated in 2015 by Advanced LIGO \cite{LIGO:14} and Advanced Virgo \cite{Virgo:15} and represented a breakthrough in physics and astrophysics.
Since then, we have entered the era of regular GW observations. 
The 11 confident detections of GW events from LIGO/Virgo collaboration (LVC) during O1 and O2 observing runs were published in the catalog GWTC-1 \cite{LVCgwtc1}. 
The final candidate GW events from the initial O3 period (O3a) were published in the GWTC-2.1 catalog, containing 44 high-significance events \cite{LVCgwtc2.1}.
The 35 LVC  high-significance  events from O3b were published in GWTC-3 \cite{LVCgwtc3}.
The early GW observations motivated the extension of data taking of the ANTARES detector from 2016 until 2022, see \S \ref{sec:ANTAintro}.
Fig. \ref{fig:GWt} presents the cumulative number of the GW candidates in O3 in which a neutrino counterpart in the ANTARES detector was searched for.

For each event in the catalogs, the LVC provides a Flexible Image Transport System (\textit{FITS}) file\footnote{\url{https://fits.gsfc.nasa.gov/}},   with the timing of the merger ($t_{\rm GW}$), the constraints on the source direction ($\Omega$) as a sky map $P(\Omega)$, as well as posterior samples containing all the correlations between source direction $\Omega$, luminosity distance estimate ($D_L$),  masses of the two merging objects ($m_1,m_2$) with the convention $m_1 > m_2$, the energy radiated ($E_{\rm GW}$), defined as the difference between the estimated mass of the final object and the sum of the masses of the initial objects, and the inclination between the total angular momentum and the line-of-sight ($\theta_{jn}$). 
The classification among the different categories is made based on the mass estimates:  binary mergers of two neutron stars  (BNS) if $m_2 < m_1 < 3M_\odot$; of a neutron star and a black hole (NSBH) if $m_2 < 3M_\odot < m_1$, binary mergers of black holes (BBH) otherwise.

Binary  BBH, BNS, and NSBH mergers could be also possible sites of neutrino production. 
Neutrinos in the TeV--PeV energy range (HE$\nu$) can be secondary particles arising in relativistic outflows resulting from these compact object mergers.
The expected neutrino flux from a jet can vary depending on the jet angle with respect to the observer and is expected to be significantly higher than the flux from an isotropic emission. 
Lower-energy (GeV) neutrinos can additionally be produced by accelerated proton scattering in the dense environment.
The detection of GWs and HE$\nu$ from common sources would be of fundamental importance to establish the connection between the dynamics of the progenitor and the properties of the outflow.
A possible joint HE$\nu$--GW detection could also be used in targeted electromagnetic follow-up observations, given the significantly better angular resolution of neutrino events compared to gravitational waves.

The possibility to correlate HE$\nu$ and GWs mainly relies on two quantities: the region of the GW signal and the timescale for observing neutrinos from binary mergers.
The region containing 90\% of the source localization can be built directly from the GW sky map $P(\Omega)$ and it is usually called ${\cal R}_{90}$. The ${\cal R}_{90}$ size  is extremely variable (depends on the position on the sky and the number of interferometers that observed the signal) ranging from few tens of deg$^2$ up to $\sim~10^4$ deg$^2$. 
The time reference scale for observing neutrinos from binary mergers is estimated to be $(t_{\nu} - t_{\rm GW}) \sim~\pm 500$ s, as derived from models yielding the evolution of neutrino emission from GRBs \cite{BARET:11}. 
Other models predict longer timescales for the neutrino emission, in particular from BNS and NSBH mergers \cite{Fang:17}.

A first HE$\nu$--GW search using ANTARES data was performed before the first  observation looking for possible GW  bursts associated with high energy neutrinos \cite{ANTARES:30}.
The discovery of the first GW transient, GW150914 observed on September 14, 2015 triggered a joint search from data recorded by the IceCube and ANTARES  detectors \cite{ANTARES:49}. No neutrino candidates were found in both temporal and spatial coincidence with the GW event within $\pm 500$ s. 
Similar results were jointly obtained by ANTARES, IceCube and the LVC for other BBH mergers reported during O1 \cite{ANTARES:61}.

Associated HE$\nu$-GW  emission from astrophysical transients with minimal assumptions were also searched for using data collected by Advanced LIGO during O1 run and ANTARES/IceCube telescopes \cite{ANTARES:77}. This study focused on sub-threshold candidate events (not entering in the catalog of firmly established GWs) whose astrophysical origins could not be determined from a single messenger. No significant coincident candidate was found and the rate density of astrophysical sources dependent on their gravitational-wave and neutrino emission processes was set.

Using ANTARES data, the quest for a neutrino counterpart was done for all GW coalescences of compact object (about 80) reported by the LVC during the observing runs O2 \cite{ANTARES:85} and O3 \cite{ANTARES:101}.
In both cases, the search was conducted using upward-going track events produced by $\nu_\mu$ CC interactions and showering events induced by interactions of neutrinos of any flavor. In addition, the severe spatial and time coincidence provided by the GW alert allowed to probe regions above the detector horizon, extending the ANTARES sensitivity over the entire sky for all $\nu$ flavors with $ E_\nu\gtrsim 100$ GeV. 
The searches for prompt neutrino emission within $\pm 500$ s around the GW time and a reconstructed direction compatible with the ${\cal R}_{90}$ localization yielded no neutrino counterpart. 
Then, using the information from the GW catalogs and assuming isotropic emission, upper limits on the total energy $E^{\rm tot}_\nu$ emitted as neutrinos of all flavours and on the ratio $f_\nu = E^{\rm tot}_\nu/E_{\rm GW}$ were computed. 
In \cite{ANTARES:101}, the stacked analysis of all the 72 BBH mergers and of 7 NSBH merger candidates was performed to constrain the typical neutrino emission within these populations, leading to the limits: 
$E^{\rm tot}_\nu < 4.0 \times 10^{53}$ erg and $f_\nu< 0.15$ for BBH and 
$E^{\rm tot}_\nu < 3.2 \times 10^{53}$ erg and $f_\nu< 0.88$ for NSBH, assuming isotropic neutrino emission with $E_\nu^{-2}$ spectrum. Other assumptions including softer spectra and non-isotropic scenarios where the neutrino emission can be constrained down to opening angles as small as  $10^\circ-30^\circ$
were also tested.
Similar analyses performed by the IceCube collaboration  produced null results for both the HE$\nu$ \cite{ICgw:23} and for the sub-TeV neutrino sample detected using the DeepCore infill \cite{ICgwlow:23}.

\begin{figure}
  \centering
\includegraphics[width=0.75\textwidth]%
    {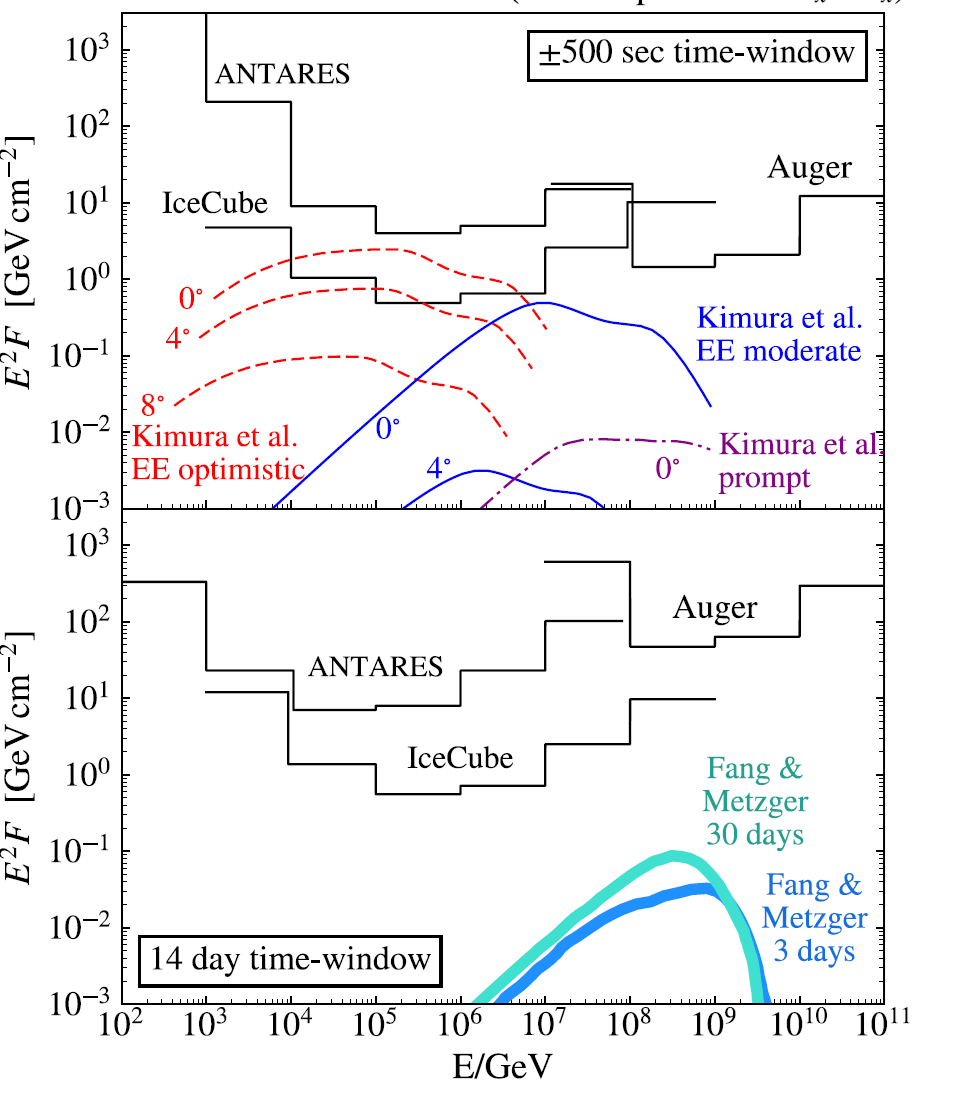}
  \caption{\small   90\% CL upper limits on the $\nu$ spectral fluence from GW170817 during a $\pm 500$ s window centered on the GW trigger time (top), and a 14 days window after the GW (bottom). For each detector, limits are calculated separately for each energy decade. The predictions for HE$\nu$ emission models are from   \cite{Kimura:17} (upper) scaled to a distance of the merger $D_L=40$ Mpc and shown for the case of on-axis and selected off-axis angles. GW data and the redshift of the host galaxy constrain the viewing angle to $\sim~30^\circ$, justifying the non-observation. In the lower plot, models from \cite{Fang:17} are scaled to a distance of 40 Mpc.  Plot from \cite{ANTARES:68}.}
\label{fig:GWnu}
\end{figure}
On  August 17, 2017 the first BNS coalescence candidate (GW170817)  was observed through gravitational waves.
The GBM detector onboard Fermi satellite independently observed a $\gamma$-ray burst (GRB 170817A) with a time delay of $\sim$1.7 s with respect to the GW merger time. From the GW signal, the source was initially localized in a sky region of 31 deg$^2$ at a $D_L=40\pm 8$ Mpc and with component masses measured in the range 0.86 to 2.26 $M_\odot$. 
This joint GW/GRB detection reported through GCN was followed by the most extensive worldwide observational campaign ever performed, with the use of space- and ground-based telescopes, to scan the region of the sky where the event was detected. The results were reported in a paper signed by almost 70 collaborations and 3614 co-authors \cite{ANTARES:67}.
The extensive observing campaign led to the discovery (independently by multiple teams) of a bright optical transient in the NGC 4993 Galaxy (at $\sim$40 Mpc) less than 11 hours after the merger. Subsequent observations targeted the object and its environment. Early ultraviolet observations revealed a blue transient that faded within 48 hours. Optical and infrared observations showed a redward evolution over about 10 days. Following early non-detections, X-ray and radio emission were discovered at the transient’s position 9 and 16 days, respectively, after the merger.
Both the X-ray and radio emission arise from a physical process that is distinct from the one that generates the UV/optical/near-infrared emission which are consistent with being powered by the radioactive decay of $r$-process nuclei synthesized in the merger ejecta, a kilonova \cite{Met:17}. 

The ANTARES and IceCube detectors (together with the Pierre Auger Observatory) participated searching for ultra-high-energy neutrinos.
No neutrino directionally coincident with the source was detected within $\pm 500$ s around the merger time. Additionally, no MeV neutrino burst signal was detected coincident with the merger. An extended search in the direction of the source for high-energy neutrinos within the 14-day period following the merger was further carried out \cite{ANTARES:68},  finding no evidence of emission. The obtained limits are shown in Fig. \ref{fig:GWnu}. These results were used to probe dissipation mechanisms in relativistic outflows driven by the binary neutron star merger.

All these observations were in agreement with the hypothesis that GW170817 was produced by a BNS merger in NGC4993 yielding the short GRB 170817A and a kilonova/macronova powered by the radioactive decay of $r$-process nuclei synthesized in the ejecta. The non-detection of neutrinos was consistent with model predictions of short GRBs observed at a large ($\sim 30^\circ$) off-axis angle.

\section{Legacy towards future water experiments\label{sec:legacy}}

The ANTARES detector has been operated successfully for more than a decade and a half. All challenges related to the operation in the deep sea have been successfully addressed by the collaboration. Deployment and connection operations have become smoother over time, and the constant re-calibration of the detector due to the variable environmental conditions was fully automated. A wealth of physics results has been obtained, despite the relative modest size of the  detector,  confirming the need and feasibility for several high-energy neutrino detectors around the world.
The interest for the  study of cosmic neutrinos with detectors in the deep sea has further increased in recent years, after the discovery of the diffuse cosmic neutrino flux by the IceCube collaboration and the possibility of wider multi-messenger studies following the observation of GWs.

At the end of the dismantling operations in June 2022, \S \ref{sec:construction}, all the 885 optical modules (OM) were recovered. Some of them were not  functioning anymore (the failure rate was on average one OM per month).  A few of these have been distributed to different institutions of the collaboration for exposition purposes. Most of the OM failures were related to the active base or motherboard of the offshore electronics, and not to the phototubes.
Many components of the ANTARES detector have been reused after the dismantling. The anchors and buoys are now in use by our partners from sea science, the various titanium elements have been recycled to industry and there are several future projects which have expressed interest in reusing either the entire ANTARES optical modules or the 10$^{\prime\prime}$ PMTs, most of which are still in excellent shape.

The ANTARES collaboration was one of the founders of the  Global Neutrino Network (GNN)\footnote{\url{https://www.globalneutrinonetwork.org/}}. 
The GNN aims for a closer collaboration and a coherent strategy among the neutrino telescope projects. It serves as a forum to formalize and further develop  meetings among collaborations and organize an international workshop on Very Large Volume Neutrino Telescopes (VLVNT).  
The goals of GNN include the coordination of alert and multi-messenger policies, exchange and mutual checks of software,  establishing a common legacy of public documents, cross-checks of results with different systematics, the organization of schools, and  other forms of exchanging expertise.

\begin{figure}[tbh]
  \centering
\includegraphics[width=0.75\textwidth]%
    {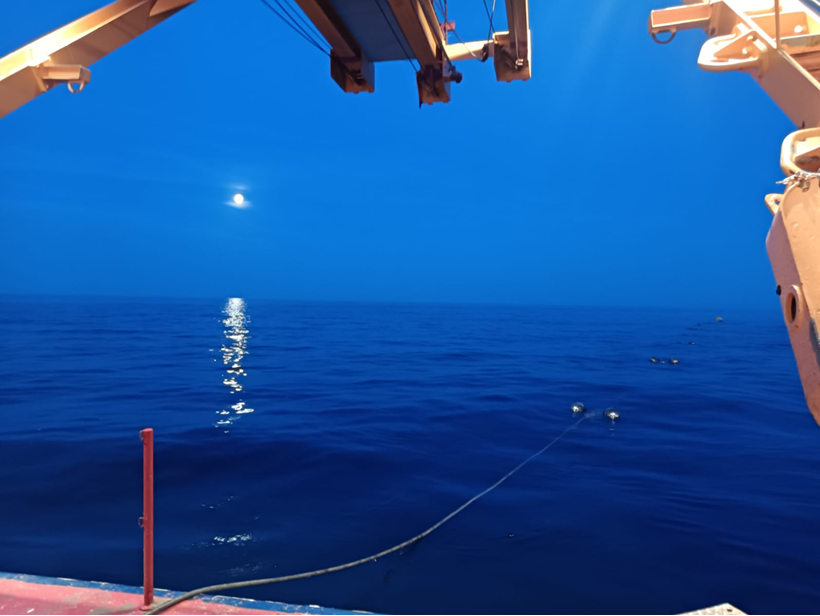}
  \caption{\small   The recovery of one of the last ANTARES detector line on May 2022.}
\label{fig:sunset}
\end{figure}

The most advanced  project for the construction of a larger detector is KM3NeT in the Mediterranean Sea \cite{KM3NeT:LoI}. At the end of 2021 the photo-cathode area of the installed KM3NeT optical modules superseded that of the ANTARES one, and this motivated the end of ANTARES operations, Fig. \ref{fig:sunset}.
The KM3NeT collaboration benefits of the large experience acquired with the ANTARES detector, as a significant part of the ANTARES collaboration is now involved in the  construction, operation, and data analysis of KM3NeT.

The KM3NeT infrastructure is based on a phased and distributed implementation which maximizes the access to regional funds, the availability of human resources and the synergistic opportunities for the Earth and sea sciences communities. 
The infrastructure will consist of three so-called \textit{building blocks} in two deep-sea selected sites. A building block comprises 115 detection lines, each with 18 optical modules and each optical module comprises 31  small PMTs of 3$^{\prime\prime}$ each.  As such, a single optical module of KM3NeT offers a similar amount of photocathode area as the triplet of OMs in the ANTARES storey (inset of Fig. \ref{fig:ANTARES}). 
The first selected site is offshore Toulon (France), close to the ANTARES location. Here, the deployment of strings for one building block (KM3NeT/ORCA) with an instrumented mass of about 7 Mtons is in progress.
KM3NeT/ORCA corresponds to a geometrical configuration optimized for studies of neutrino properties through the observation of atmospheric neutrino oscillations in the energy range between a few GeV and a few hundred GeV. 
The second site is Capo Passero (Sicily, Italy) which will consist of two building blocks (KM3NeT/ARCA) in a configuration suited for detecting high-energy cosmic neutrinos.
Its effective volume will go beyond a cubic kilometer. The construction of both KM3NeT detectors should be completed by the end of this decade. 

The enormous potential of the KM3NeT detector was recently demonstrated with the detection of an exceptionally high-energy muon of estimated energy of $120_{-60}^{+110}$ PeV \cite{KM3NeT:25} by KM3NeT/ARCA with 21 detection lines  in operation.
In light of its huge energy and near-horizontal direction, the muon most probably originated from the interaction of a neutrino of even higher energy in the vicinity of the detector. 
The energy of this event is larger by more than one order of magnitude than that of any neutrino ever detected, opening the possibility that the neutrino may have originated in a different cosmic accelerator than the lower-energy neutrinos detected so far by the IceCube detector, or represents the first detection of a cosmogenic neutrino, resulting from the interactions of UHECRs with background photons in the Universe.

The possibility to have a neutrino underwater detector near the Earth equator is an interesting option. Because of the Earth's rotation, it would see the entire neutrino sky within 24 hours. The properties of water and environmental conditions towards  a next generation neutrino telescope have been studied and led to a preliminary design for a detector located in the  South China Sea, at a latitude of about +17$^\circ$ and a depth of $\sim~3500$ m.
The \textit{TRopIcal DEep-sea Neutrino Telescope} (TRIDENT) \cite{Trident:23} plans to employ state-of-the-art technologies such as SiPMs and long waveform readout electronics inside a hybrid optical module. The project considers the construction of an array of 1200 strings, monitoring a volume of 7.5 km$^3$.
Additional proposals for very large area experiments in the Chinese Sea are NEON \cite{NEON:24} and HUNT \cite{HUNT:23}, the latter studying also other possible locations.
An additional initiative towards constructing a multi-km$^3$ neutrino telescope is the \textit{Pacific Ocean Neutrino Experiment} (P-ONE) \cite{PONE:20}. The project is planned to be immersed in the Pacific Ocean using the underwater infrastructure of Ocean Networks Canada. 

All these next generation of neutrino telescopes in the deep sea plan to have an improved angular resolution and sensitivity with respect to the ANTARES detector. 
They will complement the IceCube-Gen2 experiment \cite{ICgen2:21} which plans to construct  an array of about 10,000 optical sensors, embedded within $\sim$8 km$^3$ of ice, having a sensitivity five times greater than that of IceCube. In addition IceCube-Gen2 will include buried radio antennas distributed over an area of more than 400 km$^2$ to enhance the sensitivity to neutrino sources beyond EeV, as well as a surface array targeting air showers.

The ultimate objective of all these efforts is to identify the sources of the not yet resolved diffuse neutrino flux and monitor the entire neutrino sky permanently.
The scientific experience gained by the 16 years of continued operations of the ANTARES experiment is a solid guide for future projects, and available for the whole community.

\vspace{2.0 cm}

\noindent \textbf{{\large Acknowledgments }}
\vspace{0.5 cm}

The ANTARES detector was built, operated and maintained by many scientists, including those inserted in this list of authors according to our internal rules for authorship and others who retired or have left the collaboration.
We warmly acknowledge the contribution of all the former members who dedicated part of the scientific career to such an endeavor adventure. 
In particular, we would remember those who significantly contributed to the development of this experiment and passed away: Luciano Moscoso (APC), Giorgio Giacomelli (Bologna), Carlo de Marzo (Bari), Vlad Popa (Bucarest).


The authors acknowledge the financial support of the funding agencies:
Centre National de la Recherche Scientifique (CNRS), Commissariat \`a
l'\'ener\-gie atomique et aux \'energies alternatives (CEA),
Commission Europ\'eenne (FEDER fund and Marie Curie Program),
LabEx UnivEarthS (ANR-10-LABX-0023 and ANR-18-IDEX-0001),
R\'egion Alsace (contrat CPER), R\'egion Provence-Alpes-C\^ote d'Azur,
D\'e\-par\-tement du Var and Ville de La
Seyne-sur-Mer, France;
Bundesministerium f\"ur Bildung und Forschung
(BMBF), Germany; 
Istituto Nazionale di Fisica Nucleare (INFN), Italy;
Nederlandse organisatie voor Wetenschappelijk Onderzoek (NWO), the Netherlands;
Romanian Ministry of Research, Innovation and Digitalisation (MCID), Romania;
MCIN for PID2021-124591NB-C41, -C42, -C43 and PDC2023-145913-I00 funded by MCIN/AEI/10.\-13039/\-501100011033 and by “ERDF A way of making Europe”, for ASFAE/2022/014 and ASFAE/2022 /023 with funding from the EU NextGenerationEU (PRTR-C17.I01) and Generalitat Valenciana, for Grant AST22-6.2 with funding from Consejer\'{\i}a de Universidad, Investigaci\'on e Innovaci\'on and Gobierno de Espa\~na and European Union - NextGenerationEU, for CSIC-INFRA23013 and for CNS2023-144099, Generalitat Valenciana for CIDEGENT/2018/034, /2019/043, /2020/049, /2021/23, for CIDEIG/2023/20 and for GRISOLIAP/2021/192 and EU for MSC/101025085, Spain;
Ministry of Higher Education, Scientific Research and Innovation, Morocco, and the Arab Fund for Economic and Social Development, Kuwait.
We also acknowledge the technical support of Ifremer, AIM and Foselev Marine
for the sea operation and the CC-IN2P3 for the computing facilities.

\bibliographystyle{unsrt}
\bibliography{bibliography}

\end{document}